\documentclass[letterpaper,11pt]{article}

\usepackage{dynkin-diagrams}
\usepackage{cite}
\usepackage{amsmath}
\usepackage{authblk}
\usepackage{ytableau}
\usepackage[vcentermath]{youngtab}

\usepackage[normalem]{ulem} %defines \sout{word} which strikes out {word}

\usepackage{amssymb}
\usepackage{amsthm}
\usepackage{bbm}
\usepackage{mathrsfs}
\usepackage{xypic}
\usepackage{tikz-cd}
\usepackage{chngcntr}
\pgfkeys{/Dynkin diagram, edge length=.5cm,
fold radius=.6cm,
root-radius=.08cm,
backwards=false,
*/.append style={fill=lightgray},
indefinite edge/.style={
draw=black, fill=white, dotted,
thin}}

\usepackage{hyperref}
\usepackage{color}
\definecolor{darkred}{rgb}{0.65,0.15,0}
\hypersetup{pdfborder={0 0 0},colorlinks=true,urlcolor=darkred,citecolor=blue,linkcolor=darkred,linktocpage=true}

\usepackage{float}

\usepackage{geometry}

\newgeometry{vmargin={35mm}, hmargin={30mm}}

\usepackage{keyval}
\makeatletter
\define@key{setpar}{left}[0pt]{\leftmargin=#1}
\define@key{setpar}{right}[0pt]{\rightmargin=#1}
\define@key{setpar}{both}{\leftmargin=#1\relax\rightmargin=#1}
\makeatother

\makeatletter
\renewcommand*\env@matrix[1][\arraystretch]{%
  \edef\arraystretch{#1}%
  \hskip -\arraycolsep
  \let\@ifnextchar\new@ifnextchar
  \array{*\c@MaxMatrixCols c}}
\makeatother

\def\eg{{\it e.g.}}
\def\ie{{\it i.e.}}
\def\Ie{{\it I.e.}}

\def\EWeight#1#2#3#4{\bigl({}^{\mathstrut}_{#1\mathstrut}{}_{#2\mathstrut}^{#4\mathstrut}{}_{#3\mathstrut}^{\mathstrut}\bigr)}

\def\DWeight#1#2#3{\bigl(\raise2.5pt\hbox{${}_{#1}$}{}^{#2}_{#3}\bigr)}

\def\AAWeight#1#2{\bigl(\raise0pt\hbox{${}^{#1}_{#2}$}\bigr)}

\def\fg{{\mathfrak g}}
\def\fh{{\mathfrak h}}

\def\so{{\mathfrak{so}}}
\def\sl{{\mathfrak{sl}}}

\def\be{\begin{equation}}
\def\ee{\end{equation}}
\def\bea{\begin{eqnarray}}
\def\eea{\end{eqnarray}}
\def\tr {\mathop{\rm Tr}}

\def\nn{\nonumber}
\def\so{\mathfrak{so}}

\def\*{\partial}

\def\RR{{\mathbb R}}
\def\CC{{\mathbb C}}
\def\ZZ{{\mathbb Z}}
\def\NN{{\mathbb N}}

\numberwithin{equation}{section}
\newlength\symlength
\newlength\pluslength
\def\Sym{\mathrm{Sym}}

\makeatletter
\newcommand{\oset}[3][0ex]{%
  \mathrel{\mathop{#3}\limits^{
    \vbox to#1{\kern-3\ex@
    \hbox{$\scriptscriptstyle#2$}\vss}}}}
\makeatother

\def\c{\chi}
\def\d{\delta}

\def\g{\gamma}

\def\l{\lambda}
\def\m{\mu}
\def\n{\nu}

\def\s{\sigma}

\def\J{\Psi}

\def\O{\Omega}

\def\Q{\Theta}
\def\S{\Sigma}
\def\U{\Upsilon}

\newcommand{\sSpin}{\mathsf{Spin}}

\newcommand{\ve}{\varepsilon}

\newcommand{\hf}{\frac12}

\newcommand{\vf}{\varphi}

\newcommand{\bsubeq}{\begin{subequations}}
\newcommand{\esubeq}{\end{subequations}}

\def\double #1{#1{\hbox{\kern-2pt $#1$}}}

\def\tr{{\rm tr}}

\begin{document}

\frenchspacing

\title{{\huge\bf Some remarks on invariants}\vspace{24pt}}

\author[1]{{\Large Martin Cederwall\thanks{{\tt martin.cederwall@chalmers.se}, corresponding author}}}

\author[2]{{\Large Jessica Hutomo\thanks{\tt jessica.hutomo@pd.infn.it}}}

\author[3]{\\\vspace{6pt}\Large Sergei M. Kuzenko\thanks{\tt sergei.kuzenko@uwa.edu.au}}

\author[4,2]{{\Large Kurt Lechner\thanks{\tt kurt.lechner@pd.infn.it}}}

\author[2,4]{{\Large Dmitri P. Sorokin\thanks{\tt dmitri.sorokin@pd.infn.it}}\vspace{12pt}}

\affil[1]{{\small Dept. of Physics, Chalmers University of Technology, 412 96 Gothenburg, Sweden}\vspace{8pt}}
\affil[2]{{\small INFN, Sezione di Padova, Via Marzolo 8, 35131 Padova, Italy}\vspace{8pt}}
\affil[3]{{\small Department of Physics M013, The University of Western Australia,
\par\vspace{-6pt}
35 Stirling Highway,
Perth W.A. 6009, Australia}\vspace{8pt}}
\affil[4]{{\small Dipartimento di Fisica e Astronomia ``Galileo Galilei'', Universit\`a degli Studi di Padova,\par\vspace{-6pt} Via Marzolo 8, 35131 Padova, Italy}%\vspace{18pt}
}

%\date{\today}
\date{\vspace{-5ex}}

\maketitle

\begin{abstract}
\noindent
The demand to know the structure of functionally independent invariants of tensor fields arises in many problems of theoretical and mathematical physics, for instance for the construction of interacting higher-order tensor field actions. In mathematical terms the problem can be formulated as follows.
Given a semi-simple finite-dimensional Lie algebra $\fg$ and a $\fg$-module $V$, one may ask about the structure of the sub-ring of $\fg$-invariants inside the ring freely generated by the module. We point out how some information about the ring of invariants may be obtained by studying an extended Lie algebra. Numerous examples are given, with particular focus on the difficult problem of classifying invariants of a self-dual 5-form in 10 dimensions.
\end{abstract}

\newpage

\tableofcontents

\section{Introduction}

The problem we are dealing with is a very generic one: given a (semi-simple) Lie algebra $\fg$ and a highest weight $\fg$-module $V$, what are the most general polynomials in $V$ which are invariant under $\fg$?
This is the subject of invariant theory, with a long history in mathematics
(see \eg\ refs. \cite{bams/1183548690,Olver,Goodman:1998}).

The question is of course relevant in physics and met regularly by any mathematical physicist; it may be the question of finding possible terms in an action, or many other applications.
In spite of this, the awareness in the physics community of the possible structures of the rings of invariants thus arising is rather low, to our knowledge. In particular, the possibility of having relations among invariants has received very little attention in physics.
A main purpose of the present notes is to try to improve this situation, and also to present a few simple criteria that can help to determine the nature of the ring of invariants.

The concrete question that led to these investigations was the search for independent $\so(1,9)$-invariants of a self-dual 5-form, naturally arising \eg\ in higher-derivative corrections to string effective actions \cite{Paulos:2008tn,Melo:2020amq} and other interacting chiral 4-form models \cite{Buratti:2019guq,Bandos:2020hgy,Avetisyan:2022zza,Hutomo:2025dfx}. More generally, there exist generating formulations for interacting duality-invariant or chiral gauge $p$-form theories in $d = 2p + 2$ space-time dimensions, in which the self-interaction is described by an arbitrary real function of a self-dual $(p+1)$-form \cite{Kuzenko:2019nlm},    \cite{Avetisyan:2022zza,Ferko:2024zth}.\footnote{In the $d=4n$ case, the self-coupling $L_{\rm int}(F_+,F_-)$ is a Lorentz-invariant function of a self-dual $2n$-form $F_+$ and its conjugate $F_-$, $\star F_\pm = \pm i F_\pm$, which is subject to the condition $
L_{\rm int} (e^{i \varphi} F_+, e^{-i \varphi} F_-)  = L_{\rm int} (F_+, F_-) $, with
$\varphi \in {\mathbb R}$, see \cite{Kuzenko:2019nlm} for the technical details. }
To construct the most general self-interactions, one has to address the problem of classifying the functionally independent invariants of a self-dual form. Unfortunately, we will not be able to give a full answer to the question; it turns out to be extremely much more complicated than for self-dual forms in lower dimensions.

In Section \ref{RingSection}, the general framework is formulated.
We discuss the use of partition functions of invariants and their meaning.
With the given data, an extended Lie algebra $\fg^+$ can be defined, and we investigate the implications of the properties of $\fg^+$ for the ring of invariants. In Section \ref{ExamplesSection} numerous examples are given, with a particular focus on cases which we find physically interesting, such as antisymmetric tensor fields. Some more complicated examples are also displayed in full in terms of their partition functions. We discuss how the rings of invariants relate to the findings in Section \ref{RingSection}.
Finally, Section \ref{SDTenSection}
deals with the difficult problem that prompted this investigation, the classification of the invariants of a self-dual 5-form in 10 dimensions. Partial results are derived concerning the partition function (up to order 20) and the tensorial structure of low-order invariants (up to order 10). A list of useful identities to handle self-dual 5-forms and construct invariants thereof are given in Appendices \ref{id5f} and \ref{spin10d-app}.

We hope that our presentation, in particular the criteria given in Sections \ref{WhenDim1Section} and \ref{WhenFreeSection}, and the properties of partition functions following from the Gorenstein property, can be useful for physicists faced with the problem of finding invariants for particular $\fg$ and $V$.

\section{The ring of invariants\label{RingSection}}

\subsection{Generalitites}

Let $\fg$ be a finite-dimensional\footnote{The focus on finite-dimensional $\fg$ is no restriction; tensor products of highest weight representations of infinite-dimenional Kac--Moody algebras never contain the trivial representation.} simple (or semi-simple) Lie algebra and $V=R(\lambda)$ an irreducible %representation
module with integral dominant highest weight\footnote{Meaning that the Dynkin labels of $\lambda$ are natural numbers.} $\lambda$.
Let $G$ denote a Lie group with Lie algebra $\fg$ acting on $V$.
We want to investigate the polynomial expressions in $X\in V$ which are invariant under $\fg$.
Let $S=\Sym^\bullet V$ be the commutative $\NN$-graded ring freely generated by $X\in V$ at level 1. Every level $S_k$ contains a completely reducible $\fg$-module $\vee^k V$, spanned by monomials of degree $k$ in $V$.
Invariants are in the subring $S^\fg$.

We work over the field $\CC$, which can be replaced by $\RR$ whenever a real form of $\fg$ is chosen such that $R(\lambda)$ is real (\eg\ the split real form for any $\lambda$).
We will often denote a module by its highest weight, which then is expressed as Dynkin labels (coefficients in the basis of fundamental weights).

A very common question in physical applications is to find a basis of independent invariants that generate the elements in $S^\fg$. Our interest in it arose from considerations of possible additional (\eg\ higher derivative) terms in an action, like powers of some field strength.

In many simple situations, the answer is straightforward. If $\lambda=\theta$, the highest weight, $S^\fg=U(\fg)^\fg$ is the ring of polynomial invariants
inside the universal enveloping algebra of $\fg$, generated by elementary Casimir invariants, the number of which equals the rank of $\fg$.
The invariants are found by the Harish-Chandra homomorphism.
This ring is freely generated.
Is this statement true in general? \Ie, is $S^\fg$, generated by a number of monomials, free of relations?
Most physicists would probably spontaneously guess an answer in the positive, by some kind of intuition.
However, as is a well known fact in invariant theory, the generic answer is that $S^\fg$ is {\it not} freely generated.
A criterion helping to decide the answer is given in Section \ref{WhenFreeSection}.

It has been shown that all rings of the type we are discussing are Gorenstein\footnote{This and other ring-theoretic concepts can be skipped by readers who are not interested in them; note however the implications explained in Section \ref{PartitionSection}. Gorenstein is for rings what Calabi--Yau is for manifolds; the spaces of invariants are in fact (non-compact) Calabi-Yau varieties.} \cite{HOCHSTER1974115,Khinich_1978}. This can be of concrete help in characterising the ring, see Section \ref{PartitionSection}, and examples.
It also guarantees (through ``Gorenstein $\Rightarrow$ Cohen-Macaulay'') that the following concepts are well defined. We refer the reader to the book \cite{Passman:1991} for an introduction to the concepts of ring theory.

Consider a generic object $X\in V$, \ie, $X$ is in a generic $G$-orbit.
Let its stability group be $H$, with Lie algebra $\fh$.
We can now calculate the dimension  $d$ of the space of invariants
(the closure of the space of generic $G$-orbits) as
$d=\dim V-\dim\fg+\dim\fh$.
This dimension should match the (Krull) dimension of the ring $S^\fg$, the ring being seen as the coordinate ring of the space of invariants. Thus,
\begin{align}
 d=\dim S^\fg=\dim V-\dim\fg+\dim\fh\;.\label{DimensionEq}
\end{align}

Let us denote by $N$ the number of (linearly independent and non-factorizable at each level $k$) generators of the ring $S^\fg$, and let $R$ be the ring {\it freely} generated by them. Then $S^\fg=R/I$ for some ideal generated by the relations in $S^\fg$. (Note that $R$ is not a sub-ring of $S$.)
Only if $S^\fg$ is freely generated is $N=\dim S^\fg$.
If there are polynomial relations between the generators, the number of these need to be subtracted. There may then be further ``syzygies'', relations between relations etc. This is handled by finding a resolution of the ring. Two resolutions are of interest:
\begin{itemize}
\item{The (minimal) Tate resolution\footnote{For physicists, this is the same as a BRST operator enforcing the relevant constraints on the generators, and also introducing the appropriate ghosts to eliminate any reducibility between constraints, etc., so that the BRST cohomology consists precisely of elements in the ring one seeks to describe (see \cite{Barnich:2000zw} for a review).}, a free multiplicative resolution over $\CC$;}
\item{The minimal additive resolution over $R$ (sometimes called a Koszul resolution).}
\end{itemize}
In Section \ref{PartitionSection}, we will explain the use of these in terms of the partition function (Hilbert series) of $S^\fg$.
The Tate resolution, which in the general physics community is understood as BRST, with a differential (BRST operator) whose cohomology is $S^\fg$, is often most convenient for finding the structure of the ring in concrete situations. It however has the ``drawback'' that it generically is infinite, \ie, contains ghosts at arbitrarily high degree (degree is inherited from $S$).
The additive resolution, on the other hand, is always finite; this is the Hilbert syzygy theorem \cite{HilbertAlgForm}. It is expressed as the sequence
\begin{align}
R^{(0)}\longleftarrow R^{(1)}\longleftarrow R^{(2)}\longleftarrow\ldots\longleftarrow R^{(n)}\;,\label{AddRes}
\end{align}
where $R^{(0)}=R$, and the only cohomology is $S^\fg$ inside $R^{(0)}$. All $R^{(p)}$ are tensor products of $R$ with some vector space.
The Cohen-Macaulay property ensures that the {\it depth} $n$ of the resolution
equals the codimension of $S^\fg$ in $R$, \ie,
$\dim S^\fg=N-n$.
The Gorenstein property implies that $R^{(p)}\simeq\bar R^{(n-p)}$.

In Subsection \ref{PartitionSection}, we will explain the use of these notions in terms of the partition function (Hilbert series) of $S^\fg$, and also introduce a minimal additive form obtained from the additive resolution.

\subsection{Partition functions\label{PartitionSection}}

The partition function, or Hilbert series, $P(t)$ of $S^\fg$ is a formal power series in a variable (``fugacity'') $t$, counting the dimension of each of the vector spaces $(S^\fg)_k$ at level $k$ making up $S^\fg$:
\begin{align}\label{PF1}
P(t)=\sum_{k=0}^\infty\dim(S^\fg)_k\, t^k\;.
\end{align}
For rings with some level-preserving action of an algebra, the definition can be refined to lift the coefficients from dimensions to representations, but this is not the case here, since we are counting singlets.

The dimension of the ring is read off from the partition function as the power of its divergence as $t\rightarrow1$. Its behaviour close to $t=1$ is
\begin{align}
P(t)={\frac c{(1-t)^d}}+O\left({\frac 1{(1-t)^{d-1}}}\right)
\end{align}
for some constant $c$,
where $d=\dim S^\fg$.

A multiplicative form of $P(t)$, reflecting the Tate resolution \cite{Tate1957HomologyON}, is
\begin{align}\label{PF2}
P(t)=\prod_{k=1}^\infty(1-t^k)^{-m_k}\;.
\end{align}
Here, $m_k\in\ZZ$ are integers which are positive for even
and negative for odd variables (generators, ghosts, second order ghosts,...)
in the resolution.
 See ref. \cite{Henneaux:1992ig} for an introduction to such resolutions for physicists.

One aspect to bear in mind, which limits the usefulness of the partition function for more ``complicated'' rings, is that there is no guarantee that even and odd variables do not appear at the same level. The partition function will then only know the difference.\footnote{In simple terms, the coefficients of the power series partition function (2.3) are the numbers of the invariants at order $k$ in the symmetric product of the considered group representation module denoted as $dim (S^g)_k$ (these numbers were computed with the help of the LiE program \cite{LiE}). The multiplicative partition function (2.5) is reconstructed in such a way that the coefficients of its Taylor expansion coincide with the coefficients of the partition function (2.3). The Taylor expansion can be computed ``by hand", or e.g. with the use of the Mathematica function ``Series". The powers $m_k$ of (2.5) count the number of polynomially independent invariants at the order $k$, i.e. the difference between all linearly independent $k$-order invariants  and the number of $k$-order invariants that can be constructed by taking the products of the invariants of low orders. For instance, if $n=dim (S^g)_2$ and $dim (S^g)_4$ are the number of the invariants at orders 2 and 4, then $m_4=dim (S^g)_4-n(n+1)/2$, etc.
}

An additive form of $P(t)$, reflecting an additive resolution
(but which requires knowledge of the set of generators) is
\begin{align}
P(t)=Q(t)\sum_{p=0}^n(-1)^p\pi_p(t)
\end{align}
Here, $p$ is the same label as in eq. \eqref{AddRes} and
$Q(t)=\prod_{i=1}^N(1-t^{k_i})^{-1}$ is the partition function of a ring $R$,
which is defined to be {\it freely} generated by the generators of $S^\fg$ at levels $\{k_i\}_{i=1}^N$ (so, any relations are ignored).
The $\pi_p(t)$ are {\it polynomials}, such that the partition function of $R^{(p)}$ in eq. \eqref{AddRes} is $Q(t)\pi_p(t)$.
%the resolution being
%\begin{align}
%R\leftarrow R^{(1)}\leftarrow R^{(2)}\leftarrow\ldots
%\leftarrow R^{(n)}\;.
%\end{align}
% \\ \MC{I tried to reformulate a bit.} {\color{red} Thanks, Martin. This sounds clearer.}

Finally, from either form of the partition function, one can
identify a minimal set of algebraically independent generators, the number of course being equal to $d=\dim S^\fg$, so that the factors $(1-t^k)^{-m_k}$ in the partition function from the {\it dependent} generators cancel against those from odd variables (ghosts) in the Tate resolution, or against factors in the numerator $\sum_{p=0}^n(-1)^p\pi_p(t)$ in the additive resolution.
Let $\{k_i\}_{i=1}^d$ be the degrees of the independent generators.
Independency means that the subring $T$ generated by these
generators is free.
The result is a rational function
\begin{align}
P(t)={\frac {\varrho(t)}{\prod_{i=1}^d(1-t^{k_i})}}\;
\end{align}
with no common factor in numerator and denominator.
Reflecting the Gorenstein property, the numerator $\varrho(t)=1+\ldots+t^m$ is a ``symmetric'' polynomial of some order $m$, \ie, it obeys the relation
$\varrho(t^{-1})=t^{-m}\varrho(t)$, with non-negative integer coefficients.
This minimal form of the partition function reflects the partition function for the freely generated ring $T$ (the denominator) together with degree-shifted copies (modules) of $T$ (the numerator). The numerator can be thought of as the partition function of the $0$-dimensional ring $S^\fg/\langle T\rangle$.

The aim is concrete understanding of $S^\fg$ for concrete choices of $\fg$
and $V$. %, let us illustrate calculations for a few examples already here.
The main method is to use computer-aided calculation of the number of elements in $S^\fg$ to a degree which is high enough to give a clear indication of the form of the partition function. This is often practically possible, but not always (with limited access to processing power and memory).
We have been using the program LiE \cite{LiE}, simply asking it for the number of singlets in a symmetric tensor product. In view of the limitations, it would be of interest to find an even more efficient method; we are unfortunately not aware of one.

%Also: limitations (later example: $A_1$, (7))

%Check C-M later in some example ($A_1$, (8) \cite{Stanley:1978}).

\subsection{The extended Lie algebra}\label{extendedLa}

Given $\fg$ and the highest weight $\fg$-module $V$, one may define an extended simple Kac--Moody Lie algebra $\fg^+$, which in a grading contains $\fg\oplus\CC$ at degree 0, $V$ at degree $-1$ and the conjugate module $\bar V$ at degree $1$.
This means that the Dynkin diagram of $\fg^+$ consists of that for $\fg$ together with one more node connected in a way that reflects the highest weight
%{\color{red}
%$\lambda=\lambda_i\Lambda^i$ %WHAT IS $\Lambda^i$ HERE?}
of $V$ ($V=R(\lambda)$).
%[TO EXPLAIN THIS FORMULA and relate it to the next paragraph]}
The algebra $\fg^+$ is not completely determined by these definitions.
Let us number the simple roots of $\fg$ by $i=1,\ldots,r$, and let the extending node be number 0. Then, the definitions
imply that the components $A_{i0}$ of the Cartan matrix of $\fg^+$ are
$A_{0i}=-\lambda_i$, while the components $A_{0i}$ are undetermined, where $\lambda_i$ are the coefficients of $\lambda$ in the basis of fundamental weights $\Lambda^i$, $\lambda=\lambda_i\Lambda^i$.
For example, starting from $\fg =A_2=\sl(3)$ and $V=(01)$, a 3-dimensional module,
both the extension to $A_3$ (by a root of the same length) and to $B_3$ (by a shorter root) contain $V$ at level $-1$.
We complement the definitions by the following prescriptions:
\begin{itemize}
    \item If there is a choice of $\fg^+$ which is finite-dimensional, choose it. \item If there is more than one finite-dimensional choice for $\fg^+$, choose any (or \eg\ the smallest one).
    \item If there is no finite-dimensional choice of $\fg^+$, but an affine one, choose it.
    \item If none of the above exists, choose any.
\end{itemize}
In what follows, we are only interested in the question whether $\fg^+$ is finite-dimensional or affine.
Kac \cite{KAC1980190} also considers extending the algebra in this way for the study of nilpotent orbits.

In the following two Subsections, we state sufficient criteria for $S^\fg$ to be at most one-dimensional, and/or for it to be freely generated.
Unfortunately, we are not aware of a way to show this using the properties of $\fg^+$, which is somewhat frustrating since this is what the criteria involve. Instead we rely on examining all possible cases.
It should be stressed that the criteria are sufficient but not necessary.

\subsection{When is the dimension 0 or 1?\label{WhenDim1Section}}

By inspecting the cases when $\fg^+$ is finite-dimensional, we arrive at the statement

\underline{\it If $\dim\fg^+<\infty$, $S^\fg$ is either $\{1\}$ or (freely) generated by a single invariant.}

\noindent All such instances are listed in Table \ref{FiniteTable}.
Most cases are straightforward and well known.
Among the less trivial cases are the 3-forms under $\sl(6)$, $\sl(7)$ and $\sl(8)$.
The stability algebras are $\sl(3)\oplus\sl(3)$, $G_2$ and $\sl(3)$, respectively \cite{Hitchin:2000,Midoune}, yielding the dimension 1 according to
eq. \eqref{DimensionEq} in all cases:
$20-35+16=1$, $35-48+14=1$, $56-63+8=1$.
The order of the invariants are then obtained using LiE.
The stability algebra of a generic chiral spinor under $\so(12)$ is $\sl(6)$ \cite{Igusa:1970,Cederwall:2025tmc} and under $\so(14)$ it is $G_2\oplus G_2$ \cite{Igusa:1987}. The dimensions are obtained as
$32-66+35=1$, $64-91+28=1$.

%\begin{center}
\begin{table}
\centering
\renewcommand{\arraystretch}{1.25}
\begin{tabular}{c|c|c|c|c}
$\fg^+$&$\fg$&Dynkin diagram&$\lambda$&order of inv's\\ \hline
$A_{r}$&$A_{r-1}$
    &\dynkin[*/.append style={fill=lightgray}]{A}{*oo.oo}
    &$(10\ldots0)$&---\\ \hline
$B_r$&$B_{r-1}$
    &\dynkin[*/.append style={fill=lightgray}]{B}{*oo.oo}
    &$(10\ldots0)$&2\\
	&$A_{r-1}$
    &\dynkin[*/.append style={fill=lightgray}]{B}{oo.oo*}
    &$(10\ldots0)$&---\\ \hline
$C_r$&$C_{r-1}$
    &\dynkin[*/.append style={fill=lightgray}]{C}{*oo.oo}
    &$(10\ldots0)$&---\\
	&$A_{r-1}$
    &\dynkin[*/.append style={fill=lightgray}]{C}{oo.oo*}
    &$(20\ldots0)$&$r$\\ \hline
    &&&\\[-12pt]
$D_r$&$D_{r-1}$
    &\dynkin[*/.append style={fill=lightgray}]{D}{*oo.ooo}
    &$\DWeight{10\ldots0}00$&2\\[24pt]
	&$A_{r-1}$
    &\dynkin[*/.append style={fill=lightgray}]{D}{ooo.o*o}
    &$(010\ldots0)$&${\frac r 2}$ ($r$ even)\\
	&&&&--- ($r$ odd)\\ \hline
$G_2$&$A_1$
    &\begin{dynkinDiagram}[mark=o]G2
       \dynkinRootMark{*}2
    \end{dynkinDiagram}
    &$(1)$&---\\
	&$A_1$
    &\begin{dynkinDiagram}[mark=o]G2
       \dynkinRootMark{*}1
    \end{dynkinDiagram}
    &$(3)$&4\\ \hline
$F_4$&$C_3$
    &\begin{dynkinDiagram}[mark=o]F4
       \dynkinRootMark{*}1
    \end{dynkinDiagram}
    &$(001)$&4\\
	&$B_3$
    &\begin{dynkinDiagram}[mark=o]F4
       \dynkinRootMark{*}4
    \end{dynkinDiagram}
    &$(001)$&2\\ \hline
&&&\\[-12pt]
$E_6$&$D_5$
    &\begin{dynkinDiagram}[mark=o,backwards]E6
       \dynkinRootMark{*}1
    \end{dynkinDiagram}
    &$\DWeight{000}10$&---\\
	&$A_5$
    &\begin{dynkinDiagram}[mark=o,backwards]E6
       \dynkinRootMark{*}2
    \end{dynkinDiagram}
    &$(00100)$&4\\ \hline
&&&\\[-12pt]
$E_7$&$E_6$
    &\begin{dynkinDiagram}[mark=o,backwards]E7
      \dynkinRootMark{*}7
    \end{dynkinDiagram}
    &$\EWeight{10}0{00}0$&3\\
	&$D_6$
    &\begin{dynkinDiagram}[mark=o,backwards]E7
       \dynkinRootMark{*}1
    \end{dynkinDiagram}
    &$\DWeight{0000}10$&4\\
	&$A_6$
    &\begin{dynkinDiagram}[mark=o,backwards]E7
       \dynkinRootMark{*}2
    \end{dynkinDiagram}
    &$(001000)$&7\\ \hline
&&&\\[-12pt]
$E_8$&$E_7$
    &\begin{dynkinDiagram}[mark=o,backwards]E8
      \dynkinRootMark{*}8
    \end{dynkinDiagram}
    &$\EWeight{100}0{00}0$&4\\
	&$D_7$
    &\begin{dynkinDiagram}[mark=o,backwards]E8
       \dynkinRootMark{*}1
    \end{dynkinDiagram}
    &$\DWeight{00000}10$&8\\
	&$A_7$
    &\begin{dynkinDiagram}[mark=o,backwards]E8
       \dynkinRootMark{*}2
    \end{dynkinDiagram}
    &$(0010000)$&16\\ \hline
\end{tabular}
\caption{ A list of irreps of simple algebras $\fg$ with Dynkin labels associated with the weight $\lambda$ (modulo outer automorphisms) such that the extended algebra $\fg^+$ is finite-dimensional. The extending node is colored gray. The last column lists orders at which one independent invariant appears.
\label{FiniteTable}
}
\end{table}
%\end{center}

\subsection{When is the ring freely generated?\label{WhenFreeSection}}

By inspecting the cases in which $\fg^+$ is an affine Kac--Moody algebra,
listed in Table \ref{AffineTable}, we arrive at the conjecture

\underline{\it If $\fg^+$ is an affine Kac--Moody algebra, $S^\fg$ is freely generated.}

\noindent The table excludes cases in which $\fg^+$ is the untwised affine extension of $\fg$. Then the Casimir invariants are obtained, and $S^\fg\simeq U[\fg]^\fg$, freely generated by the Casimirs.
Some cases are straightforward and correspond to traces of powers of matrices.
In some other cases we can prove the result. For example, the invariants of a spinor under $Spin(16)$ provides an intermediary step in the Harish-Chandra homomorphism for $E_8$ \cite{Cederwall:2007qb}. Analogous statements hold for a 4-form under $SL(8)$ and for an irreducible 4-form $(0001)$ under $Sp(8)$. This explains the orders of invariants being identical to Casimirs of $E_8$, $E_7$ and $E_6$ in Table \ref{AffineTable}.
The stability algebra is trivial for a generic 3-form  under $SL(9)$ \cite{Vinberg1978}, so the dimension of the ring is $84-80=4$. \hbox{In this $\lambda=(00100000)$ case,} with our limited computer power, the LiE program has been able to calculate the orders 12 and 18 of the two lowest invariants.
The remaining two should have degrees at least 24.

%\begin{center}
\begin{table}
\centering
\renewcommand{\arraystretch}{1.35}
\begin{tabular}{c|c|c|c|c}
$\fg^+$&$\fg$&Dynkin diagram&$\lambda$&order of inv's\\ \hline
&&&\\[-12pt]
$A_{2k-1}^{(2)}$&$C_{k-1}$
    &\begin{dynkinDiagram}[mark=o] A[2]{odd}
    \dynkinRootMark{*}1
    \end{dynkinDiagram}
    &$(010\ldots0)$&$2, 3, \ldots, 2k-1$\\[12pt] \hline
&&&\\[-12pt]
$A_{2k-1}^{(2)}$&$D_{k-1}$
    &\begin{dynkinDiagram}[mark=o] A[2]{odd}
    \dynkinRootMark{*}7
    \end{dynkinDiagram}
    &$\DWeight{20\ldots0}00$&
$2,3,\ldots,2(k-1)$\\[12pt] \hline
$A_{2k}^{(2)}$&$B_{k-1}$
    &\begin{dynkinDiagram}[mark=o] A[2]{even}
    \dynkinRootMark{*}6
    \end{dynkinDiagram}
    &$(20\ldots0)$&$2,3,\ldots,2k-1$\\ \hline
$A_2^{(2)}$&$A_1$
    &\begin{dynkinDiagram}[mark=o] A[2]2
    \dynkinRootMark{*}1
    \end{dynkinDiagram}
    &$(4)$&$2, 3$\\ \hline
$D_4^{(3)}$&$A_2$
    &\begin{dynkinDiagram}[mark=o] D[3]4
    \dynkinRootMark{*}2
    \end{dynkinDiagram}
    &$(30)$&$4, 6$\\ \hline
$F_4^{(1)}$&$B_4$
    &\begin{dynkinDiagram}[mark=o] F[1]4
    \dynkinRootMark{*}4
    \end{dynkinDiagram}
    &$(0001)$&2\\ \hline
$E_6^{(2)}$&$F_4$
    &\begin{dynkinDiagram}[mark=o,affine mark=*] E[2]6
    \end{dynkinDiagram}
    &$(0001)$&$2, 3$\\
	&$C_4$
    &\begin{dynkinDiagram}[mark=o] E[2]6
    \dynkinRootMark{*}4
    \end{dynkinDiagram}
    &$(0001)$&$2, 5, 6, 8, 9, 12$\\ \hline
&&&\\[-12pt]
$E_7^{(1)}$&$A_7$
    &\begin{dynkinDiagram}[mark=o] E[1]7
    \dynkinRootMark{*}2
    \end{dynkinDiagram}
    &$(0001000)$&$2, 6, 8, 10, 12, 14, 18$\\ \hline
&&&\\[-12pt]
$E_8^{(1)}$&$A_8$
    &\begin{dynkinDiagram}[mark=o,backwards] E[1]8
    \dynkinRootMark{*}2
    \end{dynkinDiagram}
    &$(00100000)$&$12, 18,\ldots$\\
	&$D_8$
    &\begin{dynkinDiagram}[mark=o,backwards] E[1]8
    \dynkinRootMark{*}1
    \end{dynkinDiagram}
    &$\DWeight{000000}10$&$2, 8, 12, 14, 18, 20, 24, 30$
    \\ \hline
\end{tabular}
\caption{A list of irreps of simple algebras $\fg$ with Dynkin labels associated with the weight $\lambda$ (modulo outer automorphisms of $\fg^+$) such that the extended algebra $\fg^+$ is affine, except the cases where $\fg^+$ is the untwisted affine extension of $\fg$.
Cases where finite-dimensional extensions with identical $\fg$-modules at levels $\pm1$ exist are omitted.
The extending node is colored gray. The last column lists the orders at which single independent invariants appear.
\label{AffineTable}
}
\end{table}
%\end{center}

%   \subsection{Strictly semi-simple $\fg$}

\section{Examples\label{ExamplesSection}}

Many examples refer to entries in Tables \ref{FiniteTable} and \ref{AffineTable}.

\subsection{Modules of $\sl(2)$}\label{sl2}

Take $\fg=A_1$. This is a good testing area, and has been a main subject of
invariant theory for a long time.

We will write the partition functions $P_n(t)$ of $S^\fg$ for $\lambda=(n)$, with fugacity $t$ measuring the level.
We reproduce known results, see \cite{Sylvester:1879,Olver}, with corrections in \cite{Brouwer} (see also \cite{Campoamor:2007}).

When $\lambda=(1)$ (\ie, the fundamental weight), $P(t)=1$. This of course happens when $R(\lambda)$ contains a single $G$-orbit, so that $S=\oplus_{k=0}^\infty R(k\lambda)$ and $S^\fg=\{1\}$.
In this case $\fg^+=A_2$.

For $\lambda=(2)$, $\fg^+=C_2$. We get the ring generated by the quadratic Casimir at level 2, and
$P_2(t)=(1-t^2)^{-1}$.

For $\lambda=(3)$, $\fg^+=G_2$, and there is a single quartic invariant, which is straightforward to construct.
The partition function is $P_3(t)=(1-t^4)^{-1}$.

For $\lambda=(4)$, $\fg^+$ is the twisted affine Kac--Moody algebra $A_2^{(2)}$, and there are two elementary invariants at degrees 2 and 3, freely generating $S^\fg$.
$P_4(t)=(1-t^2)^{-1}(1-t^3)^{-1}$.

At $\lambda=(5)$, something happens. Now, $\fg^+$ is neither finite-dimensional nor affine. We have checked  the total number of invariants up to level 80, and they match the partition function
\begin{align}
P_5(t)=(1 - t^4)^{-1} (1 - t^8)^{-1} (1 - t^{12})^{-1} (1 - t^{18})^{-1} (1 - t^{36})\;\label{P5Tate}
\end{align}
corresponding to generators of $S^\fg$ at levels 4, 8, 12 and 18, and a relation between them at level 36.
These invariants and the relation are explicitly given in Appendix \ref{A15Appendix}.
The form \eqref{P5Tate} reflects the Tate resolution, as described in Section \ref{PartitionSection}, but also the additive resolution, since this is a complete intersection and the depth is 1.
Canceling the common factor $1-t^{18}$ gives the canonical minimal form
\begin{align}
P_5(t)={\frac{1+t^{18}}{(1 - t^4)(1 - t^8)(1 - t^{12})}}\;,\label{P5TateMinimal}
\end{align}
It reflects the fact that this $S^\fg$ has three elementary generators and hence $\dim S^\fg=3$ in accordance with the formula \eqref{DimensionEq} in which $\dim \fh=0$.

Similarly, for $\lambda=(6)$
\begin{align}\label{pf6}
P_6(t)&=(1 - t^2)^{-1} (1 - t^4)^{-1} (1 - t^6)^{-1} (1 - t^{10})^{-1} (1 -
    t^{15})^{-1} (1 - t^{30})\nn\\
    &={1+t^{15}\over(1-t^2)(1-t^4)(1-t^6)(1-t^{10})}\;,
\end{align}
corresponding to generators of $S^\fg$ at levels 2, 4, 6, 10 and 15, and an identity at level 30. Details are given in Appendix
\ref{A16Appendix}.

For higher $\lambda$, the situation is more complicated. The Tate resolution is infinite. For $\lambda=(7)$, up to level 42,
\begin{align}
P_7(t)&=(1 - t^4)^{-1} (1 - t^8)^{-3} (1 - t^{12})^{-6} (1 - t^{14})^{-4} (1 -
    t^{16})^{-2} (1 - t^{18})^{-9}\nn\\
    &\times(1 - t^{20})(1 - t^{22})^{-1} (1 -
    t^{24})^{10} (1 - t^{26})^{25} (1 - t^{28})^{20} (1 - t^{30})^{49} (1 -
    t^{32})^{37} \label{A17Tate}\\
&\times   (1 - t^{34})^{19}  (1 - t^{36})^{27} (1 - t^{38})^{-107} (1 -
    t^{40})^{-130} (1 - t^{42})^{-277}\times\ldots\nn
\end{align}
Note the mixture of negative and positive exponents.
It is known that the partition function is \cite{Brouwer}
\begin{align}
P_7(t)&=(1+t^2)^2(1-t^2+t^4)(1-t^2+t^4-t^6+t^8)\\
&\times{1 - t^6 + 2 t^8 - t^{10} + 5 t^{12} + 2 t^{14} + 6 t^{16} + 2 t^{18} + 5 t^{20} -
 t^{22} + 2 t^{24} - t^{26} + t^{32}\over(1–t^4)(1–t^8)(1–t^{12})^2(1–t^{20})}\;.\nn
\end{align}
Multiplying in the prefactor gives the minimal form with nominator
$1 + 2 t^8 + 4 t^{12} + 4 t^{14} + 5 t^{16} + 9 t^{18} + 6 t^{20} + 9 t^{22} + 8 t^{24} + 9 t^{26} + 6 t^{28} + 9 t^{30} + 5 t^{32} + 4 t^{34} + 4 t^{36} + 2 t^{40} + t^{48}$.
One observation here is that there is a generator at degree 20, which is not seen in the multiplicative form of the partition function. Indeed, there are also relations at degree 20, and eq. \eqref{A17Tate} only counts the difference. Unfortunately, this is a feature, probably quite generic, that limits the power of using the partition function for more complicated rings.

$\lambda=(8)$ has a more elegant structure. Up to level 29,
\begin{align}
P_8(t)&=(1 - t^2)^{-1} (1 - t^3)^{-1} (1 - t^4)^{-1} (1 - t^5)^{-1} (1 - t^6)^{-1} (1 -
    t^7)^{-1} (1 - t^8)^{-1} \nn\\
    &\times(1 - t^9)^{-1} (1 - t^{10})^{-1} (1 - t^{16}) (1 -
   t^{17}) (1 - t^{18}) (1 - t^{19}) (1 -
   t^{20}) \\
   &\times(1 - t^{25})^{-1} (1 - t^{26})^{-1} (1 - t^{27})^{-1} (1 - t^{28})^{-1} (1 -
    t^{29})^{-1}\times\ldots\nn\\
&={1+t^8+t^9+t^{10}+t^{18}\over(1-t^2)(1-t^3)(1-t^4)(1-t^5)(1-t^6)(1-t^7)}\;.
\nn
\end{align}
There are generators at degrees 2, 3, 4, 5, 6, 7, 8, 9 and 10.
Knowing that $\dim S^\fg=6$, the Cohen--Macaulay property implies that the additive resolution has depth 3. Indeed, this is seen in the form of the partition function reflecting the additive resolution, which is
\begin{align}
P_8(t)=
{1-(t^{16}+t^{17}+t^{18}+t^{19}+t^{20})+(t^{25}+t^{26}+t^{27}+t^{28}+t^{29})-t^{45}\over(1-t^2)(1-t^3)(1-t^4)(1-t^5)(1-t^6)(1-t^7)(1-t^8)(1-t^9)(1-t^{10})}\;.
\end{align}

\subsection{Forms under $\sl(n)$}

One needs only to consider form degree $\leq{n\over2}$.

1-forms have no invariants. The extended algebra is $\sl(n+1)$.

For 2-forms, the extended algebra is $\so(2n)$, and there is a single invariant (the Pfaffian) at degree ${n\over2}$ when $n$ is even.

3-forms under $\sl(6)$, $\sl(7)$ and $\sl(8)$ give the extended algebras $E_6$, $E_7$ and $E_8$. There is a single invariant at degree 4, 7 and 16, respectively.

3-forms under $\sl(9)$ have 4 independent invariants, the first two occurring at degrees 12 and 18, and we conjecture that $S^\fg$, whose dimension is 4, is freely generated. The extended algebra is $E_9=E_8^{(1)}$, which is affine.

4-forms under $\sl(8)$ have 7 independent invariants, and $S^\fg$ is freely generated. $\fg^+=E_7^{(1)}$, and the generators appear at the same degrees as the $E_7$ Casimirs.

In other cases, none of our criteria are applicable, and there is reason to expect rings with complicated structures. Computer-aided calculations are difficult in many cases, since a $p$-form under $\sl(n)$ only can have invariants at degrees $m$ such that $mp=kn$ for some $k\in\NN$.

\subsection{Forms under $\so(n)$}

We need only consider form degree $\leq{n\over2}$.

1-forms (vectors) have a single invariant, the norm squared. The extended algebra is $\so(n+2)$.

2-forms span the adjoint, so $\fg^+$ is the untwisted affine extension of $\fg$, and the invariants are freely generated by Casimirs.

For 3-forms under $\so(7)$, $\fg^+$ is already neither finite-dimensional nor affine.
The stability group is trivial, which means that the dimension of $S^\fg$ is $35-21=14$. The beginning of the multiplicative form of the partition function is
\begin{align}
    P(t)&=(1 - t^2)^{-1} (1 - t^3)^{-1} (1 - t^4)^{-2} (1 -
   t^5)^{-1} (1 - t^6)^{-2} (1 - t^7)^{-2} \nn\\
   &\times
   (1 - t^8)^{-3} (1 - t^9)^{-1} (1 -
    t^{10})^{-2} (1 - t^{11})^{-1} (1 - t^{12})^{-3} (1 - t^{13})^{-1}
     \nn\\
    &\times(1 - t^{14})^{-2}(1 - t^{15})^{-1} (1 - t^{16})^{-3} (1 - t^{17})^{-1} (1 - t^{18})^{-2} (1 -
   t^{19})^{-1}
   \\
   &\times(1 - t^{20})^{-3} (1 - t^{21})^{-2} (1 - t^{22})^{-2} (1 - t^{23})^{-1} (1 -
   t^{24})^{-1} (1 -
   t^{25})^{-1}
   \nn\\
   &\times(1 - t^{26}) (1 - t^{27})^2 (1 - t^{28})^6 (1 - t^{29})^6 (1 -
    t^{30})^{13} (1 - t^{31})^{13}
    \nn\\
   &\times(1 - t^{32})^{25} (1 - t^{33})^{23}\times\ldots\nn
\end{align}
This shows that $S^\fg$ is not freely generated, but does not allow for guessing the precise content of linearly independent generators, except that they {\it at least} (due to possible overlap with relations) correspond to the initial factors with negative exponents, so their number is at least the sum of these negative exponents, i.e 40.

3-forms, %$\lambda=(00100)$
$\lambda=\DWeight{001}00$, under $\so(10)$ appearing as NS and RR 3-form field strengths in D=10 supergravities. The dimension of the ring is $75=120-45$ and the partition function up to order 18 is
\begin{align}
P(t)&=1+t^2+3t^4+9t^6+33\,t^8+121\,t^{10}+524\,t^{12}+2496\,t^{14}+13006\,t^{16}+70909\,t^{18}\ldots\nonumber\\
&=(1-t^2)^{-1}(1-t^4)^{-2}(1-t^6)^{-6}(1-t^8)^{-21}(1-t^{10})^{-76}\nn\\
&\times (1-t^{12})^{-336}(1-t^{14})^{-1676}(1-t^{16})^{-9041}(1-t^{18})^{-50379}\ldots\nn
\end{align}
Already by order 10 the total number of generators exceeds the dimension of the ring, while relations between them have not shown up yet.

\subsection{Self-dual forms under $\so(4)$, $\so(6)$, $\so(8)$ and $\so(10)$}

Self-dual forms under  $\so(4)\simeq\sl(2)\oplus\sl(2)$ are in the adjoint of $\sl(2)$, and there is a single (Casimir) invariant.

Self-dual forms under $\so(6)$ have a single quartic invariant. In Table \ref{FiniteTable} they occur as $(200)$ under $A_3$ ($r=4$).
$\fg^+=C_4$.

Self-dual 4-forms, $\lambda=\DWeight{00}20$, under $\so(8)$ have 35 components and are equivalent via triality to traceless symmetric $8\times 8$ matrices $M$. The dimension of $\so(8)$ is 28 and there are $7=35-28$ independent invariants $\tr M^p$, $p=2,\ldots8$. The ring $S^\fg$ is freely generated. $\fg^+$ is the twisted affine algebra $A_9^{(2)}$.

%{\color{red} In applications to duality-invariant theories of a field-strength $F_4=dA_3$ in $D=8$ space-time of Lorentz signature (see e.g. \cite{Buratti:2019cbm}), one is interested in the number of independent {\it real} invariants that one can construct from a complex self-dual $F^+_4=F_4+iF_4^*=i(F^+_4)^*$ and its complex conjugate anti-self-dual $F^-_4$. This number is equal to the dimension of the ring of $\lambda=(0020)^++(0002)^-$, with $\pm$ being the $U(1)$ charges. In this case there is no stability subalgebra and the dimension of the ring is 41=70-28-1, where 70=35+35 is the dimension of the module and 1 stands for the reality condition. The ring is not freely generated.....}

For self-dual forms under $\so(10)$ and higher, $\fg^+$ is neither finite-dimensional nor affine, and an explicit structure of the rings of these invariants is presently out of reach. We will  discuss the case of a $D=10$ self-dual 5-form further in Section \ref{SDTenSection}.

\subsection{Complex self-dual forms in $D$-dimensional spaces}

Only $D$-dimensional spaces with metric signatures $(t,s)$   fulfilling $s-t\in4\ZZ$ allow for real self-dual $(D/2)$-forms. In other cases, when $\star^2=-1$, one may consider the complex self-dual form $F^+=F+i\star F=i\star F^+$  and its complex conjugate anti-self-dual $F^-$ (constructed from a generic $(D/2)$-form $F$). One can then ask for real invariants of $F^+$ and $F^-$ under $\so(D)\oplus{\mathfrak{u}}(1)$. This falls slightly outside the main scope of the paper, but can be treated with the same methods. Note that such invariants only can exist at even degrees, and that their number at degree $2n$ equals the number of irreducible representations in the symmetric $n$-fold product of a self-dual form.

In $D=4$, there is a single quartic invariant of a  self-dual 2-form and its complex conjugate.

In $D=6$, the ring of real invariants of a  self-dual 3-form and its complex conjugate is freely generated by single independent invariants of degrees 2,4,6,8. The stability group is trivial, and the dimension is  $4=\dim (F_3)- \dim (\so(6)+{\mathfrak{u}}(1))=20-15-1$. The partition function is
\be
P(t)={1\over(1-t^2)(1-t^4)(1-t^6)(1-t^8)}\,.
\ee

It is instructive to compare this case with the structure of the ring of the $\so(6)$ invariants of a generic 3-form $F_3$ without requiring the ${\mathfrak u}(1)$ (duality) invariance. The dimension of the ring is $5=20-15$ and it is freely generated by independent invariants at orders $2,4,4,4,6$.\footnote{Explicit construction of these invariants is given in Appendix \ref{AppendixE}, eqs. \eqref{I4+}-\eqref{CHt}.} The three ones at degree 4 have ${\mathfrak u}(1) $ charges $-4,0,4$ associated with powers of $F^{\pm}_{3}$. Introducing the fugacity $q$ associated with the charge, we get the following partition function
\begin{align}
P(t,q)={1\over(1-t^2)(1-t^4q^{-4})(1-t^4)(1-t^4q^4)(1-t^6)}\,.
\end{align}
The generators at $t^4q^{\pm4}$ go away when we demand $U(1)$-invariance.
There are many elements at degree 8 in the ring of the $\so(6)$-invariants, and fewer in the ring of $(\so(6)\oplus{\mathfrak u}(1))$-invariants. However, the former has no {\it independent generators} at degree 8. The latter does have, because there is one element obtained as the product of the generators at $t^4q^{-4}$ and $t^4q^4$ that is composite in the ring of $\so(6)$-invariants but not in the ring of $(\so(6)\oplus{\mathfrak u}(1))$-invariants.

In $D=8$ space-time of (e.g.) Lorentz signature there is no stability subalgebra and the dimension of the ring of the real invariants of $F^+_4$ and $F_4^-$ is $41=70-28-1$, where $70=35+35$ is the dimension of the module, 28 is the dimension of $\so(1,7)$ and 1 stands for the reality condition under ${\mathfrak{u}}(1)$ duality. The ring is not freely generated; we have not been able to find its partition function.

\subsection{Weyl tensors of $\so(n)$}

A Weyl tensor belongs to the irreducible module $(020\ldots0)$ of $\so(n)$ ($n\geq7$). When $n=4$, the module is reducible, and falls outside our considerations (self-dual Weyl tensors are covered by the $A_1$ examples: invariants of degrees 2 and 3).
Already when $n=5$, $\lambda=(04)$ and $\fg^+$ is beyond affine.
For $n=6$, $\fg=A_3$ and $\lambda=(202)$.
For algebraic invariants of the Weyl tensor in 5 dimensions
the ring $S^\fg$ has the partition function
\begin{align}
P(t)&=(1 - t^2)^{-1} (1 - t^3)^{-1} (1 - t^4)^{-3} (1 - t^5)^{-4} (1 - t^6)^{-9} (1 -
    t^7)^{-15} (1 - t^8)^{-33} \nn\\
    &\times(1 - t^9)^{-63} (1 - t^{10})^{-136} (1 -
    t^{11})^{-276} (1 - t^{12})^{-569} (1 - t^{13})^{-1135} \nn\\
    &\times(1 - t^{14})^{-2243} (1 -
    t^{15})^{-4265} (1 - t^{16})^{-7823} (1 - t^{17})^{-13709} (1 -
    t^{18})^{-22580} \\
    &\times(1 - t^{19})^{-34144} (1 - t^{20})^{-44522} (1 -
    t^{21})^{-40658} (1 - t^{22})^{12835} (1 - t^{23})^{199009} \nn\\
    &\times(1 -
    t^{24})^{702382} (1 - t^{25})^{1903228} (1 - t^{26})^{4557099} (1 -
    t^{27})^{10085268} \nn\\
    &\times(1 - t^{28})^{21050933} (1 - t^{29})^{41801879} \times\ldots\nn
\end{align}
pointing towards a very complicated ring structure with a very large number of generators and relations among them. The dimension of the ring is 69. For Weyl invariant models of 8-dimensional gravity the explicit form of Weyl tensor invariants at order 4 were given e.g. in \cite{Fulling:1992vm,Boulanger:2004zf}.

\subsection{Other examples}

We recall that the criteria of Sections \ref{WhenDim1Section} and \ref{WhenFreeSection} are sufficient, but not necessary.
Even if they are not ubiquitous, there are examples of free generation even if neither criterion is fulfilled.
One such case is a spinor under $B_6=\so(13)$.
The stability algebra of a generic spinor \cite{Gatti1978} is $\sl(3)\oplus\sl(3)$, so $\dim S^\fg=64-78+16=2$. $S^\fg$ is freely generated by generators of degrees 4 and 8.

%Examples of free generation, although $\fg^+$ is not finite-dimensional or affine.

The situations where our computing power is sufficient for safely deducing the full partition function of $S^\fg$ when neither of the criteria are fulfilled are rare.
One such example, which is not a complete intersection, is $\fg=B_2$, $\lambda=(11)$ (with no obvious physical applications). The multiplicative form of the partition function yields
\begin{align}
P(t)&=(1 - t^4)^{-1} (1 - t^8)^{-4} (1 - t^{12})^{-4} (1 - t^{14})^{-3} (1 - t^{16})^{-1} (1 - t^{18})^{-6}\nn\\
&\times(1 - t^{24})^3 (1 - t^{26})^{12} (1 - t^{28})^9 (1 - t^{30})^{18} (1 - t^{32})^{21} (1 - t^{34})^6 (1 - t^{36})^{19} \\
&\times(1 - t^{38})^{-24} (1 - t^{40})^{-42} (1 - t^{42})^{-59} (1 - t^{44})^{-132} (1 - t^{46})^{-96} \nn\\
&\times(1 - t^{48})^{-125} (1 - t^{50})^{-78}\times\ldots\nn
\end{align}
The ring is 6-dimensional ($16-10=6$), but has 19 generators subject to certain relations.
An independent set of generators is found to be one at degree 4, three at degree 8 and two at degree 12. The partition function has the minimal form
\begin{align}
P(t)={1 + t^8 + 2 t^{12} + 3 t^{14} + 2 t^{16} + 6 t^{18} + 2 t^{20} + 3 t^{22} +
 2 t^{24} + t^{28} + t^{36}\over(1-t^4)(1-t^8)^3(1-t^{12})^2}\;.
\end{align}

Another case is $\fg=A_3$, $\lambda=(011)$.
$\fg^+$ is the hyperbolic Kac--Moody algebra $A_2^{++}$.
The partition function is
\begin{align}
P(t)&=(1 - t^8)^{-2} (1 - t^{12})^{-2} (1 - t^{16})^{-1} (1 - t^{20})^{-1} (1 - t^{24})^{-2} (1 - t^{28})^{-1} \nn\\
&\times(1 - t^{32}) (1 - t^{36}) (1 - t^{40})^2 (1 - t^{44}) (1 - t^{48})^2 \\
&\times(1 - t^{56})^{-1}\times\ldots\nn
\end{align}
The ring has 9 generators and dimension $20-15=5$.
The minimal form of the partition function is
\begin{align}
P(t)={1 + t^{16} + t^{20} + t^{24} + t^{28} + t^{44}
\over(1-t^8)^2(1-t^{12})^2(1-t^{24})}
\end{align}
Note that in this example as well as in the previous one (and all other) the numerator has the symmetric form  reflecting the Gorenstein property.

\subsection{Non-simple $\fg$}

Although this should be possible, we have not conducted a complete survey
of situations in which $\fg$ is semi-simple but not simple.
A few examples lead to the
tentative conjecture that the criteria hold also in these cases:

\noindent
$\fg=A_1\oplus A_1$, $\lambda=(1)(2)$.
$\fg^+=C_3$.
$S^\fg$ is generated by a single invariant at degree 4.

\noindent
$\fg=A_1\oplus A_1$, $\lambda=(1)(3)$.
$\fg^+$ is the twisted affine algebra $D_4^{(3)}$.
$S^\fg$ is freely generated by two generators at degrees 2 and 6.

\noindent
$\fg=A_1\oplus A_1$, $\lambda=(1)(4)$.
Neither criterion is fulfilled. $S^\fg$ is a complete intersection, with generators of degrees 4, 4, 8, 12 and 18, and a relation at degree 36.
\begin{align}
P(t)={1+t^{18}\over(1-t^4)^2(1-t^8)(1-t^{12})}.
\end{align}

\noindent
$\fg=A_1\oplus A_1$, $\lambda=(1)(5)$.
Neither criterion is fulfilled. $S^\fg$ has generators at degrees 2, 4, 6, 6, 8, 10, 10, 12 and 14. Out of these an independent subset is at degrees 2, 4, 6, 6, 8 and 10.
\begin{align}
P(t)={1+t^{10}+t^{12}+t^{14}+t^{24}\over(1-t^2)(1-t^4)(1-t^6)^2(1-t^8)(1-t^{10})}.
\end{align}

\noindent
$\fg=A_1\oplus A_1\oplus A_1$, $\lambda=(1)(1)(1)$. $\fg^+=D_4$.
$S^\fg$ is generated by a single invariant at degree 4.

\noindent
$\fg=A_1\oplus A_1\oplus A_1\oplus A_1$, $\lambda=(1)(1)(1)(1)$. $\fg^+$ is the untwisted affine extension of $D_4$.
$S^\fg$ is freely generated by generators with the same degrees as the Casimirs of $D_4$ (2, 4, 4, 6).

\noindent
$\fg=A_2\oplus A_2\oplus A_2$, $\lambda=(10)(10)(10)$. $\fg^+$ is the untwisted affine extension of $E_6$. There are generators at degrees 6, 9 and 12, and $S^\fg$ is freely generated.

\noindent
$\fg=A_3\oplus A_3$, $\lambda=(010)(010)$. $\fg^+$ is the untwisted affine extension of $D_6$, and $S^\fg$ is freely generated by generators with the same degrees as the Casimirs of $D_6$ (2, 4, 6, 6, 8, 10).

\section{Self-dual 5-forms in 10 dimensions\label{SDTenSection}}
Aiming at possible applications to type IIB $D=10$ supergravity and string theory, as well as to other non-linear generalizations of the self-dual 5-form theory (see the accompanying paper \cite{Hutomo:2025dfx} for more details),
in this Section we will consider and derive an explicit form of some of) invariants under the Lorentz group $SO(1,9)$ of a real self-dual rank-5 tensor
 $$F_{\mu_1\mu_2\mu_3\mu_4\mu_5}=\star F_{\mu_1\mu_2\mu_3\mu_4\mu_5}=\frac 1{5!}\varepsilon_{\mu_1\mu_2\mu_3\mu_4\mu_5\nu_1\nu_2\nu_3\nu_4\nu_5}F^{\nu_1\nu_2\nu_3\nu_4\nu_5}$$
 in a $10D$ Minkowski space-time with a metric of almost plus signature. The Greek letters now stand for the $D=10$ vector indices. The dimension of this $SO(1,9)$ module is $\bf 126$ and its Dynkin label is $(00002).$

Let us  compute the number of functionally independent invariants in this case using the formula \eqref{DimensionEq}. First, let us prove that there is no a stability subgroup of $SO(1,9)$ that leaves $F_5$ invariant. To this end we construct a traceless symmetric matrix
 \be\label{M}
 M_{\mu\nu} :=F_{\mu\lambda_1\ldots\lambda_4} F_\nu{}^{\lambda_1\ldots\lambda_4}\,. \qquad
 \ee
Using an $SO(1,9)$ transformation this matrix can be brought to the diagonal form ${\rm diag}\,M_{\mu}{}^{\nu}=(\lambda_1,\cdots,\lambda_{10})$, where generically all eigenvalues are different (but $\sum\lambda_i=0$). We require that $F_5$ be invariant under infinitesimal $SO(1,9)$-rotations with parameters $\omega^{\mu\nu}=-\omega^{\nu\mu}$, this  implies the stability condition on the matrix $\omega^\alpha{}_\mu M^{\mu\beta}+ \omega^\beta{}_\mu M^{\mu\alpha}=0$. Since all matrix eigenvalues are different, this equation implies $\omega^{\mu\nu}=0$. Therefore the stability group of $F_5$ is trivial.
Thus, the number of functionally independent $SO(1,9)$ invariants which one can construct with powers of the components of the self-dual tensor $F_5$ is $81=126-45$, where {\bf 45} is the dimension of $SO(1,9)$.

Up to order 22 the partition function characterizing the ring of these invariants looks as follows
\begin{align}\label{pf}
P(t)&=1 + t^4 + 2 t^6 + 7 t^8 + 14 t^{10} + 72 t^{12} + 247 t^{14}\nonumber\\
&+ 1364 t^{16} + 6851 t^{18}+{ 40170} t^{20}+{227979 t^{22}}\ldots\nn\\
&=(1 - t^4)^{-1} (1 - t^6)^{-2} (1 - t^8)^{-6} (1 - t^{10})^{-12} (1 -
    t^{12})^{-62} (1 - t^{14})^{-221}\\
&\qquad\times(1 - t^{16})^{-1247} (1 - t^{18})^{-6404}(1 - t^{20})^{-37896}(1 - t^{22})^{-216486}\times\ldots\nn
 \end{align}

The number $n$ of linearly independent invariants (given by negative powers of $(1-t^p)^{-n}$)  at higher levels is huge, and we do not know when non-linear relations between them start to appear. One can notice that already at orders $p=4$, 6, 8, 10 and 12 the total number of unfactorisable and linearly independent (at each level) invariants is 83 which exceeds the dimension 81 of the ring. Since the complete classification of these invariants is beyond our present reach, in what follows we present a number of invariants appearing at lower orders of $F^n$. To construct such invariants we will use two methods, the first one will be based on the tensor structure of the self-dual $F_5$ and the second one on its spin-tensor realization.

\subsection{Tensor form of $F_5$ invariants}

To construct and identify independent invariants of powers of $F_5$ one notices that in all the invariants at least one index of each $F_5$ is contracted with an index of another $F_5$. So the building blocks of the invariants are the following tensor structures
\be\label{id3ccf*-2}
F_{\mu_1\mu_2\mu_3\mu_4\lambda}F^{\nu_1\nu_2\nu_3\nu_4\lambda}=4\,\delta^{[\nu_1}_{[\mu_1}N_{\mu_2\mu_3\mu_4]}{}^{\nu_2\nu_3\nu_4]}-\delta^{[\nu_1}_{[\mu_1}\delta^{\nu_2}_{\mu_2}\delta^{\nu_3}_{\mu_3}M_{\mu_4]}{}^{\nu_4]}\,,
\ee
where $M_{\mu}{}^{\nu}$ has already been introduced in \eqref{M} and
\be\label{N}
N_{\mu_1\mu_2\mu_3,}{}^{\nu_1\nu_2\nu_3}:=F_{\mu_1\mu_2\mu_3\lambda_1\lambda_2}F^{\nu_1\nu_2\nu_3\lambda_1\lambda_2}\,.
\ee
The indices in the square brackets are anti-symmetrized such that
$$T_{[\mu_1\ldots\mu_p]}=\frac 1{p!}(T_{\mu_1\mu_2\ldots\mu_p}-T_{\mu_2\mu_1\ldots\mu_p}+\ldots).$$
The relation between the l.h.s. of \eqref{id3ccf*-2} and the tensors $N$ and $M$ is a consequence of the self-duality of $F_5$ (see the list of identities in Appendix \ref{id5f} which were used to derive all the expressions given in this Section).
The tensor $M_{\mu\nu}$ is symmetric and traceless, and has dimension {\bf 54}.
\if{}
Its Young tableau is \hbox{
%\ytableausetup{centertableaux}
{\tiny
\begin{ytableau}
~&~
\end{ytableau}
}
.
}
\fi
The tensor $N_{\mu_1\mu_2\mu_3}{}^{\nu_1\nu_2\nu_3}$ splits into the irreducible representations of dimensions {\bf 54}, {\bf 1050} and {\bf 4125}. Different (but equivalent) forms of this splitting are
\bea\label{4125irrep}
%N_{\mu_1\mu_2\mu_3,\nu_1\nu_2\nu_3}&=&2N^{^{(1050)}}_{[\mu_1\mu_2\mu_3,{\color{red}[}\nu_1]\nu_2\nu_3{\color{red}]}}+N^{(4125)}_{\mu_1\mu_2\mu_3,\nu_1\nu_2\nu_3}+\frac{9}{28}\delta^{[\alpha_1}_{[\mu_1}\delta^{\alpha_2}_{\mu_2}M_{\mu_3]}{}^{\alpha_3]}\eta_{\alpha_1\nu_1}\eta_{\alpha_2\nu_2}\eta_{\alpha_3\nu_3}\nonumber\\
N_{\mu_1\mu_2\mu_3,\nu_1\nu_2\nu_3}&=&5N^{^{(1050)}}_{[\mu_1\mu_2\mu_3,{\color{red}[}\nu_1\nu_2]\nu_3{\color{red}]}}+N^{(4125)}_{\mu_1\mu_2\mu_3,\nu_1\nu_2\nu_3}+\frac{9}{28}\delta^{[\alpha_1}_{[\mu_1}\delta^{\alpha_2}_{\mu_2}M_{\mu_3]}{}^{\alpha_3]}\eta_{\alpha_1\nu_1}
\eta_{\alpha_2\nu_2}\eta_{\alpha_3\nu_3}
\nonumber\\
N_{\mu_1\mu_2\mu_3,\nu_1\nu_2\nu_3}&=&
-3N_{{\color{red}[}\nu_1[\mu_1\mu_2,\mu_3]\nu_2\nu_3{\color{red}]}}+2N^{(4125)}_{\mu_1\mu_2\mu_3,\nu_1\nu_2\nu_3}+\frac{9}{14}\delta^{[\alpha_1}_{[\mu_1}\delta^{\alpha_2}_{\mu_2}M_{\mu_3]}{}^{\alpha_3]}\eta_{\alpha_1\nu_1}
\eta_{\alpha_2\nu_2}\eta_{\alpha_3\nu_3}\,\nonumber\\
\eea
where in the second equality the indices $[\mu_1,\mu_2,\mu_3]$ and ${\color{red}[}\nu_1,\nu_2,\nu_3{\color{red}]}$ are anti-symmetrized separately, while in the first one the anti-symmetrization of the three indices $\nu_i$ with the ``red" brackets is performed upon the anti-symmetrization of the five indices\linebreak $[\mu_1,\mu_2,\mu_3,\nu_1,\nu_2]$ in the ``black" brackets.

The irrep $N^{^{(1050)}}_{[\mu_1\mu_2\mu_3,\mu_4\mu_5]\nu}$ is self-dual with respect to the five anti-symmetric indices $\mu_i$, and the contraction of $\nu$ with $\mu_i$ is zero.
\if{}
Its Young tableau is
%\ytableausetup{centertableaux}
{\tiny
\begin{ytableau}
~&~\\
~\\
~\\
~\\
~
\end{ytableau}
}.
\fi

In the irrep $N^{(4125)}_{\mu_1\mu_2\mu_3,\nu_1\nu_2\nu_3}$ the groups of three indices $\mu_i$ and $\nu_i$ are anti-symmetric, the tensor is symmetric under the exchange of these anti-symmetric groups of indices and traceless. The anti-symmetrization of any four indices is zero.

The Young tableaux of the irreps $\bf 54$, $\bf 1050$ and $\bf 4125$ are \if{}
%\ytableausetup{centertableaux}
{\tiny
\begin{ytableau}
~&~\\
~&~\\
~&~
\end{ytableau}
}.
\fi
\\
\\
\Yboxdim8pt
\begin{equation}
    {\bf54}\,:\;\yng(2)\;,
     \qquad{\bf1050}\,:\;\yng(2,1,1,1,1)\;,
     \qquad{\bf4125}\,:\;\yng(2,2,2)
\end{equation}
Therefore the building blocks of the $F_5$ invariants are the irreducible $SO(1,9)$ tensors $M^{^{(54)}}$, $N^{^{(1050)}}$ and $N^{(4125)}$.

The simplest choice of nine invariants of $F_5$ is to construct them as traces of products of the matrix $M^{^{(54)}}$. Since the $10\times 10$ matrix $M^{^{(54)}}$ is symmetric and traceless, all its invariants are freely generated by a basis of 9 invariants which can be chosen as
\be\label{Mn}
I^{^{(1)}}_{2n}=\tr M^n\,, \qquad n=2,\ldots 10,
\ee
where the subscript $2n$ indicates the order of $F_5$ in these invariants.

\subsubsection{4th order invariant}
The unique independent fourth-order invariant of $F_5$ is
\be\label{M2}
I_4=M_{\mu\nu}M^{\nu\mu}=\tr M^2\,.
\ee
This is because the scalar product of two self-dual $N^{^{(1050)}}$ is zero, while the scalar product of two $N^{^{(4125)}}$ composed of $FF$ is proportional  to $\tr M^2$ due to the identity \eqref{NNscalar} and the relation \eqref{4125irrep}. This invariant was considered in the context of non-linear self-dual 5-form theories in \cite{Buratti:2019guq,Avetisyan:2022zza}.
\subsubsection{6th order invariants}

The LiE program \cite{LiE} tells us that at the sixth order there are two independent invariants (see eq. \eqref{pf}). The analysis shows that one can choose them as follows
\be\label{6-order 1}
I_{6}^{^{(1)}}=\tr M^3,
\ee
\bea\label{6-order 2}
I^{^{(2)}}_6 &=& N^{^{(1050)}}_{[\rho_1\rho_2\rho_3,\rho_4\rho_5]\kappa}\Big(N_{_{(1050)}}^{[\rho_1 \rho_2 \rho_3,}{}_{\alpha_1 \alpha_2 ]\alpha_3}{}N_{_{(1050)}}^{[\rho_4\rho_5\kappa,\alpha_1 \alpha_3 ]\alpha_2}\Big)\nonumber\\
&\equiv & -\frac 32 N^{^{(1050)}}_{[\rho_1\rho_2\rho_3,\rho_4\rho_5]\kappa}\Big(N_{_{(1050)}}^{[\rho_1 \rho_2 \rho_3,}{}_{\alpha_1 \alpha_2 ]\alpha_3}{}N_{_{(1050)}}^{[\rho_4\rho_5\kappa,{\color{red}[}\alpha_1 \alpha_2 ]\alpha_3{\color{red}]}}\Big)\,,
\eea
where in the last equality the three indices $\alpha_i$ are anti-symmetrized upon the anti-symmetrization of five  indices $\rho_4,\rho_5,\kappa,\alpha_1,\alpha_2$. The tensor structure of $(NN)$ in the brackets with the anti-symmetrized indices $\rho_1,\ldots, \rho_5$  is an anti-self-dual $\overline{\bf 1050}$ irrep.

All other six-order invariants that one can construct with the use of the tensors $M^{^{(54)}}$, $N^{^{(1050)}}$ and $N^{^{(4125)}}$ are linear combinations of \eqref{6-order 1} and \eqref{6-order 2}, which one can check using the fact that $M$ and $N$ are composed of $FF$ and applying identities given in Appendix \ref{id5f}. For instance, all the invariants of the form $MNN$ are proportional to $\tr M^3$ (due to identities like \eqref{NN=I-MM*} and \eqref{hat M=MM}), while e.g.
the invariant\footnote{Note the difference in the order of the indices $\alpha_i$ in the last term of this invariant and that of \eqref{6-order 2}.}
\bea\label{NNN1*}
I_6&=&N^{^{(4125)}}_{\mu_1 \mu_2 \mu_3}{}^{\rho_1 \rho_2 \rho_3}\Big(N_{^{(1050)}}^{[\mu_1 \mu_2 \mu_3,\alpha_1\alpha_2]\alpha_3}N^{^{(1050)}}_{[\rho_1\rho_2\rho_3,\alpha_1\alpha_2]\alpha_3}\Big)\nonumber\\
%&\equiv & N_{\mu_1 \mu_2 \mu_3}{}^{\rho_1 \rho_2 \rho_3}\Big(N_{^{(1050)}}^{[\mu_1 \mu_2 \mu_3,\alpha_1\alpha_2]\alpha_3}N^{^{(1050)}}_{[\rho_1\rho_2\rho_3,\alpha_1\alpha_2]\alpha_3}\Big)\nonumber\\
&=& -\frac{75}{2} I^{^{(2)}}_6 + x \,\tr M^3,
\eea
with some coefficient $x$ which we did not compute explicitly because its value is not important.

\subsubsection{8th order invariants }
At the 8th order in $F$, according to the LiE program (see \eqref{pf}), there are six linearly independent invariants, the seventh one being the square of the 4th order invariant \eqref{M2}.
\noindent
\\
{\bf One can choose a basis of these invariants as follows.}
Two invariants can be constructed as products of four tensors $M$s or three $M$s and one $N$, namely
\be\label{trM4}
I^{^{(1)}}_8=\tr M^4\,, \qquad I^{^{(2)}}_8=M_{\mu_1}{}^{\nu_1}M_{\mu_2}{}^{\nu_2}M_{\mu_3}{}^{\nu_3}N^{^{(4125)}}_{\nu_1 \nu_2 \nu_3}{}^{\mu_1 \mu_2 \mu_3}\,.
\ee
Other three invariants can be constructed with two $M$s and two $N$s.
One of these is constructed by taking the inner product of the {\bf 210} irrep $M^{\mu\nu}N^{^{(1050)}}_{[\alpha_1\alpha_2\alpha_3\alpha_4\mu]\nu}$ (the only independent {\bf 210} irrep in the symmetric product of $F^4$) with itself
\be\label{MMNN1}
I^{^{(3)}}_8=M^{\mu\nu}N^{^{(1050)}}_{[\alpha_1\alpha_2\alpha_3\alpha_4\mu]\nu}N_{_{(1050)}}^{[\alpha_1\alpha_2\alpha_3\alpha_4\rho]\lambda}M_{\rho\lambda}\,.
\ee
In \eqref{MMNN1} the indices $\mu,\nu,\rho,\lambda$ are automatically totally symmetrized due to the symmetry properties of the product of two self-dual $N_{(1050)}$ (see eq. \eqref{NNsim}). So this invariant can also be seen as the product of the {\bf 660} irrep in $N^{^{(1050)}}_{[\alpha_1\alpha_2\alpha_3\alpha_4{\color{red}(}\mu]\nu}N_{_{(1050)}}^{[\alpha_1\alpha_2\alpha_3\alpha_4}{}_{\rho]\lambda\color{red})}$ with $M^{\mu\nu}M^{\rho\lambda}$ (upon subtracting all traces).\footnote{The indices within the round brackets are symmetrized such that  $$T_{(\mu_1\ldots\mu_p)}=\frac 1{p!}(T_{\mu_1\mu_2\ldots\mu_p}+T_{\mu_2\mu_1\ldots\mu_p}+\ldots).$$} Note that the symmetric product of $F^4$ contains two {\bf 660} irreps, one of which can be chosen as the traceless part of $(MM)_{(\mu\nu\lambda\rho)}$, and another one is contained in $(N_{_{(1050)}})^2$. So the fourth invariant can be chosen as
\be\label{MMNN2}
I^{^{(4)}}_8=N^{^{(1050)}}_{[\alpha_1\alpha_2\alpha_3\alpha_4{\color{red}(}\mu]\nu}N_{_{(1050)}}^{[\alpha_1\alpha_2\alpha_3\alpha_4}{}_{\rho]\lambda\color{red})}N_{_{(1050)}}^{[\beta_1\beta_2\beta_3\beta_4\mu]\nu}N^{^{(1050)}}_{[\beta_1\beta_2\beta_3\beta_4}{}^{\rho]\lambda}\,,
\ee
where the indices $\mu,\nu,\rho,\lambda$ are totally symmetrized, i.e. this invariant is constructed by taking the inner product of the two copies of the {\bf 660} irrep in $(N_{_{(1050)}})^2$ (upon subtracting all traces).

The product of $M_{\mu\nu}M_{\rho\lambda}$ splits into two tensors, the totally symmetric one $M_{(\mu\nu}M_{\rho\lambda)}$ which takes values in the 660 irrep of $SO(1,9)$ (it has appeared in \eqref{MMNN1}) and the tensor $M_{[\mu}{}^{[\nu}M_{\rho]}{}^{\lambda]}$ which takes values in the {\bf 770} irrep (upon subtracting traces). The contraction of the latter with two $N^{^{(1050)}}$ produces the following invariant
\be\label{MNNN1*}
I_{8}^{^{(5)}}=M_{[\nu_1}{}^{[\mu_1}M_{\nu_2]}{}^{\mu_2]}N^{^{(1050)}}_{[\rho_1\rho_2\rho_3,\rho_4\mu_1]\mu_2}N_{_{(1050)}}^{[\rho_1\rho_2\rho_3,\rho_4\nu_1]\nu_2}.
\ee
Finally, the contraction of $MM$ with the single {\bf 770} irrep in the product of $N^{^{(1050)}}$ and $N^{^{(4125)}}$ produces the sixth invariant
\be\label{MMNN3}
I_{8}^{^{(6)}}=M_{\nu_1}{}^{\mu_1}M_{\nu_2}{}^{\mu_2}N^{^{(1050)}}_{[\mu_1\mu_2\rho_1,\rho_2\rho_3]\rho_4}N_{_{(4125)}}^{\rho_1\rho_2\rho_3,\rho_4\nu_1\nu_2}\,.
\ee
Note that the symmetric product of $F^4$ contains three {\bf 770} irreps. They can be associated with
$$M_{[\mu_1}{}^{[\nu_1}M_{\mu_2]}^{\nu_2]}, \qquad  N^{^{(1050)}}_{[\rho_1\rho_2\rho_3,\rho_4{\color{red}[}\mu_1]\mu_2{\color{red}]}}N_{_{(1050)}}^{[\rho_1\rho_2\rho_3,\rho_4{\color{red}[}\nu_1]\nu_2{\color{red}]}}$$
and
$$N^{^{(1050)}}_{[\mu_1\mu_2\rho_1,\rho_2\rho_3]\rho_4}N_{_{(4125)}}^{\rho_1\rho_2\rho_3,\rho_4\nu_1\nu_2}+N_{_{(1050)}}^{[\nu_1\nu_2\rho_1,\rho_2\rho_3]\rho_4}N^{^{(4125)}}_{\rho_1\rho_2\rho_3,\rho_4\mu_1\mu_2}.$$
\\
\noindent
{\bf An alternative choice of the basis of the 8th order invariants.}
\medskip
\\
Instead of the basis of six $I_8$ given above, one can choose it using the results of \cite{Paulos:2008tn,Melo:2020amq} on higher-order string corrections to the effective action of type IIB supergravity produced by $10D$ curvature tensor and the five-form. It was shown there that supersymmetry singles out five independent 8th order terms in the action involving exclusively the $F_5$ form and its derivative $\partial_{\nu} F_{\mu_1\ldots\mu_5}$. \footnote{Other terms are constructed by taking products of components of the Weyl tensor with  themselves or with $F_5$\,.} In the assumption that the self-dual $F_5$ is the external derivative of a four form $(F_5=dA_4)$ and satisfies a free equation of motion ($\partial_{\mu_1} F^{\mu_1\ldots\mu_5}=0$), the tensor $\partial_{\nu} F_{\mu_1\ldots\mu_5}$ forms an irreducible $\bf 1050$ module of $SO(1,9)$. The five independent invariants individualized in \cite{Paulos:2008tn,Melo:2020amq} (see  the last five lines in the tables of those papers) form a basis of the invariants in the symmetric product of four {\bf 1050} irreps of the form
\be\label{cT}
\mathcal T^{^{(1050)}}_{[\mu_1\ldots\mu_5]\nu}=\partial_{\nu} F_{\mu_1\ldots\mu_5}+N^{^{(1050)}}_{[\mu_1\mu_2\mu_3,\mu_4\mu_5]\nu}\,.
\ee
The LiE program tells us that there are exactly five independent invariants in $(\mathcal T^4)_{sym}$. However, in this paper we are interested exclusively in $F_5$ invariants that do not contain derivatives of $F_5$. So we need to check that the five invariants found in \cite{Paulos:2008tn,Melo:2020amq}  remain independent if we set $\partial_{\nu} F_{\mu_1\ldots\mu_5}=0$. This is {\it a priori} not guaranteed because of the composite nature of $N^{(1050)}=FF$, but it turns out that this is indeed the case. These invariants have the following structure (in which we skip the superscript $(1050)$ from $N^{(1050)}$)
\bea\label{F8IIB1}
\hat I^{^{(1)}}_{8}&=&-\Big(N_{[\mu_1\mu_2\mu_3{\color{red}[}\mu_4\rho_1]\rho_2{\color{red}]}}N^{[\mu_1\mu_2\mu_3{\color{red}[}\mu_4\nu_1]\nu_2{\color{red}]}}\Big)
\Big(N_{[\nu_1}{}^{\rho_1\lambda_1{\color{red}[}\lambda_2\lambda_3]\lambda_4{\color{red}]}}N^{[\rho_2}{}_{\lambda_1\lambda_2{\color{red}[}\nu_2\lambda_3]\lambda_4{\color{red}]}}\Big)\nonumber\\
&=&\frac 1{3^4}\Big(\frac 56 N_{[\mu_1\mu_2\mu_3\mu_4{\color{red}[}\rho_1]\rho_2{\color{red}]}}N^{[\mu_1\mu_2\mu_3\mu_4{\color{red}[}\nu_1]\nu_2{\color{red}]}}N^{[\lambda_1\lambda_2\lambda_3\lambda_4\rho_1]\rho_2}N_{[\lambda_1\lambda_2\lambda_3\lambda_4\nu_1]\nu_2}\nonumber\\
&&+\frac{1}{7200}\tr M^4- \frac{70}{1200^2}(\tr M^2)^2\Big)\,,
\eea
\bea\label{F8IIB2}
\hat I^{^{(2)}}_8 &=&\Big(N_{[\mu_1\mu_2\mu_3{\color{red}[}\mu_4\rho_1]\rho_2{\color{red}]}}N^{[\mu_1\mu_2\mu_3{\color{red}[}\mu_4\nu_1]\nu_2{\color{red}]}}\Big)
\Big(N^{[\rho_1}{}_{\lambda_1\lambda_2{\color{red}[}\nu_1\lambda_3]\lambda_4{\color{red}]}}N^{[\rho_2\lambda_1\lambda_3{\color{red}[}\lambda_2\lambda_4]}{}_{\nu_2{\color{red}]}}\Big)\\
&=&\frac 1{3^4} \Big(N_{[\mu_1\mu_2\mu_3\mu_4{\color{red}[}\rho_1]\rho_2{\color{red}]}}N^{[\mu_1\mu_2\mu_3\mu_4{\color{red}[}\nu_1]\nu_2{\color{red}]}}N^{[\lambda_1\lambda_2\lambda_3\lambda_4\rho_1]\rho_2}N_{[\lambda_1\lambda_2\lambda_3\lambda_4\nu_1]\nu_2}\nonumber\\
&&+ \frac{1}{240^2}\tr M^4- \frac{1}{10 \cdot 240^2}(\tr M^2)^2 \Big),
\eea
where the indices in the red brackets are antisymmetrized after the antisymmetrization of the indices in black brackets.
These two invariants are composed of $\tr M^4$, $(\tr M^2)^2$ and the scalar product of the $\bf 770$ irrep $(N_{[\mu_1\mu_2\mu_3\mu_4{\color{red}[}\rho_1]\rho_2{\color{red}]}}N^{[\mu_1\mu_2\mu_3\mu_4{\color{red}[}\nu_1]\nu_2{\color{red}]}})$ with itself,  as one finds using identities of Appendix \ref{id5f}.
\bea
\hat I^{^{(3)}}_8 =
\Big(N_{[\mu_1 \mu_2 \mu_3 {\color{red}[}\rho_1 \rho_2] \rho_3 {\color{red}]}} N^{[\mu_1 \mu_2 \mu_3 {\color{red}[}\nu_1 \nu_2] \nu_3 {\color{red}]}}\Big)
\Big(N^{[\rho_1 \rho_2 \lambda_1 {\color{red}[}\lambda_2 \lambda_3]}{}_{ \nu_1{\color{red}]}}  N^{[\rho_3}{}_{\nu_2 \lambda_2 {\color{red}[}\lambda_1 \lambda_3] \nu_3 {\color{red}]}}\Big)\,,
\eea
\bea
{\hat I}^{(4)}_8=\Big(N_{[\mu_1 \mu_2 \mu_3 {\color{red}[}\rho_1 \rho_2] \rho_3 {\color{red}]}} N^{[\mu_1 \mu_2 \mu_3 {\color{red}[}\nu_1 \nu_2] \nu_3 {\color{red}]}}\Big)\Big(N_{[\nu_1}{}^{\rho_1\lambda_1 {\color{red}[}\rho_2 \lambda_2] \lambda_3 {\color{red}]}} N^{[\rho_3}{}_{\nu_2\lambda_2 {\color{red}[}\nu_3 \lambda_1] \lambda_3 {\color{red}]}}\Big)
\eea
and
\bea\label{F8IIB3}
\hat I^{^{(5)}}_8 &=&\Big(N_{[\nu_1}{}^{\rho_1\mu_1{\color{red}[}\mu_2\mu_3]\mu_4{\color{red}]}}N^{[\rho_2}{}_{\mu_1\mu_2{\color{red}[}\nu_2\mu_3]\mu_4{\color{red}]}}\Big)\Big(N_{[\rho_2}{}^{\nu_1\lambda_1{\color{red}[}\lambda_2\lambda_3]\lambda_4{\color{red}]}}N^{[\nu_2}{}_{\lambda_1\lambda_2{\color{red}[}\rho_1\lambda_3]\lambda_4{\color{red}]}}\Big)\nonumber\\
&=& -\frac{5}{6} \hat I^{^{(1)}}_8 + \frac{25}{4\cdot 9^3}I^{^{(4)}}_8 + x \tr M^4 + y (\tr M^2)^2~\,,
\eea
where $I^{^{(4)}}_8$ is given in \eqref{MMNN2}, and $x$ and $y$ are coefficients whose explicit values we did not compute, since they are not important.

To these five invariants we can add e.g. $\hat I^{^{(6)}}_8=I^{^{(2)}}_8$ in \eqref{trM4} to complete the basis.
It is not an easy computational problem to establish the full relation between the two bases $I^i_8$ and ${\hat I}^i_8$ $(i=1,\ldots, 6)$, which we leave a part.

\subsubsection{10th and higher order invariants}

At the 10th order there are 12 linearly independent invariants and at order 12 there are 64.
As we have already mentioned, together with the lower-order invariants their number is 83 which is by two units higher than the dimension of the ring. Therefore, there should be at least 2 non-linear relations between these 83 invariants.  The problem of individualizing the linearly independent invariants at order 10 and higher and to find  functional relations between them (similar to those considered in Appendices \ref{A15Appendix} and \ref{A16Appendix} for $sl(2)$ invariants of the modules of dimension 6 and 7) becomes highly involved (if at all doable) without using a computer program, a challenge which we leave to experts. Below we only give 12 possible candidates for a basis of the 10th-order invariants constructed by choosing and contracting different irreps in the decompositions of symmetric products of $M^{^{(54)}}$, $N^{^{(1050)}}$ and $N^{^{(4125)}}$:
\bea\label{I101}
I^{^{(1)}}_{10}&=&\tr M^5\,,\nonumber\\
\\
\label{I102}
I^{^{(2)}}_{10}&=&(MM)_{\mu_1}{}^{\nu_1}M_{\mu_2}{}^{\nu_2}M_{\mu_3}{}^{\nu_3}N^{^{(4125)}}_{\nu_1 \nu_2 \nu_3}{}^{\mu_1 \mu_2 \mu_3}\,,\nonumber\\
\nonumber\\
\label{I103}
I^{^{(3)}}_{10}&=&M_{\mu_1}{}^{\nu_1}M_{\mu_2}{}^{\nu_2}M_{\mu_3}{}^{\nu_3}\Big(N_{^{(1050)}}^{[\mu_1 \mu_2 \mu_3,\alpha_1\alpha_2]\alpha_3}N^{^{(1050)}}_{[\nu_1\nu_2\nu_3,\alpha_1\alpha_2]\alpha_3}\Big)\,,\nonumber\\
\nonumber\\
\label{I104}
I^{^{(4)}}_{10}&=&(MM)^{\mu\nu}M^{\rho\lambda}\Big(N^{^{(1050)}}_{[\alpha_1\alpha_2\alpha_3\alpha_4{\color{red}(}\mu]\nu}N_{_{(1050)}}^{[\alpha_1\alpha_2\alpha_3\alpha_4}{}_{\rho]\lambda\color{red})}\Big),\nonumber\\
\nonumber\\
\label{I105}
I_{10}^{^{(5)}}&=&(MM)_{[\nu_1}{}^{[\mu_1}M_{\nu_2]}{}^{\mu_2]}N^{^{(1050)}}_{[\rho_1\rho_2\rho_3,\rho_4\mu_1]\mu_2}N_{_{(1050)}}^{[\rho_1\rho_2\rho_3,\rho_4\nu_1]\nu_2},
\nonumber
\eea
\bea
\label{I106}
I_{10}^{^{(6)}}&=&(MM)_{\nu_1}{}^{\mu_1}M_{\nu_2}{}^{\mu_2}N^{^{(1050)}}_{[\mu_1\mu_2\rho_1,\rho_2\rho_3]\rho_4}N_{_{(4125)}}^{\rho_1\rho_2\rho_3,\rho_4\nu_1\nu_2}\,,\nonumber\\
\nonumber\\
\label{I107}
I_{10}^{^{(7)}}&=& N^{^{(1050)}}_{[\rho_1\rho_2\rho_3,\rho_4\rho_5]\mu}(MM)^{\mu}{}_{\kappa}\Big(N_{_{(1050)}}^{[\rho_1 \rho_2 \rho_3,}{}_{\alpha_1 \alpha_2 ]\alpha_3}{}N_{_{(1050)}}^{[\rho_4\rho_5\kappa,\alpha_1 \alpha_3 ]\alpha_2}\Big),\nonumber
\eea
\bea
\label{I108}
I_{10}^{^{(8)}}&=& N^{^{(1050)}}_{[\rho_1\rho_2\rho_3,\rho_4\nu]\mu}M^{[\nu}{}_{\rho_5} M^{\mu]}{}_{\kappa}\Big(N_{_{(1050)}}^{[\rho_1 \rho_2 \rho_3,}{}_{\alpha_1 \alpha_2 ]\alpha_3}{}N_{_{(1050)}}^{[\rho_4\rho_5\kappa,\alpha_1 \alpha_3 ]\alpha_2}\Big),\nonumber\\
\nonumber\\
\label{I109}
I_{10}^{^{(9)}}&=& N^{^{(1050)}}_{[\alpha_1\alpha_2\alpha_3\alpha_4\kappa]{\color{red}(}\nu}M^\kappa{}_\mu N_{_{(1050)}}^{[\alpha_1\alpha_2\alpha_3\alpha_4}{}_{\rho]\lambda\color{red})}N_{_{(1050)}}^{[\beta_1\beta_2\beta_3\beta_4\mu]\nu}N^{^{(1050)}}_{[\beta_1\beta_2\beta_3\beta_4}{}^{\rho]\lambda}\,,
\nonumber
\eea
\bea
\label{I1010}
I_{10}^{^{(10)}}&=&\Big (N^{^{(1050)}}_{[\rho_1\rho_2\rho_3,\rho_4{\color{red}[}\mu_1]\mu_2{\color{red}]}}N_{_{(1050)}}^{[\rho_1 \rho_2 \rho_3,}{}_{\alpha_1 \alpha_2 ]\alpha_3}{}N_{_{(1050)}}^{[\rho_4\nu_1\nu_2,\alpha_1 \alpha_3 ]\alpha_2}\Big)
\Big(N^{^{(1050)}}_{[\beta_1\beta_2\beta_3,\beta_4\nu_1]\nu_2}N_{_{(1050)}}^{[\beta_1\beta_2\beta_3,\beta_4\mu_1]\mu_2}\Big)\,,\nonumber\\
\nonumber\\
\label{I1011}
I_{10}^{^{(11)}}&=&\Big(N^{^{(1050)}}_{[\rho_1 \rho_2 \rho_3,}{}^{\alpha_1 \alpha_2 ]\alpha_3}{}N^{^{(1050)}}_{[\mu_1\mu_2\mu_3,\alpha_1 \alpha_2 ]\alpha_3}\Big)
\Big(N_{_{(1050)}}^{[\mu_1 \mu_2 \mu_3 {\color{red}[}\nu_1 \nu_2] \nu_3 {\color{red}]}}
N_{_{(1050)}}^{[\rho_1 \rho_2 \lambda_1 {\color{red}[}\lambda_2 \lambda_3]}{}_{ \nu_1{\color{red}]}}  N_{_{(1050)}}^{[\rho_3}{}_{\nu_2 \lambda_2 {\color{red}[}\lambda_1 \lambda_3] \nu_3 {\color{red}]}}\Big)\,,\nonumber\\
\nonumber\\
\label{I1012}
I_{10}^{^{(12)}}&=&\Big(N^{^{(1050)}}_{[\rho_1 \rho_2 \rho_3,}{}^{\alpha_1 \alpha_2 ]\alpha_3}{}N^{^{(1050)}}_{[\mu_1\mu_2\mu_3,\alpha_1 \alpha_2 ]\alpha_3}\Big)\Big(N_{_{(1050)}}^{[\mu_1 \mu_2 \mu_3 {\color{red}[}\nu_1 \nu_2] \nu_3 {\color{red}]}}N^{^{(1050)}}_{[\nu_1}{}^{\rho_1\lambda_1 {\color{red}[}\rho_2 \lambda_2] \lambda_3 {\color{red}]}} N_{_{(1050)}}^{[\rho_3}{}_{\nu_2\lambda_2 {\color{red}[}\nu_3 \lambda_1] \lambda_3 {\color{red}]}}\Big)\,,\nonumber
\eea
where the symmetrization and/or anti-symmetrization of the indices within the red brackets is made upon the anti-symmetrization within the black brackets.

12th-order invariants that have appeared in certain models of the non-linear self-dual 5-form theory \cite{Hutomo:2025dfx} are
\bea
I_{12}^{^{(1)}}=\tr M^6, \qquad I_{12}^{^{(2)}}&=&(M^3)^{\mu \nu} \Big(N^{^{(4125)}}_{\mu \alpha_1 \alpha_2, \nu \beta_1 \beta_2} M^{\alpha_1 \beta_1} M^{\alpha_2 \beta_2}\Big)\equiv (M^3)^{\mu \nu} (N^{^{^{(4125)}}} MM)_{\mu \nu}\,,\qquad\nonumber\\
I_{12}^{^{(3)}}&=&(N^{^{(4125)}} MM)_{\mu \nu} (N^{^{(4125)}}MM)^{\mu \nu}\,.
\eea

An alternative approach to try for the construction of independent invariants is the spin-tensor description of the self-dual 5-form which we consider below.

\subsection{$F_5$ invariants in spinor formalism}\label{spinorform}
We refer the reader to Appendix \ref{spin10d-app} which is devoted to mathematical conventions utilized in this Subsection and various useful identities concerning the spin-tensor formalism in ten-dimensional spacetime. Given a self-dual five-form $F_{\m(5)}= F_{\mu_1 \mu_2 \mu_3 \mu_4 \mu_5}$, it is equivalently described by a symmetric rank-two spinor $F^{ab}$ defined by
\bea
F^{ab} = \frac{1}{5!} F_{\m(5)} (\tilde{\s}^{\m(5)} )^{ab} \quad \implies \quad F^{ab} (\s_\m)_{ab} =0~\, \qquad a,b=1,\ldots, 16\,.
\label{Fab*}
\eea
Since there is no metric to lower the upper indices, the invariants of products of $F^{ab}$ are obtained by contracting them with an invariant tensor carrying only lower spinor indices, $I_{a_1 a_2, a_3 a_4}$.
Such invariant tensors may be constructed from
\begin{subequations}\label{I-property1}
\bea
I_{a_1 a_2, b_1b_2} := (\s^\m)_{a_1a_2} (\s_\m)_{b_1 b_2} = I_{b_1 b_2, a_1a_2} = I_{(a_1 a_2), (b_1b_2)}~,
\label{I-property1-a}
\eea
and its analogue with upper indices
\bea
\tilde{I}^{a_1a_2,b_1b_2}  := (\tilde{\s}^\m)^{a_1a_2} (\tilde{\s}_\m)^{b_1 b_2} = \tilde{I}^{b_1 b_2, a_1a_2}
= \tilde{I}^{(a_1 a_2), (b_1b_2)}~.
\label{I-property1-b}
\eea
 There also exists the invariant tensor with upper and lower indices
\bea
J^{a_1a_2}_{b_1b_2}  := (\tilde{\s}^\m)^{a_1a_2} ({\s}_\m)_{b_1 b_2}
%= \tilde{I}^{b_1 b_2, a_1a_2}
= J^{(a_1 a_2)}_{(b_1b_2)}~.
\label{I-property1-c}
\eea
\end{subequations}
It turns out that the role of a basic invariant tensor to contract the indices of $F$s can be played by products
of the structure given in
\eqref{I-property1-a}. Of course, there exist more general invariant tensors, for example
 \bea
 \tilde{I}^{f_1 f_2, f_3 f_4} I_{f_1 a_1, a_2 a_3} I_{f_2 b_1, b_2b_3}
 I_{f_3 c_1,c_2c_3} I_{f_4d_1, d_2d_3}~,
 \eea
 where $\tilde{I}^{f_1 f_2, f_3 f_4}$ is defined in \eqref{I-property1-b}.
 However, each of them may be reduced to an invariant tensor that involves only products
  of \eqref{I-property1-a}, by making use of the identity \eqref{IIJJ}.
  \if{}
  \footnote{Life is much easier in six dimensions, see Appendix \ref{AppendixE}.} {\color{red}[TO MOVE THIS IDENTITY TO APPENDIX D]}
  \fi

It also follows from the definitions \eqref{I-property1-a} - \eqref{I-property1-c} and Fierz identities that
\begin{subequations}
\bea
\tilde{I}^{c_1c_2, a_1a_2} I_{c_1 c_2, b_1 b_2} &=& - 16 J^{a_1a_2}_{b_1b_2}~,
\qquad  \tilde{I}^{c_1c_2, a_1a_2} I_{c_1 b_1, c_2 b_2} = 8 J^{a_1a_2}_{b_1b_2}~, \\
\tilde{I}^{a_1[c_1, c_2]a_2} I_{b_1 c_1, c_2 b_2} &=& 16 \d^{[a_1}_{b_1} \d^{a_2]}_{b_2}~,
\qquad \tilde{I}^{a_1(c_1, c_2)a_2} I_{b_1 c_1, c_2 b_2} = -4 J^{a_1a_2}_{b_1b_2}~.
\eea
\end{subequations}

For constructing invariants of $F^{ab}$, eq. \eqref{Fab*}, there are four building blocks:
\begin{subequations}
\bea
H_{a_1 a_2}{}^{b_1 b_2} &:=& I_{a_1 a_2, c_1 c_2} F^{c_1 b_1} F^{c_2 b_2} = H_{(a_1 a_2)}{}^{(b_1 b_2)}~, \\
G_{a_1 a_2}{}^{b_1 b_2} &:=& I_{a_1 [ c_1 , c_2 ] a_2} F^{c_1 b_1} F^{c_2 b_2}
= I_{c_1 [ a_1 , a_2 ] c_2} F^{c_1 b_1} F^{c_2 b_2} = G_{[a_1 a_2]}{}^{[b_1 b_2]}~,~~~ \\
\Theta_a{}^{b_1,b_2b_3}  &:=& I_{a c_1, c_2 c_3} F^{c_1 b_1} F^{c_2 b_2} F^{c_3 b_3}
= \Q_a{}^{b_1,(b_2b_3)} ~, \label{Theta}\\
\O^{a_1 a_2, a_3 a_4}&:=& I_{b_1b_2, c_1c_2} F^{b_1a_1} F^{b_2a_2} F^{c_1a_3} F^{c_2a_4}
%\non \\
%\O^{a_1 a_2, a_3 a_4}&:=&
%&=&I_{b_1b_2, c_1c_2} F^{b_1a_1} F^{b_2a_2} F^{c_1a_3} %F^{c_2a_4}
=\O^{(a_1 a_2), (a_3 a_4)}
%= \O^{a_1 a_2, a_3 a_4}
~.
\label{Omega}
\eea
\end{subequations}
The additional properties of $H_{a_1 a_2}{}^{b_1 b_2} $ and $G_{a_1 a_2}{}^{b_1 b_2} $ are:
\begin{subequations}
\bea
 H_{a c}{}^{b c} &=&0~, \qquad
F^{a_1 a_2} H_{a_1 a_2}{}^{b_1 b_2} =0 ~,
\\
G_{a c}{}^{b c} &=&0~.
\eea
\end{subequations}
The important algebraic properties of $\Theta_a{}^{b_1,b_2b_3}  $ are
\bea
\Theta_c{}^{c,b_1b_2} = 0~, \qquad \Q_c{}^{a,cb} =0~, \qquad
\Theta_a{}^{(b_1,b_2b_3)} =0~
\eea
and
\bea
\O^{a_1 a_2, a_3 a_4} = \O^{a_3 a_4, a_1 a_2}~, \qquad \O^{a_1 (a_2, a_3 a_4) }=0~.
\label{Omega-properties}
\eea
They follow from the analogous properties of $I_{b_1b_2, c_1c_2} $
given in eqs. \eqref{I-property1} and \eqref{I-property2}.

The building blocks $H_{a_1 a_2}{}^{b_1 b_2}$ and $G_{a_1 a_2}{}^{b_1 b_2}$ are related to $M_{\mu \nu}$, \linebreak $N^{^{(1050)}}_{[\mu_1 \mu_2 \mu_3, \nu_1 \nu_2] \nu_3}$ and $N^{^{(4125)}}_{\mu_1 \mu_2 \mu_3, \nu_1 \nu_2 \nu_3}$ which were used to construct invariants in the tensor form as follows:
\bsubeq
\bea
M_{\mu \nu} &:=& (\tilde{\sigma}_{\mu})^{a_1 a_2} (\sigma_{\nu})_{b_1 b_2} H_{a_1 a_2}{}^{b_1 b_2} \nonumber\\
&=& 8 \,\tr \big[ \sigma_{\mu} \tilde{\sigma}_{\rho(5)} \sigma_{\nu} \tilde{\sigma}_{\delta(5)}\big] F^{\rho(5)} F^{\delta(5)}~.
\eea
One can see that $M_{\mu \nu} = M_{\nu \mu}$ and it is traceless
\bea
\eta^{\mu \nu} M_{\mu \nu} = 0~,
\eea
which follows from the fact that $H_{a_1 a_2}{}^{b_1 b_2}$ has vanishing contractions.
\esubeq
Furthermore,
\bsubeq
\bea
N^{(1050)}_{[\mu_1 \mu_2 \mu_3 \mu_4 \mu_5] \nu} := (\tilde{\sigma}_{\nu})^{a_1 a_2} (\sigma_{\mu_1 \mu_2 \mu_3 \mu_4 \mu_5})_{b_1 b_2} H_{a_1 a_2}{}^{b_1 b_2}~,
\eea
which has vanishing trace
\bea
\eta^{\nu \rho} N^{(1050)}_{[\rho \mu_1 \dots \mu_4] \nu} = 0~,
\eea
since $H_{a_1 a_2}{}^{b_1 b_2}$ has vanishing contractions.
\esubeq
Defining
\bsubeq
\bea
N^{(4125)}_{\mu_1 \mu_2 \mu_3, \nu_1 \nu_2 \nu_3} &:=& (\tilde{\sigma}_{\mu(3)})^{a_1 a_2} (\sigma_{\nu(3)})_{b_1 b_2} G_{a_1 a_2}{}^{b_1 b_2} \nonumber\\
&=& 4 \, \tr \big[ \sigma_{\mu(3)} \tilde{\sigma}_{\rho(5)} \sigma_{\nu(3)} \tilde{\sigma}_{\delta(5)}\big] F^{\rho(5)} F^{\delta(5)}~,
\eea
we see that
\bea
N^{(4125)}_{\mu_1 \mu_2 \mu_3, \nu_1 \nu_2 \nu_3} = N^{(4125)}_{\nu_1 \nu_2 \nu_3, \mu_1 \mu_2 \mu_3}~.
\eea
\esubeq

\subsubsection{4th order invariant}
Every $F^{2n}$ invariant contains an $F^4$ building block of the form $\O^{a_1 a_2, a_3 a_4}$, eq. \eqref{Omega}.
Indeed, given such an $F^{2n}$ invariant, the $4n$ indices of $2n$ $F$s should be contracted with $n$ invariant tensors $I$s. Let us pick one of the tensors $I$s. Its four indices must be contracted with four different $F$s, due to \eqref{D41}.
% {\color{red}[Kurt: can we add a short explanation for this basic statement, or is it obvious?]}
Using this observation, it is easy to see that there is a unique $F^4$ invariant. Indeed, there are only two Lorentz-invariant $F^4$ scalars:
\begin{subequations}
\bea
\U_1&=& \O^{a_1 a_2, a_3 a_4} I_{a_1a_2, a_3 a_4}  ~, \\
\U_2&=& \O^{a_1 a_2, a_3 a_4} I_{a_1a_3, a_2 a_4}  ~.
\eea
\end{subequations}
However, $\U_2$ proves to be proportional to $\U_1$, since
\bea
\U_2=  -\O^{a_1 a_2, a_3 a_4} \Big( I_{a_1a_2, a_4 a_3}   + I_{a_1 a_4, a_3 a_2} \Big)
= - \U_1 - \U_2~,
\eea
and therefore $\U_2 = -\hf \U_1$.

\subsubsection{6th order invariants}
Let us now identify all independent $F^6$ invariants. There are several  Lorentz-invariant $F^6$ structures:
\begin{subequations}
\bea \label{Sigma1}
\S_1 &=& \O^{a_1 a_2, a_3 a_4} I_{b_1 b_2, a_1 a_2} I_{c_1 c_2, a_3 a_4} F^{b_1 c_1} F^{b_2 c_2}~, \\ \label{S2}
\S_2 &=& \O^{a_1 a_2, a_3 a_4} I_{b_1 a_1, a_3 b_2} I_{c_1 a_2, a_4 c_2} F^{b_1 c_1} F^{b_2 c_2} ~, \\
\S_3 &=& \O^{a_1 a_2, a_3 a_4} I_{b_1 a_1, a_2 b_2} I_{c_1 a_3, a_4 c_2} F^{b_1 c_1} F^{b_2 c_2} ~, \\
\S_4 &=& \O^{a_1 a_2, a_3 a_4} I_{b_1 b_2, a_1 a_3} I_{c_1 a_2, a_4 c_2} F^{b_1 c_1} F^{b_2 c_2} ~, \\
\S_5 &=& \O^{a_1 a_2, a_3 a_4} I_{b_1 a_1, a_3 b_2} I_{c_1 a_4, a_2 c_2} F^{b_1 c_1} F^{b_2 c_2} ~.
\eea
\end{subequations}
Here $\S_5$ differs from $\S_2$ by swapping $a_2 \leftrightarrow a_4$ in the factor
$ I_{c_1 a_2, a_4 c_2}$.
The structure $\S_1$ can be rewritten as
\bea
\S_1 = H_{d_1 d_2}{}^{a_1 a_2} H_{a_1 a_2}{}^{ c_1 c_2} H_{c_1c_2}{}^{d_1 d_2}
= \tr (H^3)~.
\eea

The structure $\S_2$ can be reshuffled as follows:
\bea
\S_2 &=& { I_{d_1 d_2, e_1 e_2}} { F^{d_1 a_1}} F^{d_2 a_2}
{ F^{e_1 a_3} F^{e_2 a_4}}
{ I_{b_1 a_1, a_3 b_2}} I_{c_1 a_2, a_4 c_2} {F^{b_1 c_1}} F^{b_2 c_2} \nonumber\\
&=& {H_{d_1d_2}{}^{a_3 a_4}} { H_{a_3 b_2}{}^{d_1 c_1}}
I_{c_1 a_2, a_4 c_2} F^{d_2 a_2} F^{b_2 c_2}~.
\eea
%\MC{Are the colours explained somewhere?}\\
Here $I_{c_1 a_2, a_4 c_2} F^{d_2 a_2} F^{b_2 c_2}$ can be further rewritten as
\bea
&&I_{c_1 a_2, a_4 c_2} F^{a_2 d_2} F^{c_2 b_2}
= I_{c_1 (a_2, c_2) a_4} F^{a_2 d_2} F^{c_2 b_2} + G_{c_1 a_4}{}^{d_2 b_2} \nonumber\\
&=& - \hf I_{a_2c_2, c_1 a_4} F^{a_2 d_2} F^{c_2 b_2}  + G_{c_1 a_4}{}^{d_2 b_2}
= -\hf H_{c_1a_4}{}^{d_2 b_2} + G_{c_1 a_4}{}^{d_2 b_2} ~.
\eea
The outcome of these transformations is
\bea
\S_2 = - \hf { H_{d_1d_2}{}^{a_3 a_4} } {H_{a_3 b_2}{}^{d_1 c_1} } H_{c_1a_4}{}^{d_2 b_2}
+ { H_{d_1d_2}{}^{a_3 a_4} } G_{c_1 a_4}{}^{d_2 b_2} {H_{a_3 b_2}{}^{d_1 c_1} }
\eea
Relabelling the indices gives
\bea
\S_2 = - \hf
H_{a_1a_2}{}^{b_1 b_2} H_{b_1 c_1}{}^{a_2 c_2} H_{b_2c_2}{}^{a_1c_1}
+ {H_{a_1a_2}{}^{b_1c_1} } G_{c_1 c_2}{}^{a_1 b_2} {H_{b_1 b_2}{}^{c_2 a_2} }~.
\label{Sigma2}
\eea

The structures $\S_3$ and $\Sigma_4$ are proportional to $\Sigma_1$. Indeed
\bea
\S_3 &=& \O^{a_1 a_2, a_3 a_4} I_{b_1 (a_1, a_2) b_2} I_{c_1 a_3, a_4 c_2} F^{b_1 c_1} F^{b_2 c_2} \nonumber \\
&=& -\hf \O^{a_1 a_2, a_3 a_4} I_{a_1 a_2, b_1 b_2} I_{c_1 a_3, a_4 c_2} F^{b_1 c_1} F^{b_2 c_2} \nonumber \\
 &=& -\hf \O^{a_1 a_2, a_3 a_4} I_{a_1 a_2, b_1 b_2} I_{c_1 (a_3, a_4) c_2} F^{b_1 c_1} F^{b_2 c_2} \nonumber \\
 &=& \frac 14 \O^{a_1 a_2, a_3 a_4} I_{a_1 a_2, b_1 b_2} I_{a_3 a_4, c_1 c_2} F^{b_1 c_1} F^{b_2 c_2}
 = \frac 14 \S_1~.
\eea
and
\bea
\S_4 &=& \O^{a_2 (a_1, a_3) a_4} I_{b_1 b_2, a_1 a_3} I_{c_1 a_2, a_4 c_2} F^{b_1 c_1} F^{b_2 c_2}
\nonumber \\
 &=& -\hf \O^{a_1 a_3, a_2 a_4} I_{b_1 b_2, a_1 a_3} I_{c_1 (a_2, a_4) c_2} F^{b_1 c_1} F^{b_2 c_2} \nonumber \\
  &=& \frac 14 \O^{a_1 a_3, a_2 a_4} I_{b_1 b_2, a_1 a_3} I_{a_2 a_4, c_1 c_2} F^{b_1 c_1} F^{b_2 c_2}
  = \frac 14 \S_1~.
\eea

Applying the second property in \eqref{Omega-properties} to
the structure $\S_5$ we get
\bea
\S_5 &=& - \Big( \O^{a_1 a_3, a_4 a_2} + \O^{a_1 a_4, a_2 a_3} \Big)
I_{b_1 a_1, a_3 b_2} I_{c_1 a_4, a_2 c_2} F^{b_1 c_1} F^{b_2 c_2} ~.
\label{Sigma5-2}
\eea
The first contribution on the right is
\bea
&&-  \O^{a_1 a_3, a_4 a_2} I_{b_1 a_1, a_3 b_2} I_{c_1 a_4, a_2 c_2} F^{b_1 c_1} F^{b_2 c_2}
= - \O^{a_1 a_3, a_2 a_4} I_{b_1 (a_1, a_3) b_2} I_{c_1 (a_4, a_2) c_2} F^{b_1 c_1} F^{b_2 c_2} \nonumber \\
&=& - \frac14 \O^{a_1 a_3, a_2 a_4} I_{b_1 b_2, a_1 a_3 } I_{c_1 c_2, a_2 a_4} F^{b_1 c_1} F^{b_2 c_2} = - \frac 14 \S_1~
\eea
and the second contribution in \eqref{Sigma5-2} is
\bea
&&-  \O^{a_1 a_4, a_2 a_3}
I_{b_1 a_1, a_3 b_2} I_{c_1 a_4, a_2 c_2} F^{b_1 c_1} F^{b_2 c_2}
=  -  \O^{a_1 a_2, a_3 a_4}
I_{b_1 a_1, a_3 b_2} I_{c_1 a_2, a_4 c_2} F^{b_1 c_1} F^{b_2 c_2} \nonumber \\
&=& - \S_2~.
\eea
As a result,
\bea
\S_5 = - \frac 14 \S_1 - \S_2~.
\label{Sigma5-3}
\eea

% Now, let us consider the first  $F^6$ invariant in \eqref{Sigma2} constructed with $H$:
% \bea
% &&H_{a_1 a_2}{}^{b_1 b_2} H_{b_1 c_1}{}^{a_2 c_2} H_{b_2 c_2}{}^{a_1 c_1}
% = { H_{a_1 a_2}{}^{b_1 b_2}}
% I_{d_1 d_2, b_1 c_1} { F^{d_1 a_2} } F^{d_2 c_2}
% I_{e_1 e_2, b_2 c_2} { F^{e_1 a_1} }F^{e_2 c_1} \nonumber\\
% &=& \O^{e_1 d_1, b_1 b_2} I_{d_1 d_2, b_1 c_1} I_{e_1 e_2, b_2 c_2}
% F^{d_2 c_2} F^{e_2 c_1} \nonumber\\
% &=&  \O^{e_1 d_1, b_1 b_2} I_{d_2 d_1, b_1 c_1} I_{c_2 b_2 ,e_1 e_2}
% F^{d_2 c_2} F^{c_1 e_2} ~.
% \eea
% Relabelling the indices gives
% \bea
% H_{a_1 a_2}{}^{b_1 b_2} H_{b_1 c_1}{}^{a_2 c_2} H_{b_2 c_2}{}^{a_1 c_1}
% =\S_5~.
% \label{Sigma5-4}
% \eea
% Combining the results \eqref{Sigma2}, \eqref{Sigma5-3} and \eqref{Sigma5-4} we conclude that {\color{red}[Kurt: as $\Sigma_3$, $\Sigma_4$ and $\Sigma_5$ have already been shown to be linear combinations of $\Sigma_1$ and $\Sigma_2$, what does this further identity tell us?]}
% \bea
% H_{a_1 a_2}{}^{b_1 b_2} H_{b_1 c_1}{}^{a_2 c_2} H_{b_2 c_2}{}^{a_1 c_1}
% = -\hf \S_1 - 2 {H_{a_1 a_2}{}^{b_1 c_1} } G_{c_1 c_2}{}^{a_1 b_2} {H_{b_1 b_2}{}^{c_2 a_2} }~.
% \eea

Thus our analysis shows that there are two independent $F^6$ invariants, which can be chosen to be \eqref{Sigma1} and \eqref{S2}.

\subsubsection{8th order invariants}
Every $F^8$ invariant proves to contain one $F^3$ factor of the form \eqref{Theta}
and one $F^4$ factor of the form \eqref{Omega}. Due to the relation $\Omega^{b_1 b_2, b_3 b_4} = \Theta_{c}{}^{b_2, b_3 b_4} F^{b_1 c}$, along with
\bea
\Omega^{a_1 a_2, a_3 b} I_{a_1 a_2, a_3 c} = \Theta_{c}{}^{a_1, a_2 a_3} F^{b d} I_{d a_1, a_2 a_3}~,
\eea
some of the independent invariants can be expressed in terms of two $\O$-factors
 \begin{subequations}\label{OO}
 \bea
&&  \O^{a_1 a_2,a_3 a_4} \O^{b_1 b_2,b_3 b_4} I_{a_1 a_2, b_1 b_2} I_{a_3a_4, b_3 b_4} ~,\\
%&& \O^{a_1 a_2,a_3 a_4}\O^{b_1 b_2,b_3 b_4} I_{a_1 a_3, b_1 b_3} I_{a_2 a_4, b_2 b_4}~, \\
&&\O^{a_1 a_2,a_3 a_4}\O^{b_1 b_2,b_3 b_4} I_{a_1 a_2, a_3 b_4} I_{b_1 b_2, b_3 a_4}~,\\
&&  \O^{a_1 [a_2,a_3] a_4} \O^{b_1 [b_2,b_3] b_4} I_{a_1 [b_2, b_3] a_4}
I_{b_1[a_2,a_3]  b_4} ~.
 \eea
 \end{subequations}
 The other independent $F^8$ invariants may be chosen as follows
 \begin{subequations}
 \bea
 && \O^{d a_1,a_2 a_3} \Q_d{}^{b_1, b_2b_3} F^{c_1c_2} I_{c_1 b_1,  a_2 a_3 }
 I_{c_2 a_1, b_2 b_3} ~, \\
  && \O^{d a_1,a_2 a_3} \Q_d{}^{b_1, b_2b_3} F^{c_1c_2} I_{c_1 a_1,  a_2 b_3 }
 I_{c_2 b_1, b_2 a_3} ~, \\
  && \O^{d a_1,a_2 a_3} \Q_d{}^{b_1, b_2b_3} F^{c_1c_2} I_{c_1 a_1,  a_2 b_1 }
 I_{c_2 a_3, b_2 b_3} ~.
 \eea
 \end{subequations}

\section{Conclusion}
In this paper we addressed the problem of the construction and classification of functionally independent invariants of tensor fields of various Lie algebras and showed that the structure of the ring of the invariants, in particular whether or not it is freely generated, is related to the properties of the extended Lie algebra defined in Subsection
\ref{extendedLa}. This observation has been supported by the numerous examples, but its promotion to a strict theorem remains an open problem, as well as the problem of explicit computability of generating functions of most of non-freely generated rings of invariants.  On the other hand, examples considered in this paper may also be useful for applications in field theoretical contexts. In particular, we have elaborated on the complicated problem of the construction of independent higher-order invariants of the self-dual 5-form in ten space-time dimensions, some of which appeared or should appear as higher-order corrections to the effective action of type IIB String Theory.

\subsection*{Acknowledgements}
MC would like to thank Ingmar Saberi for help finding reference literature. JH, KL and DS thank Max Ba$\tilde{\rm n}$ados, Chris Ferko and Roberto Volpato for useful discussions. SMK  is grateful to J.-H. Park for discussions of the spinor formalism in ten dimensions.
Part of the research presented here was conducted by MC and DS during the program ``Cohomological Aspects of Quantum Field Theory'' at the Mittag-Leffler Institute, Djursholm, Sweden, supported by the Swedish Research Council under grant no. 2021-06594. JH acknowledges the kind hospitality of the Department of Physics, University of Western Australia during the final part of this project.
SMK acknowledges the kind hospitality of the INFN, Sezione di Padova during his visits to Padua in 2024 and 2025.
The work of JH and DS is supported by the European Union under the Marie Sklodowska-Curie grant agreement number 101107602.\footnote{Views and opinions expressed are however those of the authors only and do not necessarily reflect those of the European Union or European Research Executive Agency. Neither the European Union nor the granting authority can be held responsible for them.}
The work of SMK and DS is supported in part by the Australian Research Council, project DP230101629.
The work of DS was also partially supported by CARIPARO Foundation under grant n. 68079, the MCI, AEI, FEDER (UE) grant PID2021-125700NB-C21 “Gravity, Supergravity and Superstrings” (GRASS) and  the Basque Government Grant IT1628-22.

\begin{appendix}

\section{Invariants of the 6-dimensional module (symmetric rank-5 tensor) of $\sl(2)$\label{A15Appendix}}
A 6-dimensional module of $A_1=\sl(2)$ of weight (5) can be associated with a totally symmetric rank-5 tensor $T_{\alpha_1\alpha_2\alpha_3\alpha_4\alpha_5}$, where $\alpha_i=1,2$, and the indices are raised, lowered and contracted with the antisymmetric $sl(2)$-invariant matrices $\varepsilon^{\alpha\beta}$ and $\varepsilon_{\alpha\beta}$ ($\varepsilon^{\alpha\gamma}\varepsilon_{\gamma\beta}=\delta^\alpha_\beta$, $\varepsilon_{12}=1$).
We are interested in an explicit form of the independent invariants which one gets by contracting (symmetric) products of $n$ copies of this tensor. As we showed in Section \ref{sl2} independent invariants appear at orders $n=4$, 8, 12 and 18 with one invariant at each order, and at $n=36$ there appears an identity between these four invariants. So that the total number of independent invariants is 3.

To find an explicit form of these invariants let us individualize independent tensors in the contraction of different number of indices in the product of two tensors  $T_{(5)}$.\footnote{In the Appendices A and B the subscripts of tensors like  $T_{(5)}$, $M_{(2)}$, $N_{(6)}$ etc. indicate the number of symmetric indices, which coincides with the weight of the corresponding $\sl(2)$ module.} They correspond to the decomposition of the symmetric product $\big((5)\times (5)\big)_{sym}=(2)+(6)+(10)$ whose dimension is 21.

The contraction of all five indices obviously gives zero. The contraction of four indices produces a symmetric matrix (the module of weight (2))
\be\label{M1}
M_{\alpha\beta}=T_{\alpha\gamma_1\gamma_2\gamma_3\gamma_4}T_{\beta}{}^{\gamma_1\gamma_2\gamma_3\gamma_4}\,.
\ee
The contraction of three indices is not an independent tensor. It is expressed in terms of the components of the matrix $M$ as follows
\be\label{3}
T_{\alpha_1\alpha_2\gamma_1\gamma_2\gamma_3}T^{\beta_1\beta_2\gamma_1\gamma_2\gamma_3}=\delta^{(\beta_1}_{(\alpha_1}M_{\alpha_2)}{}^{\beta_2)}\,,
\ee
where the round brackets denote the symmetrization of $k$ indices with weight $\frac 1{k!}$\,.

The contraction of two indices produces an independent totally symmetric rank-6 tensor which is a 7-dimensional module of $sl(2)$
\be\label{N1}
N_{\alpha_1\alpha_2\alpha_3\beta_1\beta_2\beta_3}=T_{\alpha_1\alpha_2\alpha_3\gamma_1\gamma_2}T_{\beta_1\beta_2\beta_3}{}^
{\gamma_1\gamma_2}-\frac 9{10}\varepsilon_{\beta_1\rho_1}\varepsilon_{\beta_2\rho_2}\varepsilon_{\beta_3\rho_3}
\delta^{(\rho_1}_{(\alpha_1}\delta^{\rho_2}_{\alpha_2} M_{\alpha_3)}{}^{\rho_3)}\,.
\ee
The contraction of one index in the $TT$ product produces a dependent tensor
\be\label{1}
T_{\alpha_1\alpha_2\alpha_3\alpha_4\gamma}T^{\beta_1\beta_2\beta_3\beta_4\gamma}
=2\delta_{(\alpha_4}^{(\beta_4}N_{\alpha_1\alpha_2\alpha_3)}{}^{\beta_1\beta_2\beta_3)}+\frac 45\delta^{(\beta_1}_{(\alpha_1}\delta^{\beta_2}_{\alpha_2} \delta^{\beta_3}_{\alpha_3} M_{\alpha_4)}{}^{\beta_4)}\,.
\ee

Therefore the invariants can be constructed with the use of the tensors \eqref{M1} and \eqref{N1}. Obviously, the result is zero at the orders $n=2k+1$ and at order 2. At order four in $T$ we have the unique invariant
\be\label{MM}
I_4=M_{\alpha\beta}M^{\beta\alpha}=\tr M^2\,,
\ee
and this is the only independent one which can be constructed solely with powers of the symmetric $2 \times 2$ matrix $M$ (since $M_{\alpha\gamma}M^{\gamma\beta}=\frac 12 \delta_{\alpha}^\beta\tr M^2$). The $NN=N^2$ invariant is proportional to $\tr M^2$ due to the composite nature of the tensor $N=TT$ given in eq. \eqref{N1}.

At order $n=6$ in $T$ possible invariants might come from the contractions of $MMM$, $MMN$, $MNN$ and $NNN$, but they are all zero. At the 8th order one (dependent) invariant is the square of the 4th order invariant $(\tr M^2)^2$. A possible form of the independent $I_8$ invariant is
\bea\label{MNMN}
I_8&=& \Big(M^{(\mu_1}{}_{\alpha}N^{\mu_2\mu_3\mu_4\mu_5\mu_6)\alpha}\Big)\Big(M_{(\mu_1}{}^{\beta}N_{\mu_2\mu_3\mu_4\mu_5\mu_6)\beta}\Big)\nonumber\\
&=&\frac 1{12}(\tr M^2)(N^2)+\frac 23 M^{\gamma}{}_{\alpha}N^{\mu_1\mu_2\mu_3\mu_4\alpha\delta}N_{\mu_1\mu_2\mu_3\mu_4\beta\gamma}M_{\delta}{}^{\beta}\,.
\eea
A reason to choose this invariant is that the product of $M_{(2)}N_{(6)}$  contains only one irrep with $\lambda=(6)$ given in brackets in \eqref{MNMN}, which should thus be the same as the single  $\lambda=(6)$ irrep in the decomposition of the symmetric tensor product of four $T_{(5)}$\,:
\be\label{4[5]}
((5)^4)_{sym}=1\times  (20)+1\times (16) +1\times  (14 )+2\times  (12) +1\times  (10) +2\times  (8) +1\times (6) +2\times  (4) +1\times  (0)~,
\ee
where the coefficients $n \times $ count the number of independent irreps of different weights.

Thus an independent  $I_8$ invariant can be chosen as the scalar product of two copies of this $\lambda=(6)$ module (i.e. the tensor $(MN)_{(6)}$ in the first line of \eqref{MNMN}).

Another option is
\be\label{N4}
\hat I_{8}=N_{\alpha_1\alpha_2\mu_1\mu_2\mu_3\mu_4}N^{\mu_1\mu_2\mu_3\mu_4\beta_1\beta_2}N_{\beta_1\beta_2\nu_1\nu_2\nu_3\nu_4}N^{\nu_1\nu_2\nu_3\nu_4\alpha_1\alpha_2}\,.
\ee
The next independent invariant appears at the 12th order in $T_{(5)}$. A possible choice is  an invariant constructed  as a scalar product of two copies of the single rank-(2) tensor in the symmetric tensor product of three  $N_{(6)}$
 \be\label{666}
 (6)\times (6)\times (6)=1\times (18) +1\times (14) +1\times (12) +1\times(10) +1\times ( 8) +2\times ( 6) +1\times ( 2)\,.
 \ee
We thus have
\be\label{N6[2]}
I_{12}= (N^{3})_{\rho_1\rho_2}(N^{3})^{\rho_1\rho_2}\,,
\ee
where
\be\label{N32}
(N^{3})_{\rho_1\rho_2}=
N_{\alpha_1\alpha_2\alpha_3\alpha_4\alpha_5\alpha_6}N^{\alpha_1\alpha_2\alpha_3\alpha_4\nu_1\nu_2} N^{\alpha_5\alpha_6}{}_{\nu_1\nu_2{\rho_1\rho_2}}
\ee
The reason to choose this invariant is that the 6-order of $T_{(5)}$ contains two (2)-modules, one of which is factorized into $(T^2)_{(2)}\cdot (T^4)_{inv}$ and another one should coincide with \eqref{N32}.

Let us now look for a form of the $I_{18}$ invariant. It cannot be constructed as $(N_{(6)})^9$, because there is no invariants at order 9 of the rank-6 tensor. So the invariant should be constructed with powers of $N_{(6)}$ and $M_{(2)}$ or $T_{(5)}$
\be\label{18}
I_{18}=M^pN^q, \qquad  \hat I_{18}=T^{2p}N^q, \qquad 2p+2q=18\,\, (p=1,2,\ldots)
\ee
A simple guess is
$(MN)\cdot N^7 \sim T^4 \cdot N^7$. where for $MN\sim T^4$ and $N^7\sim T^{14}$ we choose (6)-modules in the decomposition of $MN$ and $N^7$. Note that the (6)-module in $MN$ is not contained in $N^2$, so the invariant $(MN)\cdot N^7$ cannot reduce to $N^2N^7=0$.
The (6)-module in $MN\sim T^4$ is
\be\label{MN=T4[6]}
M^{(\mu_1}{}_{\alpha}N^{\mu_2\mu_3\mu_4\mu_5\mu_6)\alpha}\sim T^{\mu_1\mu_2}{}_{\alpha_1\alpha_2\alpha_3}T^{\mu_3\mu_4\alpha_1\beta_1\beta_2}
T^{\mu_5\alpha_2}{}_{\beta_1\gamma_1\gamma_2}T^{\mu_6\alpha_3\gamma_1\gamma_2}{}_{\beta_2}\,
\ee
where in $T^4$ the six indices $\mu_i$ are assumed to be symmetrized.

We thus get the invariant
\be\label{18T4N7}
I_{18}=M^{(\mu_1}{}_{\alpha}N^{\mu_2\mu_3\mu_4\mu_5\mu_6)\alpha}\, (N^7)_{\mu_1\mu_2\mu_3\mu_4\mu_5\mu_6}
\ee
with
\be\label{N76}
(N^7)_{\mu_1\mu_2\mu_3\mu_4\mu_5\mu_6}=
N_{\mu_1\mu_2\,{\color{red}(2)}(2) } N^{(2){\color{blue}(2)}{\color{green}(2)}}N_{{\color{green}(2)}}{}^{{\color{violet}(2)}(2)} N_{\mu_3\mu_4(2)}{}^{(2)}N_{(2){\color{blue}(2)}(2)} N^{(2){\color{red}(2)}(2)}N_{(2){\color{violet}(2)}\mu_5\mu_6}\,
\ee
where the numbers 2 in brackets of different colors denote pairs of indices contracted with pairs of indices of the same color. The black pairs of indices are contracted with their black neighbors.

Note that that the invariant \eqref{18T4N7} is the only non-trivial one which one can construct by contracting $(MN)_{(6)}$ with $N^7_{(6)}$. Indeed in addition to the (6)-irrep of $N^7$, eq. \eqref{N76} used in \eqref{18T4N7} there are {\it six} other (6)-irreps in the decomposition of the symmetric tensor product $N^7$, but they are all factorisable:
\be\label{6irrepsN7}
N_{(6)}  ((N^2)_{inv})^3, \quad N_{(6)} (N^2)_{inv}(N^4)_{inv}; \quad (N^3)_{(6)} ((N^2)_{inv})^2, \quad N^3_{(6)} (N^4)_{inv},
\ee
$$
N^5_{(6)} (N^2)_{inv},\qquad  N^5_{(6^*)} (N^2)_{inv},\qquad N^7_{(6)},
$$
where $(6^*)$ stands for another independent (6)-module in $N^5$.

Upon opening the symmetrization brackets $(\mu_1\ldots\mu_6)$ the invariant \eqref{18T4N7} can also be viewed as the contraction $M^{\mu\nu}(N^8)_{\mu\nu}$, where $(N^8)_{\mu\nu}$ stands for the single (2)-module in the symmetric product of eight $N_{(6)}$.

According to the form of the partition function \eqref{P5Tate} and \eqref{P5TateMinimal} the square of the invariant \eqref{18T4N7} must be equal to a linear combination of 36-order monomials of products of the invariants \eqref{MM}, \eqref{MNMN} and \eqref{N6[2]}. It is however not easy to find the explicit form of this relation for the compactly-looking tensor products \eqref{MM}, \eqref{MNMN}, \eqref{N6[2]} and \eqref{18T4N7}.

To derive such a relation we will construct another set of independent invariants in a weight basis, which in spite of a very cumbersome appearance allows one to use Mathematica to find the relation (see eq. \eqref{18'2}). Let the basis for the 6-dimensional module be $\{r,s,t,u,v,w\}$. We normalise so that the $\sl(2)$ generators act as the vector fields
\begin{align}
e&=r{\*\over\*s}+2s{\*\over\*t}+3t{\*\over\*u}+4u{\*\over\*v}+5v{\*\over\*w}\;,\nn\\
h&=5r{\*\over\*r}+3s{\*\over\*s}+t{\*\over\*t}-u{\*\over\*u}-3v{\*\over\*v}-5w{\*\over\*w}\;,\\
f&=5s{\*\over\*r}+4t{\*\over\*s}+3u{\*\over\*t}+2v{\*\over\*u}+w{\*\over\*v}\;.\nn
\end{align}
We can then make a general Ansatz spanned by all polynomials with eigenvalue 0 under $h$. The requirement of their invariance under $e$ and $f$ can then be solved using Mathematica.
As a result, the following polynomials of the components of these fields at order 4, 8, 12 and 18 generate the ring of invariants $S^\fg$:

\noindent $I_4=r^2 w^2-10 r s v w+4 r t u w+16 r t v^2-12 r u^2 v+16
   s^2 u w+9 s^2 v^2-12 s t^2 w-76 s t u v+48 s u^3+48
   t^3 v-32 t^2 u^2$,

\noindent $I_8=r^3 t u w^3-r^3 t v^2 w^2-3 r^3 u^2 v w^2  +5 r^3 u v^3
   w-2 r^3 v^5-r^2 s^2 u w^3 +r^2 s^2 v^2 w^2-3 r^2 s
   t^2 w^3 +11 r^2 s t u v w^2-5 r^2 s t v^3 w +12 r^2 s
   u^3 w^2-30 r^2 s u^2 v^2 w +15 r^2 s u v^4 +12 r^2 t^3
   v w^2-21 r^2 t^2 u^2 w^2-34 r^2 t^2 u v^2 w +22 r^2
   t^2 v^4 +78 r^2 t u^3 v w-48 r^2 t u^2 v^3-27 r^2 u^5
   w +18 r^2 u^4 v^2 +5 r s^3 t w^3-5 r s^3 u v w^2-30 r
   s^2 t^2 v w^2-34 r s^2 t u^2 w^2 +133 r s^2 t u v^2
   w-54 r s^2 t v^4-18 r s^2 u^3 v w +3 r s^2 u^2 v^3 +78
   r s t^3 u w^2-18 r s t^3 v^2 w-220 r s t^2 u^2 v
   w +106 r s t^2 u v^3 +93 r s t u^4 w-30 r s t u^3
   v^2-9 r s u^5 v-27 r t^5 w^2 +93 r t^4 u v w-38 r t^4
   v^3-42 r t^3 u^3 w+8 r t^3 u^2 v^2+6 r t^2 u^4 v-2
   s^5 w^3+15 s^4 t v w^2+22 s^4 u^2 w^2-54 s^4 u v^2
   w+27 s^4 v^4-48 s^3 t^2 u w^2+3 s^3 t^2 v^2 w+106
   s^3 t u^2 v w-81 s^3 t u v^3-38 s^3 u^4 w+38 s^3 u^3
   v^2+18 s^2 t^4 w^2-30 s^2 t^3 u v w+38 s^2 t^3 v^3+8
   s^2 t^2 u^3 w+25 s^2 t^2 u^2 v^2-57 s^2 t u^4 v+18
   s^2 u^6-9 s t^5 v w+6 s t^4 u^2 w-57 s t^4 u v^2+74
   s t^3 u^3 v-24 s t^2 u^5+18 t^6 v^2-24 t^5 u^2 v+8
   t^4 u^4$,

\noindent
\bea
I_{12}&=&r^4 t^2 u^2 w^4-2 r^4 t^2 u v^2 w^3+r^4 t^2 v^4 w^2-6 r^4 t u^3 v w^3+16 r^4 t
   u^2 v^3 w^2-14 r^4 t u v^5 w +4 r^4 t v^7\nonumber\\
  && {\txt {+(more other hundreds of terms),}} \nonumber
\eea
   \if{}
  $$+4 r^4 u^5 w^3-11 r^4 u^4 v^2 w^2+10 r^4
   u^3 v^4 w-3 r^4 u^2 v^6-2 r^3 s^2 t u^2 w^4+4 r^3 s^2 t u v^2 w^3-2 r^3 s^2 t v^4
   w^2+6 r^3 s^2 u^3 v w^3-16 r^3 s^2 u^2 v^3 w^2+14 r^3 s^2 u v^5 w-4 r^3 s^2 v^7-6
   r^3 s t^3 u w^4+6 r^3 s t^3 v^2 w^3+50 r^3 s t^2 u^2 v w^3-82 r^3 s t^2 u v^3
   w^2+32 r^3 s t^2 v^5 w-36 r^3 s t u^4 w^3+30 r^3 s t u^3 v^2 w^2+30 r^3 s t u^2 v^4
   w-24 r^3 s t u v^6+28 r^3 s u^5 v w^2-50 r^3 s u^4 v^3 w+22 r^3 s u^3 v^5+4 r^3
   t^5 w^4-36 r^3 t^4 u v w^3+16 r^3 t^4 v^3 w^2+22 r^3 t^3 u^3 w^3+50 r^3 t^3 u^2 v^2
   w^2-16 r^3 t^3 u v^4 w-16 r^3 t^3 v^6-54 r^3 t^2 u^4 v w^2-46 r^3 t^2 u^3 v^3
   w+60 r^3 t^2 u^2 v^5+6 r^3 t u^6 w^2+70 r^3 t u^5 v^2 w-56 r^3 t u^4 v^4-18 r^3
   u^7 v w+14 r^3 u^6 v^3+r^2 s^4 u^2 w^4-2 r^2 s^4 u v^2 w^3+r^2 s^4 v^4 w^2+16 r^2
   s^3 t^2 u w^4-16 r^2 s^3 t^2 v^2 w^3-82 r^2 s^3 t u^2 v w^3+132 r^2 s^3 t u v^3
   w^2-50 r^2 s^3 t v^5 w+16 r^2 s^3 u^4 w^3+14 r^2 s^3 u^3 v^2 w^2-60 r^2 s^3 u^2
   v^4 w+30 r^2 s^3 u v^6-11 r^2 s^2 t^4 w^4+30 r^2 s^2 t^3 u v w^3+14 r^2 s^2 t^3 v^3
   w^2+50 r^2 s^2 t^2 u^3 w^3-168 r^2 s^2 t^2 u^2 v^2 w^2+48 r^2 s^2 t^2 u v^4 w+4 r^2
   s^2 t^2 v^6+48 r^2 s^2 t u^4 v w^2+2 r^2 s^2 t u^3 v^3 w-6 r^2 s^2 t u^2 v^5-62
   r^2 s^2 u^6 w^2+90 r^2 s^2 u^5 v^2 w-39 r^2 s^2 u^4 v^4+28 r^2 s t^5 v w^3-54 r^2
   s t^4 u^2 w^3+48 r^2 s t^4 u v^2 w^2-112 r^2 s t^4 v^4 w-82 r^2 s t^3 u^3 v
   w^2+170 r^2 s t^3 u^2 v^3 w+104 r^2 s t^3 u v^5+108 r^2 s t^2 u^5 w^2+42 r^2 s t^2
   u^4 v^2 w-298 r^2 s t^2 u^3 v^4-242 r^2 s t u^6 v w+294 r^2 s t u^5 v^3+72 r^2 s
   u^8 w-78 r^2 s u^7 v^2+6 r^2 t^6 u w^3-62 r^2 t^6 v^2 w^2+108 r^2 t^5 u^2 v
   w^2+164 r^2 t^5 u v^3 w+24 r^2 t^5 v^5-63 r^2 t^4 u^4 w^2-394 r^2 t^4 u^3 v^2
   w-194 r^2 t^4 u^2 v^4+324 r^2 t^3 u^5 v w+440 r^2 t^3 u^4 v^3-78 r^2 t^2 u^7
   w-428 r^2 t^2 u^6 v^2+180 r^2 t u^8 v-27 r^2 u^{10}-14 r s^5 t u w^4+14 r s^5 t
   v^2 w^3+32 r s^5 u^2 v w^3-50 r s^5 u v^3 w^2+18 r s^5 v^5 w+10 r s^4 t^3 w^4+30 r
   s^4 t^2 u v w^3-60 r s^4 t^2 v^3 w^2-16 r s^4 t u^3 w^3+48 r s^4 t u^2 v^2 w^2-38
   r s^4 t u v^4 w+36 r s^4 t v^6-112 r s^4 u^4 v w^2+204 r s^4 u^3 v^3 w-102 r s^4
   u^2 v^5-50 r s^3 t^4 v w^3-46 r s^3 t^3 u^2 w^3+2 r s^3 t^3 u v^2 w^2+204 r s^3
   t^3 v^4 w+170 r s^3 t^2 u^3 v w^2-42 r s^3 t^2 u^2 v^3 w-308 r s^3 t^2 u v^5+164 r
   s^3 t u^5 w^2-674 r s^3 t u^4 v^2 w+590 r s^3 t u^3 v^4+128 r s^3 u^6 v w-138 r
   s^3 u^5 v^3+70 r s^2 t^5 u w^3+90 r s^2 t^5 v^2 w^2+42 r s^2 t^4 u^2 v w^2-674 r
   s^2 t^4 u v^3 w+4 r s^2 t^4 v^5-394 r s^2 t^3 u^4 w^2+714 r s^2 t^3 u^3 v^2 w+652 r
   s^2 t^3 u^2 v^4+498 r s^2 t^2 u^5 v w-1246 r s^2 t^2 u^4 v^3-224 r s^2 t u^7 w+516
   r s^2 t u^6 v^2-48 r s^2 u^8 v-18 r s t^7 w^3-242 r s t^6 u v w^2+128 r s t^6 v^3
   w+324 r s t^5 u^3 w^2+498 r s t^5 u^2 v^2 w-136 r s t^5 u v^4-1078 r s t^4 u^4 v
   w-206 r s t^4 u^3 v^3+342 r s t^3 u^6 w+804 r s t^3 u^5 v^2-506 r s t^2 u^7 v+90 r
   s t u^9+72 r t^8 v w^2-78 r t^7 u^2 w^2-224 r t^7 u v^2 w-16 r t^7 v^4+342 r t^6
   u^3 v w+220 r t^6 u^2 v^3-106 r t^5 u^5 w-392 r t^5 u^4 v^2+222 r t^4 u^6 v-40 r
   t^3 u^8+4 s^7 u w^4-4 s^7 v^2 w^3-3 s^6 t^2 w^4-24 s^6 t u v w^3+30 s^6 t v^3
   w^2-16 s^6 u^3 w^3+4 s^6 u^2 v^2 w^2+36 s^6 u v^4 w-27 s^6 v^6+22 s^5 t^3 v w^3+60
   s^5 t^2 u^2 w^3-6 s^5 t^2 u v^2 w^2-102 s^5 t^2 v^4 w+104 s^5 t u^3 v w^2-308 s^5
   t u^2 v^3 w+234 s^5 t u v^5+24 s^5 u^5 w^2+4 s^5 u^4 v^2 w-32 s^5 u^3 v^4-56 s^4
   t^4 u w^3-39 s^4 t^4 v^2 w^2-298 s^4 t^3 u^2 v w^2+590 s^4 t^3 u v^3 w-32 s^4 t^3
   v^5-194 s^4 t^2 u^4 w^2+652 s^4 t^2 u^3 v^2 w-713 s^4 t^2 u^2 v^4-136 s^4 t u^5 v
   w+246 s^4 t u^4 v^3-16 s^4 u^7 w-4 s^4 u^6 v^2+14 s^3 t^6 w^3+294 s^3 t^5 u v
   w^2-138 s^3 t^5 v^3 w+440 s^3 t^4 u^3 w^2-1246 s^3 t^4 u^2 v^2 w+246 s^3 t^4 u
   v^4-206 s^3 t^3 u^4 v w+866 s^3 t^3 u^3 v^3+220 s^3 t^2 u^6 w-550 s^3 t^2 u^5
   v^2+56 s^3 t u^7 v+4 s^3 u^9-78 s^2 t^7 v w^2-428 s^2 t^6 u^2 w^2+516 s^2 t^6 u
   v^2 w-4 s^2 t^6 v^4+804 s^2 t^5 u^3 v w-550 s^2 t^5 u^2 v^3-392 s^2 t^4 u^5
   w-139 s^2 t^4 u^4 v^2+354 s^2 t^3 u^6 v-83 s^2 t^2 u^8+180 s t^8 u w^2-48 s t^8
   v^2 w-506 s t^7 u^2 v w+56 s t^7 u v^3+222 s t^6 u^4 w+354 s t^6 u^3 v^2-330 s t^5
   u^5 v+72 s t^4 u^7-27 t^{10} w^2+90 t^9 u v w+4 t^9 v^3-40 t^8 u^3 w-83 t^8 u^2
   v^2+72 t^7 u^4 v-16 t^6 u^6$,
\fi

\noindent
\bea
I_{18}&=& r^7 u^5 w^6-5 r^7 u^4 v^2 w^5+10 r^7 u^3 v^4 w^4-10 r^7 u^2 v^6 w^3+5 r^7 u
   v^8 w^2-r^7 v^{10} w-15 r^6 s t u^4 w^6
   \nonumber\\
  && \txt {+(more other hundreds of terms),} \nonumber
\eea
 \if{}
$
   +60 r^6 s t u^3 v^2 w^5-90 r^6 s t u^2 v^4
   w^4+60 r^6 s t u v^6 w^3-15 r^6 s t v^8 w^2+10 r^6 s u^5 v w^5-35 r^6 s u^4 v^3
   w^4+40 r^6 s u^3 v^5 w^3-10 r^6 s u^2 v^7 w^2-10 r^6 s u v^9 w+5 r^6 s v^{11}-r^6
   t^5 w^7+15 r^6 t^4 u v w^6-10 r^6 t^4 v^3 w^5-90 r^6 t^3 u^2 v^2 w^5+120 r^6 t^3 u
   v^4 w^4-40 r^6 t^3 v^6 w^3+60 r^6 t^2 u^4 v w^5+30 r^6 t^2 u^3 v^3 w^4-180 r^6 t^2
   u^2 v^5 w^3+120 r^6 t^2 u v^7 w^2-20 r^6 t^2 v^9 w-15 r^6 t u^6 w^5-110 r^6 t u^5
   v^2 w^4+265 r^6 t u^4 v^4 w^3-200 r^6 t u^3 v^6 w^2+65 r^6 t u^2 v^8 w-10 r^6 t u
   v^{10}+45 r^6 u^7 v w^4-100 r^6 u^6 v^3 w^3+81 r^6 u^5 v^5 w^2-30 r^6 u^4 v^7 w+5
   r^6 u^3 v^9+10 r^5 s^3 u^4 w^6-40 r^5 s^3 u^3 v^2 w^5+60 r^5 s^3 u^2 v^4 w^4-40
   r^5 s^3 u v^6 w^3+10 r^5 s^3 v^8 w^2+5 r^5 s^2 t^4 w^7-60 r^5 s^2 t^3 u v w^6+40
   r^5 s^2 t^3 v^3 w^5+90 r^5 s^2 t^2 u^3 w^6-90 r^5 s^2 t^2 u v^4 w^4+30 r^5 s^2 t^2
   v^6 w^3-210 r^5 s^2 t u^4 v w^5+120 r^5 s^2 t u^3 v^3 w^4+360 r^5 s^2 t u^2 v^5
   w^3-420 r^5 s^2 t u v^7 w^2+130 r^5 s^2 t v^9 w-5 r^5 s^2 u^6 w^5+195 r^5 s^2 u^5
   v^2 w^4-315 r^5 s^2 u^4 v^4 w^3+40 r^5 s^2 u^3 v^6 w^2+165 r^5 s^2 u^2 v^8 w-75
   r^5 s^2 u v^{10}-10 r^5 s t^5 v w^6-60 r^5 s t^4 u^2 w^6+210 r^5 s t^4 u v^2
   w^5-110 r^5 s t^4 v^4 w^4+60 r^5 s t^3 u^2 v^3 w^4-360 r^5 s t^3 u v^5 w^3+240 r^5
   s t^3 v^7 w^2+30 r^5 s t^2 u^5 w^5-210 r^5 s t^2 u^4 v^2 w^4-180 r^5 s t^2 u^3 v^4
   w^3+1140 r^5 s t^2 u^2 v^6 w^2-870 r^5 s t^2 u v^8 w+130 r^5 s t^2 v^{10}+310 r^5 s
   t u^6 v w^4-240 r^5 s t u^5 v^3 w^3-390 r^5 s t u^4 v^5 w^2+280 r^5 s t u^3 v^7
   w+30 r^5 s t u^2 v^9-180 r^5 s u^8 w^4+300 r^5 s u^7 v^2 w^3-120 r^5 s u^6 v^4
   w^2+30 r^5 s u^5 v^6 w-30 r^5 s u^4 v^8+15 r^5 t^6 u w^6+5 r^5 t^6 v^2 w^5-30 r^5
   t^5 u^2 v w^5-270 r^5 t^5 u v^3 w^4+196 r^5 t^5 v^5 w^3+225 r^5 t^4 u^3 v^2 w^4+615
   r^5 t^4 u^2 v^4 w^3-660 r^5 t^4 u v^6 w^2+45 r^5 t^4 v^8 w-120 r^5 t^3 u^5 v
   w^4-220 r^5 t^3 u^4 v^3 w^3-980 r^5 t^3 u^3 v^5 w^2+1320 r^5 t^3 u^2 v^7 w-260
   r^5 t^3 u v^9+60 r^5 t^2 u^7 w^4-500 r^5 t^2 u^6 v^2 w^3+2235 r^5 t^2 u^5 v^4
   w^2-1995 r^5 t^2 u^4 v^6 w+370 r^5 t^2 u^3 v^8+360 r^5 t u^8 v w^3-1320 r^5 t u^7
   v^3 w^2+1110 r^5 t u^6 v^5 w-210 r^5 t u^5 v^7-81 r^5 u^{10} w^3+270 r^5 u^9 v^2
   w^2-225 r^5 u^8 v^4 w+45 r^5 u^7 v^6-10 r^4 s^4 t^3 w^7+90 r^4 s^4 t^2 u v w^6-60
   r^4 s^4 t^2 v^3 w^5-120 r^4 s^4 t u^3 w^6+90 r^4 s^4 t u^2 v^2 w^5+110 r^4 s^4 u^4
   v w^5-50 r^4 s^4 u^3 v^3 w^4-240 r^4 s^4 u^2 v^5 w^3+280 r^4 s^4 u v^7 w^2-90 r^4
   s^4 v^9 w+35 r^4 s^3 t^4 v w^6-30 r^4 s^3 t^3 u^2 w^6-120 r^4 s^3 t^3 u v^2 w^5+50
   r^4 s^3 t^3 v^4 w^4-60 r^4 s^3 t^2 u^3 v w^5+360 r^4 s^3 t^2 u v^5 w^3-210 r^4 s^3
   t^2 v^7 w^2+270 r^4 s^3 t u^5 w^5+575 r^4 s^3 t u^4 v^2 w^4-1700 r^4 s^3 t u^3 v^4
   w^3+480 r^4 s^3 t u^2 v^6 w^2+670 r^4 s^3 t u v^8 w-315 r^4 s^3 t v^{10}-685 r^4
   s^3 u^6 v w^4+540 r^4 s^3 u^5 v^3 w^3+1515 r^4 s^3 u^4 v^5 w^2-2080 r^4 s^3 u^3 v^7
   w+705 r^4 s^3 u^2 v^9+110 r^4 s^2 t^5 u w^6-195 r^4 s^2 t^5 v^2 w^5+210 r^4 s^2 t^4
   u^2 v w^5-575 r^4 s^2 t^4 u v^3 w^4+660 r^4 s^2 t^4 v^5 w^3-225 r^4 s^2 t^3 u^4
   w^5+1350 r^4 s^2 t^3 u^2 v^4 w^3-1440 r^4 s^2 t^3 u v^6 w^2-75 r^4 s^2 t^3 v^8
   w-1965 r^4 s^2 t^2 u^5 v w^4+6000 r^4 s^2 t^2 u^4 v^3 w^3-7050 r^4 s^2 t^2 u^3 v^5
   w^2+3000 r^4 s^2 t^2 u^2 v^7 w+265 r^4 s^2 t^2 u v^9+1420 r^4 s^2 t u^7 w^4-3810
   r^4 s^2 t u^6 v^2 w^3+2310 r^4 s^2 t u^5 v^4 w^2+1795 r^4 s^2 t u^4 v^6 w-1800 r^4
   s^2 t u^3 v^8+240 r^4 s^2 u^8 v w^3+30 r^4 s^2 u^7 v^3 w^2-870 r^4 s^2 u^6 v^5
   w+615 r^4 s^2 u^5 v^7-45 r^4 s t^7 w^6-310 r^4 s t^6 u v w^5+685 r^4 s t^6 v^3
   w^4+120 r^4 s t^5 u^3 w^5+1965 r^4 s t^5 u^2 v^2 w^4-2210 r^4 s t^5 u v^4 w^3-960
   r^4 s t^5 v^6 w^2-11700 r^4 s t^4 u^3 v^3 w^3+15435 r^4 s t^4 u^2 v^5 w^2-2760 r^4
   s t^4 u v^7 w+555 r^4 s t^4 v^9-780 r^4 s t^3 u^6 w^4+14040 r^4 s t^3 u^5 v^2
   w^3-10625 r^4 s t^3 u^4 v^4 w^2-3220 r^4 s t^3 u^3 v^6 w-570 r^4 s t^3 u^2
   v^8-5840 r^4 s t^2 u^7 v w^3-540 r^4 s t^2 u^6 v^3 w^2+5550 r^4 s t^2 u^5 v^5
   w+1285 r^4 s t^2 u^4 v^7+990 r^4 s t u^9 w^3+3150 r^4 s t u^8 v^2 w^2-3600 r^4 s t
   u^7 v^4 w-615 r^4 s t u^6 v^6-945 r^4 s u^{10} v w^2+900 r^4 s u^9 v^3 w+45 r^4 s
   u^8 v^5+180 r^4 t^8 v w^5-60 r^4 t^7 u^2 w^5-1420 r^4 t^7 u v^2 w^4+25 r^4 t^7 v^4
   w^3+780 r^4 t^6 u^3 v w^4+5760 r^4 t^6 u^2 v^3 w^3-2945 r^4 t^6 u v^5 w^2+1390 r^4
   t^6 v^7 w-7020 r^4 t^5 u^4 v^2 w^3-180 r^4 t^5 u^3 v^4 w^2-1275 r^4 t^5 u^2 v^6
   w-1110 r^4 t^5 u v^8+3120 r^4 t^4 u^6 v w^3+3900 r^4 t^4 u^5 v^3 w^2+1240 r^4 t^4
   u^4 v^5 w+3155 r^4 t^4 u^3 v^7-515 r^4 t^3 u^8 w^3-2920 r^4 t^3 u^7 v^2 w^2-940
   r^4 t^3 u^6 v^4 w-4300 r^4 t^3 u^5 v^6+675 r^4 t^2 u^9 v w^2+510 r^4 t^2 u^8 v^3
   w+2940 r^4 t^2 u^7 v^5-135 r^4 t u^{10} v^2 w-990 r^4 t u^9 v^4+135 r^4 u^{11}
   v^3+10 r^3 s^6 t^2 w^7-60 r^3 s^6 t u v w^6+40 r^3 s^6 t v^3 w^5+40 r^3 s^6 u^3
   w^6-30 r^3 s^6 u^2 v^2 w^5-40 r^3 s^5 t^3 v w^6+180 r^3 s^5 t^2 u^2 w^6-360 r^3
   s^5 t^2 u v^2 w^5+240 r^3 s^5 t^2 v^4 w^4+360 r^3 s^5 t u^3 v w^5-360 r^3 s^5 t u^2
   v^3 w^4-196 r^3 s^5 u^5 w^5-660 r^3 s^5 u^4 v^2 w^4+1840 r^3 s^5 u^3 v^4 w^3-1040
   r^3 s^5 u^2 v^6 w^2-180 r^3 s^5 u v^8 w+216 r^3 s^5 v^{10}-265 r^3 s^4 t^4 u
   w^6+315 r^3 s^4 t^4 v^2 w^5+180 r^3 s^4 t^3 u^2 v w^5+1700 r^3 s^4 t^3 u v^3
   w^4-1840 r^3 s^4 t^3 v^5 w^3-615 r^3 s^4 t^2 u^4 w^5-1350 r^3 s^4 t^2 u^3 v^2
   w^4+1560 r^3 s^4 t^2 u v^6 w^2+135 r^3 s^4 t^2 v^8 w+2210 r^3 s^4 t u^5 v w^4-4100
   r^3 s^4 t u^4 v^3 w^3+6000 r^3 s^4 t u^3 v^5 w^2-4880 r^3 s^4 t u^2 v^7 w+990 r^3
   s^4 t u v^9-25 r^3 s^4 u^7 w^4+3710 r^3 s^4 u^6 v^2 w^3-10755 r^3 s^4 u^5 v^4
   w^2+9875 r^3 s^4 u^4 v^6 w-2845 r^3 s^4 u^3 v^8+100 r^3 s^3 t^6 w^6+240 r^3 s^3 t^5
   u v w^5-540 r^3 s^3 t^5 v^3 w^4+220 r^3 s^3 t^4 u^3 w^5-6000 r^3 s^3 t^4 u^2 v^2
   w^4+4100 r^3 s^3 t^4 u v^4 w^3+1340 r^3 s^3 t^4 v^6 w^2+11700 r^3 s^3 t^3 u^4 v
   w^4-15240 r^3 s^3 t^3 u^2 v^5 w^2+6960 r^3 s^3 t^3 u v^7 w-1620 r^3 s^3 t^3
   v^9-5760 r^3 s^3 t^2 u^6 w^4-16120 r^3 s^3 t^2 u^5 v^2 w^3+26700 r^3 s^3 t^2 u^4
   v^4 w^2-5240 r^3 s^3 t^2 u^3 v^6 w-1640 r^3 s^3 t^2 u^2 v^8+6560 r^3 s^3 t u^7 v
   w^3+7240 r^3 s^3 t u^6 v^3 w^2-24240 r^3 s^3 t u^5 v^5 w+11420 r^3 s^3 t u^4
   v^7-980 r^3 s^3 u^9 w^3-3420 r^3 s^3 u^8 v^2 w^2+8100 r^3 s^3 u^7 v^4 w-3880 r^3
   s^3 u^6 v^6-300 r^3 s^2 t^7 v w^5+500 r^3 s^2 t^6 u^2 w^5+3810 r^3 s^2 t^6 u v^2
   w^4-3710 r^3 s^2 t^6 v^4 w^3-14040 r^3 s^2 t^5 u^3 v w^4+16120 r^3 s^2 t^5 u^2 v^3
   w^3-540 r^3 s^2 t^5 u v^5 w^2+600 r^3 s^2 t^5 v^7 w+7020 r^3 s^2 t^4 u^5 w^4-1950
   r^3 s^2 t^4 u^3 v^4 w^2-17670 r^3 s^2 t^4 u^2 v^6 w+4170 r^3 s^2 t^4 u v^8+480 r^3
   s^2 t^3 u^6 v w^3-31040 r^3 s^2 t^3 u^5 v^3 w^2+45180 r^3 s^2 t^3 u^4 v^5 w-3160
   r^3 s^2 t^3 u^3 v^7-140 r^3 s^2 t^2 u^8 w^3+18000 r^3 s^2 t^2 u^7 v^2 w^2-12180
   r^3 s^2 t^2 u^6 v^4 w-13430 r^3 s^2 t^2 u^5 v^6-7200 r^3 s^2 t u^9 v w^2-120 r^3
   s^2 t u^8 v^3 w+9960 r^3 s^2 t u^7 v^5+1890 r^3 s^2 u^{11} w^2-540 r^3 s^2 u^{10}
   v^2 w-1710 r^3 s^2 u^9 v^4-360 r^3 s t^8 u w^5-240 r^3 s t^8 v^2 w^4+5840 r^3 s
   t^7 u^2 v w^4-6560 r^3 s t^7 u v^3 w^3+8460 r^3 s t^7 v^5 w^2-3120 r^3 s t^6 u^4
   w^4-480 r^3 s t^6 u^3 v^2 w^3-25880 r^3 s t^6 u^2 v^4 w^2-1820 r^3 s t^6 u v^6
   w-3620 r^3 s t^6 v^8+49680 r^3 s t^5 u^4 v^3 w^2+17520 r^3 s t^5 u^3 v^5 w+13500
   r^3 s t^5 u^2 v^7-120 r^3 s t^4 u^7 w^3-32280 r^3 s t^4 u^6 v^2 w^2-46880 r^3 s
   t^4 u^5 v^4 w-30040 r^3 s t^4 u^4 v^6+12860 r^3 s t^3 u^8 v w^2+32000 r^3 s t^3 u^7
   v^3 w+46160 r^3 s t^3 u^6 v^5-2700 r^3 s t^2 u^{10} w^2-8820 r^3 s t^2 u^9 v^2
   w-34620 r^3 s t^2 u^8 v^4+1080 r^3 s t u^{11} v w+12060 r^3 s t u^{10} v^3-1620
   r^3 s u^{12} v^2+81 r^3 t^{10} w^5-990 r^3 t^9 u v w^4+980 r^3 t^9 v^3 w^3+515 r^3
   t^8 u^3 w^4+140 r^3 t^8 u^2 v^2 w^3-195 r^3 t^8 u v^4 w^2-5575 r^3 t^8 v^6 w+120
   r^3 t^7 u^4 v w^3-800 r^3 t^7 u^3 v^3 w^2+22600 r^3 t^7 u^2 v^5 w+7240 r^3 t^7 u
   v^7-1260 r^3 t^6 u^5 v^2 w^2-42330 r^3 t^6 u^4 v^4 w-34340 r^3 t^6 u^3 v^6+480
   r^3 t^5 u^7 v w^2+48360 r^3 t^5 u^6 v^3 w+73828 r^3 t^5 u^5 v^5+105 r^3 t^4 u^9
   w^2-30265 r^3 t^4 u^8 v^2 w-92290 r^3 t^4 u^7 v^4+9540 r^3 t^3 u^{10} v w+69220
   r^3 t^3 u^9 v^3-1215 r^3 t^2 u^{12} w-30510 r^3 t^2 u^{11} v^2+7290 r^3 t u^{13}
   v-729 r^3 u^{15}-5 r^2 s^8 t w^7+15 r^2 s^8 u v w^6-10 r^2 s^8 v^3 w^5+10 r^2 s^7
   t^2 v w^6-120 r^2 s^7 t u^2 w^6+420 r^2 s^7 t u v^2 w^5-280 r^2 s^7 t v^4 w^4-240
   r^2 s^7 u^3 v w^5+210 r^2 s^7 u^2 v^3 w^4+200 r^2 s^6 t^3 u w^6-40 r^2 s^6 t^3 v^2
   w^5-1140 r^2 s^6 t^2 u^2 v w^5-480 r^2 s^6 t^2 u v^3 w^4+1040 r^2 s^6 t^2 v^5
   w^3+660 r^2 s^6 t u^4 w^5+1440 r^2 s^6 t u^3 v^2 w^4-1560 r^2 s^6 t u^2 v^4 w^3+960
   r^2 s^6 u^5 v w^4-1340 r^2 s^6 u^4 v^3 w^3-2440 r^2 s^6 u^3 v^5 w^2+4320 r^2 s^6
   u^2 v^7 w-1620 r^2 s^6 u v^9-81 r^2 s^5 t^5 w^6+390 r^2 s^5 t^4 u v w^5-1515 r^2
   s^5 t^4 v^3 w^4+980 r^2 s^5 t^3 u^3 w^5+7050 r^2 s^5 t^3 u^2 v^2 w^4-6000 r^2 s^5
   t^3 u v^4 w^3+2440 r^2 s^5 t^3 v^6 w^2-15435 r^2 s^5 t^2 u^4 v w^4+15240 r^2 s^5
   t^2 u^3 v^3 w^3-6480 r^2 s^5 t^2 u v^7 w+1215 r^2 s^5 t^2 v^9+2945 r^2 s^5 t u^6
   w^4+540 r^2 s^5 t u^5 v^2 w^3-795 r^2 s^5 t u^4 v^4 w^2-4180 r^2 s^5 t u^3 v^6
   w+4185 r^2 s^5 t u^2 v^8-8460 r^2 s^5 u^7 v w^3+20390 r^2 s^5 u^6 v^3 w^2-16194
   r^2 s^5 u^5 v^5 w+3765 r^2 s^5 u^4 v^7+120 r^2 s^4 t^6 v w^5-2235 r^2 s^4 t^5 u^2
   w^5-2310 r^2 s^4 t^5 u v^2 w^4+10755 r^2 s^4 t^5 v^4 w^3+10625 r^2 s^4 t^4 u^3 v
   w^4-26700 r^2 s^4 t^4 u^2 v^3 w^3+795 r^2 s^4 t^4 u v^5 w^2-10070 r^2 s^4 t^4 v^7
   w+180 r^2 s^4 t^3 u^5 w^4+1950 r^2 s^4 t^3 u^4 v^2 w^3+36510 r^2 s^4 t^3 u^2 v^6
   w+25880 r^2 s^4 t^2 u^6 v w^3-32370 r^2 s^4 t^2 u^5 v^3 w^2-12180 r^2 s^4 t^2 u^4
   v^5 w-9850 r^2 s^4 t^2 u^3 v^7+195 r^2 s^4 t u^8 w^3-43800 r^2 s^4 t u^7 v^2
   w^2+72755 r^2 s^4 t u^6 v^4 w-18750 r^2 s^4 t u^5 v^6+14115 r^2 s^4 u^9 v
   w^2-23790 r^2 s^4 u^8 v^3 w+8175 r^2 s^4 u^7 v^5+1320 r^2 s^3 t^7 u w^5-30 r^2 s^3
   t^7 v^2 w^4+540 r^2 s^3 t^6 u^2 v w^4-7240 r^2 s^3 t^6 u v^3 w^3-20390 r^2 s^3 t^6
   v^5 w^2-3900 r^2 s^3 t^5 u^4 w^4+31040 r^2 s^3 t^5 u^3 v^2 w^3+32370 r^2 s^3 t^5
   u^2 v^4 w^2+38820 r^2 s^3 t^5 u v^6 w+9310 r^2 s^3 t^5 v^8-49680 r^2 s^3 t^4 u^5 v
   w^3-91260 r^2 s^3 t^4 u^3 v^5 w-50550 r^2 s^3 t^4 u^2 v^7+800 r^2 s^3 t^3 u^7
   w^3+81840 r^2 s^3 t^3 u^6 v^2 w^2+360 r^2 s^3 t^3 u^5 v^4 w+101450 r^2 s^3 t^3 u^4
   v^6-8220 r^2 s^3 t^2 u^8 v w^2-58080 r^2 s^3 t^2 u^7 v^3 w-34300 r^2 s^3 t^2 u^6
   v^5-7590 r^2 s^3 t u^{10} w^2+41640 r^2 s^3 t u^9 v^2 w-4650 r^2 s^3 t u^8
   v^4-5580 r^2 s^3 u^{11} v w+1980 r^2 s^3 u^{10} v^3-270 r^2 s^2 t^9 w^5-3150 r^2
   s^2 t^8 u v w^4+3420 r^2 s^2 t^8 v^3 w^3+2920 r^2 s^2 t^7 u^3 w^4-18000 r^2 s^2 t^7
   u^2 v^2 w^3+43800 r^2 s^2 t^7 u v^4 w^2+5030 r^2 s^2 t^7 v^6 w+32280 r^2 s^2 t^6 u^4
   v w^3-81840 r^2 s^2 t^6 u^3 v^3 w^2-85800 r^2 s^2 t^6 u^2 v^5 w-28710 r^2 s^2 t^6
   u v^7+1260 r^2 s^2 t^5 u^6 w^3+181980 r^2 s^2 t^5 u^4 v^4 w+153480 r^2 s^2 t^5 u^3
   v^6-26700 r^2 s^2 t^4 u^7 v w^2-41360 r^2 s^2 t^4 u^6 v^3 w-306900 r^2 s^2 t^4
   u^5 v^5+14360 r^2 s^2 t^3 u^9 w^2-16170 r^2 s^2 t^3 u^8 v^2 w+243000 r^2 s^2 t^3
   u^7 v^4+2340 r^2 s^2 t^2 u^{10} v w-89550 r^2 s^2 t^2 u^9 v^3+270 r^2 s^2 t u^{12}
   w+15120 r^2 s^2 t u^{11} v^2-810 r^2 s^2 u^{13} v+945 r^2 s t^{10} v w^4-675 r^2 s
   t^9 u^2 w^4+7200 r^2 s t^9 u v^2 w^3-14115 r^2 s t^9 v^4 w^2-12860 r^2 s t^8 u^3 v
   w^3+8220 r^2 s t^8 u^2 v^3 w^2+150 r^2 s t^8 u v^5 w+6155 r^2 s t^8 v^7-480 r^2 s
   t^7 u^5 w^3+26700 r^2 s t^7 u^4 v^2 w^2+63960 r^2 s t^7 u^3 v^4 w-6660 r^2 s t^7
   u^2 v^6-180600 r^2 s t^6 u^5 v^3 w-71610 r^2 s t^6 u^4 v^5-4755 r^2 s t^5 u^8
   w^2+141240 r^2 s t^5 u^7 v^2 w+219730 r^2 s t^5 u^6 v^4-45130 r^2 s t^4 u^9 v
   w-240975 r^2 s t^4 u^8 v^3+5580 r^2 s t^3 u^{11} w+128490 r^2 s t^3 u^{10}
   v^2-34155 r^2 s t^2 u^{12} v+3645 r^2 s t u^{14}-1890 r^2 t^{11} v^2 w^3+2700 r^2
   t^{10} u^2 v w^3+7590 r^2 t^{10} u v^3 w^2+8256 r^2 t^{10} v^5 w-105 r^2 t^9 u^4
   w^3-14360 r^2 t^9 u^3 v^2 w^2-43605 r^2 t^9 u^2 v^4 w-12310 r^2 t^9 u v^6+4755
   r^2 t^8 u^5 v w^2+77790 r^2 t^8 u^4 v^3 w+59835 r^2 t^8 u^3 v^5-57060 r^2 t^7 u^6
   v^2 w-114960 r^2 t^7 u^5 v^4+19020 r^2 t^6 u^8 v w+109660 r^2 t^6 u^7 v^3-2481 r^2
   t^5 u^{10} w-56110 r^2 t^5 u^9 v^2+14895 r^2 t^4 u^{11} v-1620 r^2 t^3 u^{13}+r
   s^{10} w^7+10 r s^9 t v w^6+20 r s^9 u^2 w^6-130 r s^9 u v^2 w^5+90 r s^9 v^4
   w^4-65 r s^8 t^2 u w^6-165 r s^8 t^2 v^2 w^5+870 r s^8 t u^2 v w^5-670 r s^8 t u
   v^3 w^4+180 r s^8 t v^5 w^3-45 r s^8 u^4 w^5+75 r s^8 u^3 v^2 w^4-135 r s^8 u^2
   v^4 w^3+30 r s^7 t^4 w^6-280 r s^7 t^3 u v w^5+2080 r s^7 t^3 v^3 w^4-1320 r s^7
   t^2 u^3 w^5-3000 r s^7 t^2 u^2 v^2 w^4+4880 r s^7 t^2 u v^4 w^3-4320 r s^7 t^2 v^6
   w^2+2760 r s^7 t u^4 v w^4-6960 r s^7 t u^3 v^3 w^3+6480 r s^7 t u^2 v^5 w^2-1390
   r s^7 u^6 w^4-600 r s^7 u^5 v^2 w^3+10070 r s^7 u^4 v^4 w^2-12600 r s^7 u^3 v^6
   w+4050 r s^7 u^2 v^8-30 r s^6 t^5 v w^5+1995 r s^6 t^4 u^2 w^5-1795 r s^6 t^4 u
   v^2 w^4-9875 r s^6 t^4 v^4 w^3+3220 r s^6 t^3 u^3 v w^4+5240 r s^6 t^3 u^2 v^3
   w^3+4180 r s^6 t^3 u v^5 w^2+12600 r s^6 t^3 v^7 w+1275 r s^6 t^2 u^5 w^4+17670 r
   s^6 t^2 u^4 v^2 w^3-36510 r s^6 t^2 u^3 v^4 w^2-6075 r s^6 t^2 u v^8+1820 r s^6 t
   u^6 v w^3-38820 r s^6 t u^5 v^3 w^2+66650 r s^6 t u^4 v^5 w-19800 r s^6 t u^3
   v^7+5575 r s^6 u^8 w^3-5030 r s^6 u^7 v^2 w^2-4255 r s^6 u^6 v^4 w+2175 r s^6 u^5
   v^6-1110 r s^5 t^6 u w^5+870 r s^5 t^6 v^2 w^4-5550 r s^5 t^5 u^2 v w^4+24240 r
   s^5 t^5 u v^3 w^3+16194 r s^5 t^5 v^5 w^2-1240 r s^5 t^4 u^4 w^4-45180 r s^5 t^4
   u^3 v^2 w^3+12180 r s^5 t^4 u^2 v^4 w^2-66650 r s^5 t^4 u v^6 w-8550 r s^5 t^4
   v^8-17520 r s^5 t^3 u^5 v w^3+91260 r s^5 t^3 u^4 v^3 w^2+62100 r s^5 t^3 u^2
   v^7-22600 r s^5 t^2 u^7 w^3+85800 r s^5 t^2 u^6 v^2 w^2-148890 r s^5 t^2 u^5 v^4
   w+1850 r s^5 t^2 u^4 v^6-150 r s^5 t u^8 v w^2+15440 r s^5 t u^7 v^3 w+10350 r s^5
   t u^6 v^5-8256 r s^5 u^{10} w^2+12210 r s^5 u^9 v^2 w-7050 r s^5 u^8 v^4+225 r s^4
   t^8 w^5+3600 r s^4 t^7 u v w^4-8100 r s^4 t^7 v^3 w^3+940 r s^4 t^6 u^3 w^4+12180 r
   s^4 t^6 u^2 v^2 w^3-72755 r s^4 t^6 u v^4 w^2+4255 r s^4 t^6 v^6 w+46880 r s^4 t^5
   u^4 v w^3-360 r s^4 t^5 u^3 v^3 w^2+148890 r s^4 t^5 u^2 v^5 w+38950 r s^4 t^5 u
   v^7+42330 r s^4 t^4 u^6 w^3-181980 r s^4 t^4 u^5 v^2 w^2-220125 r s^4 t^4 u^3
   v^6-63960 r s^4 t^3 u^7 v w^2+181600 r s^4 t^3 u^6 v^3 w+159000 r s^4 t^3 u^5
   v^5+43605 r s^4 t^2 u^9 w^2-62025 r s^4 t^2 u^8 v^2 w-92500 r s^4 t^2 u^7
   v^4-20610 r s^4 t u^{10} v w+41250 r s^4 t u^9 v^3+5445 r s^4 u^{12} w-6525 r s^4
   u^{11} v^2-900 r s^3 t^9 v w^4-510 r s^3 t^8 u^2 w^4+120 r s^3 t^8 u v^2 w^3+23790
   r s^3 t^8 v^4 w^2-32000 r s^3 t^7 u^3 v w^3+58080 r s^3 t^7 u^2 v^3 w^2-15440 r
   s^3 t^7 u v^5 w-12500 r s^3 t^7 v^7-48360 r s^3 t^6 u^5 w^3+41360 r s^3 t^6 u^4
   v^2 w^2-181600 r s^3 t^6 u^3 v^4 w-18400 r s^3 t^6 u^2 v^6+180600 r s^3 t^5 u^6 v
   w^2+289800 r s^3 t^5 u^4 v^5-77790 r s^3 t^4 u^8 w^2-87000 r s^3 t^4 u^7 v^2
   w-318500 r s^3 t^4 u^6 v^4+92200 r s^3 t^3 u^9 v w+179500 r s^3 t^3 u^8 v^3-17520
   r s^3 t^2 u^{11} w-69000 r s^3 t^2 u^{10} v^2+15300 r s^3 t u^{12} v-1350 r s^3
   u^{14}+135 r s^2 t^{10} u w^4+540 r s^2 t^{10} v^2 w^3+8820 r s^2 t^9 u^2 v
   w^3-41640 r s^2 t^9 u v^3 w^2-12210 r s^2 t^9 v^5 w+30265 r s^2 t^8 u^4 w^3+16170
   r s^2 t^8 u^3 v^2 w^2+62025 r s^2 t^8 u^2 v^4 w+44225 r s^2 t^8 u v^6-141240 r s^2
   t^7 u^5 v w^2+87000 r s^2 t^7 u^4 v^3 w-129000 r s^2 t^7 u^3 v^5+57060 r s^2 t^6
   u^7 w^2-5250 r s^2 t^6 u^5 v^4-46050 r s^2 t^5 u^8 v w+122800 r s^2 t^5 u^7
   v^3+10595 r s^2 t^4 u^{10} w-88125 r s^2 t^4 u^9 v^2+27300 r s^2 t^3 u^{11} v-3375
   r s^2 t^2 u^{13}-1080 r s t^{11} u v w^3+5580 r s t^{11} v^3 w^2-9540 r s t^{10}
   u^3 w^3-2340 r s t^{10} u^2 v^2 w^2+20610 r s t^{10} u v^4 w-4350 r s t^{10}
   v^6+45130 r s t^9 u^4 v w^2-92200 r s t^9 u^3 v^3 w-25050 r s t^9 u^2 v^5-19020 r
   s t^8 u^6 w^2+46050 r s t^8 u^5 v^2 w+138750 r s t^8 u^4 v^4-178200 r s t^7 u^6
   v^3-1650 r s t^6 u^9 w+103950 r s t^6 u^8 v^2-30250 r s t^5 u^{10} v+3600 r s t^4
   u^{12}+1215 r t^{12} u^2 w^3-270 r t^{12} u v^2 w^2-5445 r t^{12} v^4 w-5580 r
   t^{11} u^3 v w^2+17520 r t^{11} u^2 v^3 w+8700 r t^{11} u v^5+2481 r t^{10} u^5
   w^2-10595 r t^{10} u^4 v^2 w-31150 r t^{10} u^3 v^4+1650 r t^9 u^6 v w+37950 r t^9
   u^5 v^3-22275 r t^8 u^7 v^2+6600 r t^7 u^9 v-800 r t^6 u^{11}-5 s^{11} v w^6+10
   s^{10} t u w^6+75 s^{10} t v^2 w^5-130 s^{10} u^2 v w^5+315 s^{10} u v^3 w^4-216
   s^{10} v^5 w^3-5 s^9 t^3 w^6-30 s^9 t^2 u v w^5-705 s^9 t^2 v^3 w^4+260 s^9 t u^3
   w^5-265 s^9 t u^2 v^2 w^4-990 s^9 t u v^4 w^3+1620 s^9 t v^6 w^2-555 s^9 u^4 v
   w^4+1620 s^9 u^3 v^3 w^3-1215 s^9 u^2 v^5 w^2+30 s^8 t^4 v w^5-370 s^8 t^3 u^2
   w^5+1800 s^8 t^3 u v^2 w^4+2845 s^8 t^3 v^4 w^3+570 s^8 t^2 u^3 v w^4+1640 s^8 t^2
   u^2 v^3 w^3-4185 s^8 t^2 u v^5 w^2-4050 s^8 t^2 v^7 w+1110 s^8 t u^5 w^4-4170 s^8
   t u^4 v^2 w^3+6075 s^8 t u^2 v^6 w+3620 s^8 u^6 v w^3-9310 s^8 u^5 v^3 w^2+8550 s^8
   u^4 v^5 w-3375 s^8 u^3 v^7+210 s^7 t^5 u w^5-615 s^7 t^5 v^2 w^4-1285 s^7 t^4 u^2
   v w^4-11420 s^7 t^4 u v^3 w^3-3765 s^7 t^4 v^5 w^2-3155 s^7 t^3 u^4 w^4+3160 s^7
   t^3 u^3 v^2 w^3+9850 s^7 t^3 u^2 v^4 w^2+19800 s^7 t^3 u v^6 w+3375 s^7 t^3
   v^8-13500 s^7 t^2 u^5 v w^3+50550 s^7 t^2 u^4 v^3 w^2-62100 s^7 t^2 u^3 v^5
   w-7240 s^7 t u^7 w^3+28710 s^7 t u^6 v^2 w^2-38950 s^7 t u^5 v^4 w+25875 s^7 t u^4
   v^6-6155 s^7 u^8 v w^2+12500 s^7 u^7 v^3 w-7375 s^7 u^6 v^5-45 s^6 t^7 w^5+615
   s^6 t^6 u v w^4+3880 s^6 t^6 v^3 w^3+4300 s^6 t^5 u^3 w^4+13430 s^6 t^5 u^2 v^2
   w^3+18750 s^6 t^5 u v^4 w^2-2175 s^6 t^5 v^6 w+30040 s^6 t^4 u^4 v w^3-101450 s^6
   t^4 u^3 v^3 w^2-1850 s^6 t^4 u^2 v^5 w-25875 s^6 t^4 u v^7+34340 s^6 t^3 u^6
   w^3-153480 s^6 t^3 u^5 v^2 w^2+220125 s^6 t^3 u^4 v^4 w+6660 s^6 t^2 u^7 v
   w^2+18400 s^6 t^2 u^6 v^3 w-73375 s^6 t^2 u^5 v^5+12310 s^6 t u^9 w^2-44225 s^6 t
   u^8 v^2 w+42500 s^6 t u^7 v^4+4350 s^6 u^{10} v w-5125 s^6 u^9 v^3-45 s^5 t^8 v
   w^4-2940 s^5 t^7 u^2 w^4-9960 s^5 t^7 u v^2 w^3-8175 s^5 t^7 v^4 w^2-46160 s^5
   t^6 u^3 v w^3+34300 s^5 t^6 u^2 v^3 w^2-10350 s^5 t^6 u v^5 w+7375 s^5 t^6
   v^7-73828 s^5 t^5 u^5 w^3+306900 s^5 t^5 u^4 v^2 w^2-159000 s^5 t^5 u^3 v^4
   w+73375 s^5 t^5 u^2 v^6+71610 s^5 t^4 u^6 v w^2-289800 s^5 t^4 u^5 v^3 w-59835 s^5
   t^3 u^8 w^2+129000 s^5 t^3 u^7 v^2 w+80500 s^5 t^3 u^6 v^4+25050 s^5 t^2 u^9 v
   w-80125 s^5 t^2 u^8 v^3-8700 s^5 t u^{11} w+19875 s^5 t u^{10} v^2-1125 s^5
   u^{12} v+990 s^4 t^9 u w^4+1710 s^4 t^9 v^2 w^3+34620 s^4 t^8 u^2 v w^3+4650 s^4 t^8
   u v^3 w^2+7050 s^4 t^8 v^5 w+92290 s^4 t^7 u^4 w^3-243000 s^4 t^7 u^3 v^2 w^2+92500
   s^4 t^7 u^2 v^4 w-42500 s^4 t^7 u v^6-219730 s^4 t^6 u^5 v w^2+318500 s^4 t^6 u^4
   v^3 w-80500 s^4 t^6 u^3 v^5+114960 s^4 t^5 u^7 w^2+5250 s^4 t^5 u^6 v^2 w-138750
   s^4 t^4 u^8 v w-1250 s^4 t^4 u^7 v^3+31150 s^4 t^3 u^{10} w+40000 s^4 t^3 u^9
   v^2-18750 s^4 t^2 u^{11} v+2250 s^4 t u^{13}-135 s^3 t^{11} w^4-12060 s^3 t^{10}
   u v w^3-1980 s^3 t^{10} v^3 w^2-69220 s^3 t^9 u^3 w^3+89550 s^3 t^9 u^2 v^2
   w^2-41250 s^3 t^9 u v^4 w+5125 s^3 t^9 v^6+240975 s^3 t^8 u^4 v w^2-179500 s^3 t^8
   u^3 v^3 w+80125 s^3 t^8 u^2 v^5-109660 s^3 t^7 u^6 w^2-122800 s^3 t^7 u^5 v^2
   w+1250 s^3 t^7 u^4 v^4+178200 s^3 t^6 u^7 v w-37950 s^3 t^5 u^9 w-37125 s^3 t^5
   u^8 v^2+17875 s^3 t^4 u^{10} v-2125 s^3 t^3 u^{12}+1620 s^2 t^{12} v w^3+30510 s^2
   t^{11} u^2 w^3-15120 s^2 t^{11} u v^2 w^2+6525 s^2 t^{11} v^4 w-128490 s^2 t^{10}
   u^3 v w^2+69000 s^2 t^{10} u^2 v^3 w-19875 s^2 t^{10} u v^5+56110 s^2 t^9 u^5
   w^2+88125 s^2 t^9 u^4 v^2 w-40000 s^2 t^9 u^3 v^4-103950 s^2 t^8 u^6 v w+37125 s^2
   t^8 u^5 v^3+22275 s^2 t^7 u^8 w-4125 s^2 t^6 u^9 v+500 s^2 t^5 u^{11}-7290 s
   t^{13} u w^3+810 s t^{13} v^2 w^2+34155 s t^{12} u^2 v w^2-15300 s t^{12} u v^3
   w+1125 s t^{12} v^5-14895 s t^{11} u^4 w^2-27300 s t^{11} u^3 v^2 w+18750 s t^{11}
   u^2 v^4+30250 s t^{10} u^5 v w-17875 s t^{10} u^4 v^3-6600 s t^9 u^7 w+4125 s t^9
   u^6 v^2+729 t^{15} w^3-3645 t^{14} u v w^2+1350 t^{14} v^3 w+1620 t^{13} u^3
   w^2+3375 t^{13} u^2 v^2 w-2250 t^{13} u v^4-3600 t^{12} u^4 v w+2125 t^{12} u^3
   v^3+800 t^{11} u^6 w-500 t^{11} u^5 v^2$.
\fi
There are choices of representatives for $I_8$ and $I_{12}$. We have chosen them so that the coefficient for $r^4w^4$ in $I_8$ and those for
$r^6w^6$ and $r^5tuw^5$ in $I_{12}$ are 0. $I_4$, $I_8$ and $I_{12}$ are even under a Chevalley involution, while $I_{18}$ is odd. These invariants are related by the identity
\begin{align}\label{18'2}
I_{18}^2+27I_{12}^3+\tfrac92I_{12}^2I_8I_4-\tfrac1{16}I_{12}^2I_4^3-\tfrac12I_{12}I_8^3+\tfrac18I_{12}I_8^2I_4^2
-\tfrac1{16}I_8^4I_4=0\;.
\end{align}
{From this relation it follows that the invariant $I_{18}$ is functionally dependent.}

\section{Invariants of the 7-dimensional module (symmetric rank-6 tensor) of $\sl(2)$ \label{A16Appendix}}
The tensor in question is now an elementary totally symmetric rank-6 tensor $N_{\mu_1\ldots\mu_6}$.
In this case, as we showed in Section \ref{sl2}, there are four functionally independent invariants which  appear at order 2, 4, 6 and 10. And at order 15 there appears an invariant which is functionally dependent of those four.
A possible choice of the independent invariants is
\be\label{N2}
I_{2}=N_{\mu_1\ldots\mu_6}N^{\mu_1\ldots\mu_6}\,,
\ee
\be\label{N4*}
I_{4}=N_{\alpha_1\alpha_2\mu_1\mu_2\mu_3\mu_4}N^{\mu_1\mu_2\mu_3\mu_4\beta_1\beta_2}N_{\beta_1\beta_2\nu_1\nu_2\nu_3\nu_4}N^{\nu_1\nu_2\nu_3\nu_4\alpha_1\alpha_2}\,,
\ee
\be\label{N6}
I_{6}= N_{\alpha_1\alpha_2\alpha_3\alpha_4\alpha_5\alpha_6}N^{\alpha_1\alpha_2\alpha_3\alpha_4\nu_1\nu_2} N_{\nu_1\nu_2\rho_1\rho_2\rho_3\rho_4}N^{\rho_1\rho_2\rho_3\rho_4\mu_1\mu_2\mu_3\mu_4} N_{\mu_1\mu_2\gamma_1\gamma_2\gamma_3\gamma_4}N^{\gamma_1\gamma_2\gamma_3\gamma_4\alpha_5\alpha_6}
\,,
\ee
instead one can also choose the invariant \eqref{N6[2]}.

The 10th-order invariant can be chosen as follows
\be\label{N10}
I_{10}=(N^7)_{(\mu_1\ldots\mu_6)}\Big(N^{\mu_1\mu_2\nu_1\nu_2\rho_1\rho_2}N^{\mu_3\mu_4}{}_{\nu_1\nu_2\lambda_1\lambda_2}N^{\mu_5\mu_6\lambda_1\lambda_2}{}_{\rho_1\rho_2}\Big)\,,
\ee
where $(N^7)_{\mu_1\ldots\mu_6}$
is the same as in \eqref{N76},

Finally, a possible form of the 15th-order invariant is
\be\label{I15}
I_{15}=(N^7)_{\mu\nu\alpha\beta}{}^{\alpha\beta}(N^7)^{(\mu\rho_1\rho_2\rho_3\rho_4\rho_5)}N_{\rho_1\rho_2\rho_3\rho_4\rho_5}{}^{\nu}:=(N^7)_{\mu\nu\alpha\beta}{}^{\alpha\beta}(N^8)^{\mu\nu}.
\ee
The choice of this invariant is related to the fact that the decomposition of the symmetric product $N^7$ contains a single non-factorizable symmetric rank-2 tensor and so does the decomposition of $N^8$.

Again, as in the case of Appendix \ref{A15Appendix}, to find an explicit relation between  invariants $I_2$, $I_4$, $I_6$. $I_{10}$ and $I_{15}$ (see eq. \eqref{15'2}) we construct them in a weight basis using Mathematica as was explained in Appendix A.
Let the basis for the 7-dimensional module be $\{q,r,s,t,u,v,w\}$. We normalise so that the $\sl(2)$ generators act as the vector fields
\begin{align}
e&=q{\*\over\*r}+2r{\*\over\*s}+3s{\*\over\*t}+4t{\*\over\*u}+5u{\*\over\*v}+6v{\*\over\*w}\;,\nn\\
h&=6q{\*\over\*q}+4r{\*\over\*r}+2s{\*\over\*s}-2u{\*\over\*u}-4v{\*\over\*v}-6w{\*\over\*w}\;,\\
f&=6r{\*\over\*s}+5s{\*\over\*r}+4t{\*\over\*s}+3u{\*\over\*t}+2v{\*\over\*u}+w{\*\over\*v}\;.\nn
\end{align}
The following polynomials of the components of these fields at order 2, 4, 6, 10 and 15  generate the ring of the invariants \nolinebreak $S^\fg$:

\noindent$I_2=q w-6 r v+15 s u-10 t^2$,

\noindent$I_4=q s u w-q s v^2-q t^2 w+2 q t u v-q u^3-r^2 u w+r^2 v^2+2 r s t w-2 r s u v-2 r
   t^2 v+2 r t u^2-s^3 w+2 s^2 t v+s^2 u^2-3 s t^2 u+t^4$,

\noindent$I_6=q^2 t^2 w^2-6 q^2 t u v w+4 q^2 t v^3+4 q^2 u^3 w-3 q^2 u^2 v^2-6 q r s t w^2+18
   q r s u v w-12 q r s v^3+12 q r t^2 v w-18 q r t u^2 w+6 q r u^3 v+4 q s^3 w^2-18 q
   s^2 t v w-24 q s^2 u^2 w+30 q s^2 u v^2+54 q s t^2 u w-12 q s t^2 v^2-42 q s t u^2
   v+12 q s u^4-20 q t^4 w+24 q t^3 u v-8 q t^2 u^3+4 r^3 t w^2-12 r^3 u v w+8 r^3
   v^3-3 r^2 s^2 w^2+30 r^2 s u^2 w-24 r^2 s u v^2-12 r^2 t^2 u w-24 r^2 t^2 v^2+60 r^2
   t u^2 v-27 r^2 u^4+6 r s^3 v w-42 r s^2 t u w+60 r s^2 t v^2-30 r s^2 u^2 v+24 r s
   t^3 w-84 r s t^2 u v+66 r s t u^3+24 r t^4 v-24 r t^3 u^2+12 s^4 u w-27 s^4 v^2-8
   s^3 t^2 w+66 s^3 t u v-8 s^3 u^3-24 s^2 t^3 v-39 s^2 t^2 u^2+36 s t^4 u-8
   t^6$,

\noindent
\bea
I_{10}&=&q^4 u^3 w^3-3 q^4 u^2 v^2 w^2+3 q^4 u v^4 w-q^4 v^6-12 q^3 r t u^2 w^3+24 q^3
   r t u v^2 w^2-12 q^3 r t v^4 w
\nonumber\\
  && \txt {+(more other hundreds of terms).} \nonumber
\eea
\if{}
$
+12 q^3 r u^3 v w^2-24 q^3 r u^2 v^3 w+12 q^3 r u
   v^5+q^3 s^3 w^4-12 q^3 s^2 t v w^3
   +18 q^3 s^2 u v^2 w^2-9 q^3 s^2 v^4 w+48 q^3 s
   t^2 v^2 w^2+60 q^3 s t u^2 v w^2-264 q^3 s t u v^3 w+132 q^3 s t v^5-87 q^3 s u^4
   w^2+228 q^3 s u^3 v^2 w-114 q^3 s u^2 v^4+17 q^3 t^4 w^3-204 q^3 t^3 u v w^2+72
   q^3 t^3 v^3 w+144 q^3 t^2 u^3 w^2+366 q^3 t^2 u^2 v^2 w-264 q^3 t^2 u v^4-516 q^3
   t u^4 v w+340 q^3 t u^3 v^3+146 q^3 u^6 w-102 q^3 u^5 v^2-3 q^2 r^2 s^2 w^4+24 q^2
   r^2 s t v w^3+18 q^2 r^2 s u^2 w^3-72 q^2 r^2 s u v^2 w^2+36 q^2 r^2 s v^4 w+48 q^2
   r^2 t^2 u w^3-96 q^2 r^2 t^2 v^2 w^2-144 q^2 r^2 t u^2 v w^2+336 q^2 r^2 t u v^3
   w-120 q^2 r^2 t v^5+57 q^2 r^2 u^4 w^2-120 q^2 r^2 u^3 v^2 w+36 q^2 r^2 u^2 v^4+12
   q^2 r s^3 v w^3+60 q^2 r s^2 t u w^3-144 q^2 r s^2 t v^2 w^2-360 q^2 r s^2 u^2 v
   w^2+756 q^2 r s^2 u v^3 w-342 q^2 r s^2 v^5-204 q^2 r s t^3 w^3+876 q^2 r s t^2 u
   v w^2-168 q^2 r s t^2 v^3 w+228 q^2 r s t u^3 w^2-1692 q^2 r s t u^2 v^2 w+780 q^2
   r s t u v^4+312 q^2 r s u^4 v w-108 q^2 r s u^3 v^3+510 q^2 r t^4 v w^2-708 q^2 r
   t^3 u^2 w^2-1560 q^2 r t^3 u v^2 w+624 q^2 r t^3 v^4+2868 q^2 r t^2 u^3 v w-996
   q^2 r t^2 u^2 v^3-924 q^2 r t u^5 w+36 q^2 r t u^4 v^2+144 q^2 r u^6 v-87 q^2 s^4
   u w^3+57 q^2 s^4 v^2 w^2+144 q^2 s^3 t^2 w^3+228 q^2 s^3 t u v w^2-216 q^2 s^3 t
   v^3 w+732 q^2 s^3 u^3 w^2-1638 q^2 s^3 u^2 v^2 w+855 q^2 s^3 u v^4-708 q^2 s^2 t^3
   v w^2-2004 q^2 s^2 t^2 u^2 w^2+3636 q^2 s^2 t^2 u v^2 w-960 q^2 s^2 t^2 v^4+2100
   q^2 s^2 t u^3 v w-2100 q^2 s^2 t u^2 v^3-957 q^2 s^2 u^5 w+855 q^2 s^2 u^4
   v^2+2085 q^2 s t^4 u w^2-528 q^2 s t^4 v^2 w-5448 q^2 s t^3 u^2 v w+2280 q^2 s t^3
   u v^3+2184 q^2 s t^2 u^4 w+210 q^2 s t^2 u^3 v^2-828 q^2 s t u^5 v+126 q^2 s
   u^7-564 q^2 t^6 w^2+1812 q^2 t^5 u v w-504 q^2 t^5 v^3-768 q^2 t^4 u^3 w-390 q^2
   t^4 u^2 v^2+480 q^2 t^3 u^4 v-84 q^2 t^2 u^6+3 q r^4 s w^4-12 q r^4 t v w^3-9 q
   r^4 u^2 w^3+36 q r^4 u v^2 w^2-18 q r^4 v^4 w-24 q r^3 s^2 v w^3-264 q r^3 s t u
   w^3+336 q r^3 s t v^2 w^2+756 q r^3 s u^2 v w^2-1296 q r^3 s u v^3 w+504 q r^3 s
   v^5+72 q r^3 t^3 w^3-168 q r^3 t^2 u v w^2+48 q r^3 t^2 v^3 w-216 q r^3 t u^3
   w^2+504 q r^3 t u^2 v^2 w-288 q r^3 t u v^4-36 q r^3 u^4 v w+72 q r^3 u^3 v^3+228
   q r^2 s^3 u w^3-120 q r^2 s^3 v^2 w^2+366 q r^2 s^2 t^2 w^3-1692 q r^2 s^2 t u v
   w^2+504 q r^2 s^2 t v^3 w-1638 q r^2 s^2 u^3 w^2+3942 q r^2 s^2 u^2 v^2 w-1728 q
   r^2 s^2 u v^4-1560 q r^2 s t^3 v w^2+3636 q r^2 s t^2 u^2 w^2+2520 q r^2 s t^2 u
   v^2 w-1944 q r^2 s t^2 v^4-8208 q r^2 s t u^3 v w+5112 q r^2 s t u^2 v^3+2520 q
   r^2 s u^5 w-1890 q r^2 s u^4 v^2-528 q r^2 t^4 u w^2+588 q r^2 t^4 v^2 w+288 q r^2
   t^3 u^2 v w+144 q r^2 t^3 u v^3-114 q r^2 t^2 u^4 w-1632 q r^2 t^2 u^3 v^2+1692 q
   r^2 t u^5 v-486 q r^2 u^7-516 q r s^4 t w^3+312 q r s^4 u v w^2-36 q r s^4 v^3
   w+2868 q r s^3 t^2 v w^2+2100 q r s^3 t u^2 w^2-8208 q r s^3 t u v^2 w+3636 q r s^3
   t v^4+1296 q r s^3 u^3 v w-972 q r s^3 u^2 v^3-5448 q r s^2 t^3 u w^2+288 q r s^2
   t^3 v^2 w+11520 q r s^2 t^2 u^2 v w-6336 q r s^2 t^2 u v^3-4464 q r s^2 t u^4
   w+3960 q r s^2 t u^3 v^2-324 q r s^2 u^5 v+1812 q r s t^5 w^2-3336 q r s t^4 u v
   w+1608 q r s t^4 v^3+1512 q r s t^3 u^3 w+648 q r s t^3 u^2 v^2-2964 q r s t^2 u^4
   v+1116 q r s t u^6-480 q r t^6 v w+156 q r t^5 u^2 w-192 q r t^5 u v^2+828 q r t^4
   u^3 v-384 q r t^3 u^5+146 q s^6 w^3-924 q s^5 t v w^2-957 q s^5 u^2 w^2+2520 q
   s^5 u v^2 w-1215 q s^5 v^4+2184 q s^4 t^2 u w^2-114 q s^4 t^2 v^2 w-4464 q s^4 t
   u^2 v w+3060 q s^4 t u v^3+1044 q s^4 u^4 w-720 q s^4 u^3 v^2-768 q s^3 t^4
   w^2+1512 q s^3 t^3 u v w-880 q s^3 t^3 v^3+816 q s^3 t^2 u^3 w-2340 q s^3 t^2 u^2
   v^2+720 q s^3 t u^4 v-45 q s^3 u^6+156 q s^2 t^5 v w-1581 q s^2 t^4 u^2 w+1260 q
   s^2 t^4 u v^2+1620 q s^2 t^3 u^3 v-840 q s^2 t^2 u^5+744 q s t^6 u w-168 q s t^6
   v^2-1524 q s t^5 u^2 v+720 q s t^4 u^4-138 q t^8 w+336 q t^7 u v-160 q t^6
   u^3-r^6 w^4+12 r^5 s v w^3+132 r^5 t u w^3-120 r^5 t v^2 w^2-342 r^5 u^2 v
   w^2+504 r^5 u v^3 w-180 r^5 v^5-114 r^4 s^2 u w^3+36 r^4 s^2 v^2 w^2-264 r^4 s
   t^2 w^3+780 r^4 s t u v w^2-288 r^4 s t v^3 w+855 r^4 s u^3 w^2-1728 r^4 s u^2 v^2
   w+738 r^4 s u v^4+624 r^4 t^3 v w^2-960 r^4 t^2 u^2 w^2-1944 r^4 t^2 u v^2 w+1188
   r^4 t^2 v^4+3636 r^4 t u^3 v w-2088 r^4 t u^2 v^3-1215 r^4 u^5 w+729 r^4 u^4
   v^2+340 r^3 s^3 t w^3-108 r^3 s^3 u v w^2+72 r^3 s^3 v^3 w-996 r^3 s^2 t^2 v
   w^2-2100 r^3 s^2 t u^2 w^2+5112 r^3 s^2 t u v^2 w-2088 r^3 s^2 t v^4-972 r^3 s^2
   u^3 v w+468 r^3 s^2 u^2 v^3+2280 r^3 s t^3 u w^2+144 r^3 s t^3 v^2 w-6336 r^3 s t^2
   u^2 v w+2304 r^3 s t^2 u v^3+3060 r^3 s t u^4 w-1080 r^3 s t u^3 v^2-504 r^3 t^5
   w^2+1608 r^3 t^4 u v w-2440 r^3 t^4 v^3-880 r^3 t^3 u^3 w+4800 r^3 t^3 u^2
   v^2-3636 r^3 t^2 u^4 v+972 r^3 t u^6-102 r^2 s^5 w^3+36 r^2 s^4 t v w^2+855 r^2
   s^4 u^2 w^2-1890 r^2 s^4 u v^2 w+729 r^2 s^4 v^4+210 r^2 s^3 t^2 u w^2-1632 r^2
   s^3 t^2 v^2 w+3960 r^2 s^3 t u^2 v w-1080 r^2 s^3 t u v^3-720 r^2 s^3 u^4 w+342
   r^2 s^3 u^3 v^2-390 r^2 s^2 t^4 w^2+648 r^2 s^2 t^3 u v w+4800 r^2 s^2 t^3
   v^3-2340 r^2 s^2 t^2 u^3 w-6696 r^2 s^2 t^2 u^2 v^2+3564 r^2 s^2 t u^4 v-729 r^2
   s^2 u^6-192 r^2 s t^5 v w+1260 r^2 s t^4 u^2 w-5724 r^2 s t^4 u v^2+7728 r^2 s t^3
   u^3 v-2574 r^2 s t^2 u^5-168 r^2 t^6 u w+1992 r^2 t^6 v^2-2952 r^2 t^5 u^2 v+1050
   r^2 t^4 u^4+144 r s^6 v w^2-828 r s^5 t u w^2+1692 r s^5 t v^2 w-324 r s^5 u^2 v
   w+480 r s^4 t^3 w^2-2964 r s^4 t^2 u v w-3636 r s^4 t^2 v^3+720 r s^4 t u^3 w+3564
   r s^4 t u^2 v^2-1224 r s^4 u^4 v+828 r s^3 t^4 v w+1620 r s^3 t^3 u^2 w+7728 r s^3
   t^3 u v^2-7596 r s^3 t^2 u^3 v+2520 r s^3 t u^5-1524 r s^2 t^5 u w-2952 r s^2 t^5
   v^2+306 r s^2 t^4 u^2 v+60 r s^2 t^3 u^4+336 r s t^7 w+2352 r s t^6 u v-1020 r s
   t^5 u^3-516 r t^8 v+240 r t^7 u^2+126 s^7 u w^2-486 s^7 v^2 w-84 s^6 t^2 w^2+1116
   s^6 t u v w+972 s^6 t v^3-45 s^6 u^3 w-729 s^6 u^2 v^2-384 s^5 t^3 v w-840 s^5
   t^2 u^2 w-2574 s^5 t^2 u v^2+2520 s^5 t u^3 v-144 s^5 u^5+720 s^4 t^4 u w+1050 s^4
   t^4 v^2+60 s^4 t^3 u^2 v-1620 s^4 t^2 u^4-160 s^3 t^6 w-1020 s^3 t^5 u v+2135 s^3
   t^4 u^3+240 s^2 t^7 v-1140 s^2 t^6 u^2+330 s t^8 u-44 t^{10}$,
\fi
\noindent
\bea
I_{15}&=& q^6 t u^3 w^5-3 q^6 t u^2 v^2 w^4+ 3 q^6 t u v^4 w^3-q^6 t v^6 w^2-3 q^6 u^4 v
   w^4+ 11 q^6 u^3 v^3 w^3\nonumber\\
&&\txt{+(more other hundreds of terms).}\nonumber
\eea
\if{}
$
  -15 q^6 u^2 v^5 w^2 + 9 q^6 u v^7 w-2 q^6 v^9-3 q^5 r s u^3
   w^5+ 9 q^5 r s u^2 v^2 w^4-9 q^5 r s u v^4 w^3+ 3 q^5 r s v^6 w^2-12 q^5 r t^2 u^2
   w^5+ 24 q^5 r t^2 u v^2 w^4-12 q^5 r t^2 v^4 w^3+ 48 q^5 r t u^3 v w^4-120 q^5 r t
   u^2 v^3 w^3+ 96 q^5 r t u v^5 w^2-24 q^5 r t v^7 w+ 15 q^5 r u^5 w^4-93 q^5 r u^4
   v^2 w^3+ 177 q^5 r u^3 v^4 w^2-135 q^5 r u^2 v^6 w+ 36 q^5 r u v^8-q^5 s^3 t w^6+ 3
   q^5 s^3 u v w^5-2 q^5 s^3 v^3 w^4+ 12 q^5 s^2 t^2 v w^5-54 q^5 s^2 t u v^2 w^4+ 33
   q^5 s^2 t v^4 w^3+ 15 q^5 s^2 u^3 v w^4+ 9 q^5 s^2 u^2 v^3 w^3-18 q^5 s^2 u v^5 w^2+ 3
   q^5 s^2 v^7 w-48 q^5 s t^3 v^2 w^4+ 60 q^5 s t^2 u^2 v w^4+ 168 q^5 s t^2 u v^3
   w^3-108 q^5 s t^2 v^5 w^2-69 q^5 s t u^4 w^4-84 q^5 s t u^3 v^2 w^3+ 6 q^5 s t u^2
   v^4 w^2+ 48 q^5 s t u v^6 w+ 132 q^5 s u^5 v w^3-201 q^5 s u^4 v^3 w^2+ 141 q^5 s u^3
   v^5 w-45 q^5 s u^2 v^7+ 64 q^5 t^4 v^3 w^3+ 24 q^5 t^3 u^3 w^4-192 q^5 t^3 u^2 v^2+
   w^3-168 q^5 t^3 u v^4 w^2+ 128 q^5 t^3 v^6 w+ 60 q^5 t^2 u^4 v w^3+ 528 q^5 t^2 u^3
   v^3 w^2-336 q^5 t^2 u^2 v^5 w-24 q^5 t u^6 w^3-300 q^5 t u^5 v^2 w^2+ 159 q^5 t
   u^4 v^4 w+ 30 q^5 t u^3 v^6+ 72 q^5 u^7 v w^2-36 q^5 u^6 v^3 w-9 q^5 u^5 v^5+ 2 q^4
   r^3 u^3 w^5-6 q^4 r^3 u^2 v^2 w^4+ 6 q^4 r^3 u v^4 w^3-2 q^4 r^3 v^6 w^2+ 3 q^4 r^2
   s^2 t w^6-9 q^4 r^2 s^2 u v w^5+ 6 q^4 r^2 s^2 v^3 w^4-24 q^4 r^2 s t^2 v w^5+ 54
   q^4 r^2 s t u^2 w^5-12 q^4 r^2 s t v^4 w^3-102 q^4 r^2 s u^3 v w^4+ 144 q^4 r^2 s
   u^2 v^3 w^3-72 q^4 r^2 s u v^5 w^2+ 12 q^4 r^2 s v^7 w+ 48 q^4 r^2 t^3 u w^5-288 q^4
   r^2 t^2 u^2 v w^4+ 144 q^4 r^2 t^2 u v^3 w^3+ 24 q^4 r^2 t^2 v^5 w^2-201 q^4 r^2 t+
   u^4 w^4+ 1152 q^4 r^2 t u^3 v^2 w^3-1188 q^4 r^2 t u^2 v^4 w^2+ 336 q^4 r^2 t u v^6
   w+ 153 q^4 r^2 u^5 v w^3-732 q^4 r^2 u^4 v^3 w^2+ 822 q^4 r^2 u^3 v^5 w-270 q^4 r^2
   u^2 v^7+ 3 q^4 r s^4 w^6-48 q^4 r s^3 t v w^5-15 q^4 r s^3 u^2 w^5+ 102 q^4 r s^3 u
   v^2 w^4-51 q^4 r s^3 v^4 w^3-60 q^4 r s^2 t^2 u w^5+ 288 q^4 r s^2 t^2 v^2 w^4-444
   q^4 r s^2 t u v^3 w^3+ 126 q^4 r s^2 t v^5 w^2+ 132 q^4 r s^2 u^4 w^4-243 q^4 r s^2
   u^3 v^2 w^3+ 270 q^4 r s^2 u^2 v^4 w^2-33 q^4 r s^2 u v^6 w-18 q^4 r s^2 v^8-288
   q^4 r s t^3 v^3 w^3+ 588 q^4 r s t^2 u^3 w^4-1224 q^4 r s t^2 u^2 v^2 w^3+1140 q^4 r
   s t^2 u v^4 w^2-48 q^4 r s t^2 v^6 w-504 q^4 r s t u^4 v w^3+1440 q^4 r s t u^3
   v^3 w^2-1188 q^4 r s t u^2 v^5 w+72 q^4 r s t u v^7-588 q^4 r s u^6 w^3+1539 q^4 r
   s u^5 v^2 w^2-1590 q^4 r s u^4 v^4 w+639 q^4 r s u^3 v^6-288 q^4 r t^4 u^2 w^4+576
   q^4 r t^4 u v^2 w^3-480 q^4 r t^4 v^4 w^2+384 q^4 r t^3 u^3 v w^3+480 q^4 r t^3 u^2
   v^3 w^2+864 q^4 r t^3 u v^5 w-768 q^4 r t^3 v^7+276 q^4 r t^2 u^5 w^3-3000 q^4 r
   t^2 u^4 v^2 w^2-480 q^4 r t^2 u^3 v^4 w+1656 q^4 r t^2 u^2 v^6+1416 q^4 r t u^6 v
   w^2+1284 q^4 r t u^5 v^3 w-1674 q^4 r t u^4 v^5-360 q^4 r u^8 w^2-324 q^4 r u^7
   v^2 w+441 q^4 r u^6 v^4-15 q^4 s^5 v w^5+69 q^4 s^4 t u w^5+201 q^4 s^4 t v^2
   w^4-132 q^4 s^4 u^2 v w^4-195 q^4 s^4 u v^3 w^3+129 q^4 s^4 v^5 w^2-24 q^4 s^3
   t^3 w^5-588 q^4 s^3 t^2 u v w^4-1160 q^4 s^3 t^2 v^3 w^3+2820 q^4 s^3 t u^2 v^2
   w^3+255 q^4 s^3 t u v^4 w^2-534 q^4 s^3 t v^6 w-660 q^4 s^3 u^4 v w^3-655 q^4 s^3
   u^3 v^3 w^2+363 q^4 s^3 u^2 v^5 w+45 q^4 s^3 u v^7+288 q^4 s^2 t^4 v w^4+1800 q^4
   s^2 t^3 u v^2 w^3+2940 q^4 s^2 t^3 v^4 w^2-2940 q^4 s^2 t^2 u^3 v w^3-8760 q^4 s^2
   t^2 u^2 v^3 w^2-36 q^4 s^2 t^2 u v^5 w+960 q^4 s^2 t^2 v^7+1674 q^4 s^2 t u^5
   w^3+5670 q^4 s^2 t u^4 v^2 w^2+1965 q^4 s^2 t u^3 v^4 w-1665 q^4 s^2 t u^2
   v^6-2082 q^4 s^2 u^6 v w^2+93 q^4 s^2 u^5 v^3 w+225 q^4 s^2 u^4 v^5-1152 q^4 s t^5
   v^2 w^3+1440 q^4 s t^4 u^2 v w^3-1920 q^4 s t^4 u v^3 w^2-3456 q^4 s t^4 v^5
   w-1200 q^4 s t^3 u^4 w^3+6960 q^4 s t^3 u^3 v^2 w^2+12240 q^4 s t^3 u^2 v^4 w-1920
   q^4 s t^3 u v^6-3972 q^4 s t^2 u^5 v w^2-17760 q^4 s t^2 u^4 v^3 w+3720 q^4 s t^2
   u^3 v^5+1128 q^4 s t u^7 w^2+8808 q^4 s t u^6 v^2 w-1575 q^4 s t u^5 v^4-1584 q^4
   s u^8 v w+180 q^4 s u^7 v^3+1536 q^4 t^6 v^3 w^2+192 q^4 t^5 u^3 w^3-3456 q^4 t^5
   u^2 v^2 w^2-1536 q^4 t^5 u v^4 w+2816 q^4 t^5 v^6+1920 q^4 t^4 u^4 v w^2+2880 q^4
   t^4 u^3 v^3 w-8640 q^4 t^4 u^2 v^5-464 q^4 t^3 u^6 w^2-792 q^4 t^3 u^5 v^2
   w+11100 q^4 t^3 u^4 v^4-336 q^4 t^2 u^7 v w-7800 q^4 t^2 u^6 v^3+144 q^4 t u^9
   w+2880 q^4 t u^8 v^2-432 q^4 u^{10} v-3 q^3 r^4 s t w^6+9 q^3 r^4 s u v w^5-6 q^3
   r^4 s v^3 w^4+12 q^3 r^4 t^2 v w^5-33 q^3 r^4 t u^2 w^5+12 q^3 r^4 t u v^2 w^4+51
   q^3 r^4 u^3 v w^4-66 q^3 r^4 u^2 v^3 w^3+24 q^3 r^4 u v^5 w^2-11 q^3 r^3 s^3
   w^6+120 q^3 r^3 s^2 t v w^5-9 q^3 r^3 s^2 u^2 w^5-144 q^3 r^3 s^2 u v^2 w^4+66 q^3
   r^3 s^2 v^4 w^3-168 q^3 r^3 s t^2 u w^5-144 q^3 r^3 s t^2 v^2 w^4+444 q^3 r^3 s t
   u^2 v w^4-48 q^3 r^3 s t v^5 w^2+195 q^3 r^3 s u^4 w^4-732 q^3 r^3 s u^3 v^2
   w^3+564 q^3 r^3 s u^2 v^4 w^2-144 q^3 r^3 s u v^6 w-64 q^3 r^3 t^4 w^5+288 q^3 r^3
   t^3 u v w^4+1160 q^3 r^3 t^2 u^3 w^4-2208 q^3 r^3 t^2 u^2 v^2 w^3+1344 q^3 r^3 t^2
   u v^4 w^2-512 q^3 r^3 t^2 v^6 w-2916 q^3 r^3 t u^4 v w^3+3952 q^3 r^3 t u^3 v^3
   w^2-1104 q^3 r^3 t u^2 v^5 w+137 q^3 r^3 u^6 w^3+1716 q^3 r^3 u^5 v^2 w^2-2862 q^3
   r^3 u^4 v^4 w+1080 q^3 r^3 u^3 v^6+93 q^3 r^2 s^4 v w^5+84 q^3 r^2 s^3 t u w^5-1152
   q^3 r^2 s^3 t v^2 w^4+243 q^3 r^2 s^3 u^2 v w^4+732 q^3 r^2 s^3 u v^3 w^3-246 q^3
   r^2 s^3 v^5 w^2+192 q^3 r^2 s^2 t^3 w^5+1224 q^3 r^2 s^2 t^2 u v w^4+2208 q^3 r^2
   s^2 t^2 v^3 w^3-2820 q^3 r^2 s^2 t u^3 w^4-2844 q^3 r^2 s^2 t u v^4 w^2+1056 q^3
   r^2 s^2 t v^6 w+2367 q^3 r^2 s^2 u^4 v w^3-708 q^3 r^2 s^2 u^3 v^3 w^2-432 q^3 r^2
   s^2 u^2 v^5 w+216 q^3 r^2 s^2 u v^7-576 q^3 r^2 s t^4 v w^4-1800 q^3 r^2 s t^3 u^2
   w^4-2688 q^3 r^2 s t^3 v^4 w^2+7248 q^3 r^2 s t^2 u^3 v w^3+6336 q^3 r^2 s t^2 u^2
   v^3 w^2-5184 q^3 r^2 s t^2 u v^5 w+1536 q^3 r^2 s t^2 v^7+4140 q^3 r^2 s t u^5
   w^3-23994 q^3 r^2 s t u^4 v^2 w^2+16440 q^3 r^2 s t u^3 v^4 w-3168 q^3 r^2 s t u^2
   v^6-363 q^3 r^2 s u^6 v w^2+4554 q^3 r^2 s u^5 v^3 w-2754 q^3 r^2 s u^4 v^5+1152
   q^3 r^2 t^5 u w^4-3168 q^3 r^2 t^4 u^2 v w^3-384 q^3 r^2 t^4 u v^3 w^2+4608 q^3
   r^2 t^4 v^5 w-2520 q^3 r^2 t^3 u^4 w^3+9984 q^3 r^2 t^3 u^3 v^2 w^2-23040 q^3 r^2
   t^3 u^2 v^4 w+6144 q^3 r^2 t^3 u v^6-504 q^3 r^2 t^2 u^5 v w^2+27720 q^3 r^2 t^2
   u^4 v^3 w-13728 q^3 r^2 t^2 u^3 v^5+1092 q^3 r^2 t u^7 w^2-22398 q^3 r^2 t u^6 v^2
   w+13356 q^3 r^2 t u^5 v^4+5364 q^3 r^2 u^8 v w-3618 q^3 r^2 u^7 v^3-132 q^3 r s^5
   u w^5-153 q^3 r s^5 v^2 w^4-60 q^3 r s^4 t^2 w^5+504 q^3 r s^4 t u v w^4+2916 q^3
   r s^4 t v^3 w^3+660 q^3 r s^4 u^3 w^4-2367 q^3 r s^4 u^2 v^2 w^3-480 q^3 r s^4 u
   v^4 w^2+54 q^3 r s^4 v^6 w-384 q^3 r s^3 t^3 v w^4+2940 q^3 r s^3 t^2 u^2 w^4-7248
   q^3 r s^3 t^2 u v^2 w^3-6600 q^3 r s^3 t^2 v^4 w^2+4560 q^3 r s^3 t u^2 v^3
   w^2+10272 q^3 r s^3 t u v^5 w-2736 q^3 r s^3 t v^7-1722 q^3 r s^3 u^5 w^3+4695 q^3
   r s^3 u^4 v^2 w^2-7110 q^3 r s^3 u^3 v^4 w+1242 q^3 r s^3 u^2 v^6-1440 q^3 r s^2
   t^4 u w^4+3168 q^3 r s^2 t^4 v^2 w^3+1920 q^3 r s^2 t^3 u v^3 w^2+6336 q^3 r s^2 t^3
   v^5 w-7980 q^3 r s^2 t^2 u^4 w^3+30960 q^3 r s^2 t^2 u^3 v^2 w^2-27720 q^3 r s^2
   t^2 u^2 v^4 w-2784 q^3 r s^2 t^2 u v^6-13032 q^3 r s^2 t u^5 v w^2-2640 q^3 r s^2
   t u^4 v^3 w+12240 q^3 r s^2 t u^3 v^5+7026 q^3 r s^2 u^7 w^2-2835 q^3 r s^2 u^6 v^2
   w-1458 q^3 r s^2 u^5 v^4+768 q^3 r s t^5 v^3 w^2+9120 q^3 r s t^4 u^3 w^3-28800
   q^3 r s t^4 u^2 v^2 w^2+34560 q^3 r s t^4 u v^4 w-13824 q^3 r s t^4 v^6+8400 q^3 r
   s t^3 u^4 v w^2-78720 q^3 r s t^3 u^3 v^3 w+43200 q^3 r s t^3 u^2 v^5-7548 q^3 r s
   t^2 u^6 w^2+109800 q^3 r s t^2 u^5 v^2 w-54840 q^3 r s t^2 u^4 v^4-48912 q^3 r s t
   u^7 v w+20412 q^3 r s t u^6 v^3+7488 q^3 r s u^9 w-1836 q^3 r s u^8 v^2-2304 q^3 r
   t^6 u^2 w^3+4608 q^3 r t^6 u v^2 w^2-12288 q^3 r t^6 v^4 w+384 q^3 r t^5 u^3 v
   w^2+37632 q^3 r t^5 u^2 v^3 w-6144 q^3 r t^5 u v^5+1536 q^3 r t^4 u^5 w^2-50400
   q^3 r t^4 u^4 v^2 w+21120 q^3 r t^4 u^3 v^4+22896 q^3 r t^3 u^6 v w-30048 q^3 r t^3
   u^5 v^3-3504 q^3 r t^2 u^8 w+26856 q^3 r t^2 u^7 v^2-12384 q^3 r t u^9 v+2160 q^3
   r u^{11}+24 q^3 s^6 t w^5+588 q^3 s^6 u v w^4-137 q^3 s^6 v^3 w^3-276 q^3 s^5 t^2
   v w^4-1674 q^3 s^5 t u^2 w^4-4140 q^3 s^5 t u v^2 w^3-753 q^3 s^5 t v^4 w^2+1722
   q^3 s^5 u^3 v w^3+4641 q^3 s^5 u^2 v^3 w^2-2781 q^3 s^5 u v^5 w+756 q^3 s^5
   v^7+1200 q^3 s^4 t^3 u w^4+2520 q^3 s^4 t^3 v^2 w^3+7980 q^3 s^4 t^2 u^2 v w^3+21480
   q^3 s^4 t^2 u v^3 w^2+1020 q^3 s^4 t^2 v^5 w-41160 q^3 s^4 t u^3 v^2 w^2+615 q^3
   s^4 t u^2 v^4 w-720 q^3 s^4 t u v^6+8610 q^3 s^4 u^5 v w^2+4635 q^3 s^4 u^4 v^3
   w-675 q^3 s^4 u^3 v^5-192 q^3 s^3 t^5 w^4-9120 q^3 s^3 t^4 u v w^3-12640 q^3 s^3
   t^4 v^3 w^2-9120 q^3 s^3 t^3 u^2 v^2 w^2-41280 q^3 s^3 t^3 u v^4 w+5760 q^3 s^3
   t^3 v^6+39900 q^3 s^3 t^2 u^4 v w^2+88880 q^3 s^3 t^2 u^3 v^3 w-6600 q^3 s^3 t^2
   u^2 v^5-16186 q^3 s^3 t u^6 w^2-58884 q^3 s^3 t u^5 v^2 w-1350 q^3 s^3 t u^4
   v^4+13428 q^3 s^3 u^7 v w+945 q^3 s^3 u^6 v^3+2304 q^3 s^2 t^6 v w^3+33984 q^3 s^2
   t^5 u v^2 w^2+19968 q^3 s^2 t^5 v^4 w-45600 q^3 s^2 t^4 u^3 v w^2+5760 q^3 s^2 t^4
   u^2 v^3 w+18912 q^3 s^2 t^3 u^5 w^2-81360 q^3 s^2 t^3 u^4 v^2 w-22800 q^3 s^2 t^3
   u^3 v^4+61764 q^3 s^2 t^2 u^6 v w+49560 q^3 s^2 t^2 u^5 v^3-14280 q^3 s^2 t u^8
   w-29340 q^3 s^2 t u^7 v^2+5400 q^3 s^2 u^9 v-9216 q^3 s t^7 v^2 w^2+11520 q^3 s
   t^6 u^2 v w^2-49152 q^3 s t^6 u v^3 w-6384 q^3 s t^5 u^4 w^2+103296 q^3 s t^5 u^3
   v^2 w+7680 q^3 s t^5 u^2 v^4-63216 q^3 s t^4 u^5 v w-4800 q^3 s t^4 u^4 v^3+12960
   q^3 s t^3 u^7 w-15720 q^3 s t^3 u^6 v^2+15120 q^3 s t^2 u^8 v-3600 q^3 s t
   u^{10}+12288 q^3 t^8 v^3 w+512 q^3 t^7 u^3 w^2-24576 q^3 t^7 u^2 v^2 w+14784 q^3
   t^6 u^4 v w-5120 q^3 t^6 u^3 v^3-2960 q^3 t^5 u^6 w+11136 q^3 t^5 u^5 v^2-7440
   q^3 t^4 u^7 v+1600 q^3 t^3 u^9+q^2 r^6 t w^6-3 q^2 r^6 u v w^5+2 q^2 r^6 v^3 w^4+15
   q^2 r^5 s^2 w^6-96 q^2 r^5 s t v w^5+18 q^2 r^5 s u^2 w^5+72 q^2 r^5 s u v^2
   w^4-24 q^2 r^5 s v^4 w^3+108 q^2 r^5 t^2 u w^5-24 q^2 r^5 t^2 v^2 w^4-126 q^2 r^5
   t u^2 v w^4+48 q^2 r^5 t u v^3 w^3-129 q^2 r^5 u^4 w^4+246 q^2 r^5 u^3 v^2 w^3-108
   q^2 r^5 u^2 v^4 w^2-177 q^2 r^4 s^3 v w^5-6 q^2 r^4 s^2 t u w^5+1188 q^2 r^4 s^2 t
   v^2 w^4-270 q^2 r^4 s^2 u^2 v w^4-564 q^2 r^4 s^2 u v^3 w^3+108 q^2 r^4 s^2 v^5
   w^2+168 q^2 r^4 s t^3 w^5-1140 q^2 r^4 s t^2 u v w^4-1344 q^2 r^4 s t^2 v^3
   w^3-255 q^2 r^4 s t u^3 w^4+2844 q^2 r^4 s t u^2 v^2 w^3+480 q^2 r^4 s u^4 v
   w^3-1782 q^2 r^4 s u^3 v^3 w^2+648 q^2 r^4 s u^2 v^5 w+480 q^2 r^4 t^4 v w^4-2940
   q^2 r^4 t^3 u^2 w^4+2688 q^2 r^4 t^3 u v^2 w^3+6600 q^2 r^4 t^2 u^3 v w^3-11808 q^2
   r^4 t^2 u^2 v^3 w^2+4608 q^2 r^4 t^2 u v^5 w+753 q^2 r^4 t u^5 w^3+1719 q^2 r^4 t
   u^4 v^2 w^2-1728 q^2 r^4 t u^3 v^4 w-4725 q^2 r^4 u^6 v w^2+6885 q^2 r^4 u^5 v^3
   w-2430 q^2 r^4 u^4 v^5+201 q^2 r^3 s^4 u w^5+732 q^2 r^3 s^4 v^2 w^4-528 q^2 r^3
   s^3 t^2 w^5-1440 q^2 r^3 s^3 t u v w^4-3952 q^2 r^3 s^3 t v^3 w^3+655 q^2 r^3 s^3
   u^3 w^4+708 q^2 r^3 s^3 u^2 v^2 w^3+1782 q^2 r^3 s^3 u v^4 w^2-216 q^2 r^3 s^3 v^6
   w-480 q^2 r^3 s^2 t^3 v w^4+8760 q^2 r^3 s^2 t^2 u^2 w^4-6336 q^2 r^3 s^2 t^2 u
   v^2 w^3+11808 q^2 r^3 s^2 t^2 v^4 w^2-4560 q^2 r^3 s^2 t u^3 v w^3-8208 q^2 r^3
   s^2 t u v^5 w-4641 q^2 r^3 s^2 u^5 w^3+3942 q^2 r^3 s^2 u^4 v^2 w^2+4212 q^2 r^3
   s^2 u^3 v^4 w-972 q^2 r^3 s^2 u^2 v^6+1920 q^2 r^3 s t^4 u w^4+384 q^2 r^3 s t^4
   v^2 w^3-1920 q^2 r^3 s t^3 u^2 v w^3-9216 q^2 r^3 s t^3 v^5 w-21480 q^2 r^3 s t^2
   u^4 w^3-18432 q^2 r^3 s t^2 u^3 v^2 w^2+58176 q^2 r^3 s t^2 u^2 v^4 w-13824 q^2
   r^3 s t^2 u v^6+76428 q^2 r^3 s t u^5 v w^2-95904 q^2 r^3 s t u^4 v^3 w+24624 q^2
   r^3 s t u^3 v^5-9855 q^2 r^3 s u^7 w^2+4860 q^2 r^3 s u^6 v^2 w+2430 q^2 r^3 s u^5
   v^4-1536 q^2 r^3 t^6 w^4-768 q^2 r^3 t^5 u v w^3+12640 q^2 r^3 t^4 u^3 w^3+192 q^2
   r^3 t^4 u^2 v^2 w^2+1536 q^2 r^3 t^4 u v^4 w-8192 q^2 r^3 t^4 v^6-20400 q^2 r^3
   t^3 u^4 v w^2+18304 q^2 r^3 t^3 u^3 v^3 w+10752 q^2 r^3 t^3 u^2 v^5-4320 q^2 r^3
   t^2 u^6 w^2-30600 q^2 r^3 t^2 u^5 v^2 w+6480 q^2 r^3 t^2 u^4 v^4+45684 q^2 r^3 t
   u^7 v w-24084 q^2 r^3 t u^6 v^3-13932 q^2 r^3 u^9 w+8586 q^2 r^3 u^8 v^2+300 q^2
   r^2 s^5 t w^5-1539 q^2 r^2 s^5 u v w^4-1716 q^2 r^2 s^5 v^3 w^3+3000 q^2 r^2 s^4
   t^2 v w^4-5670 q^2 r^2 s^4 t u^2 w^4+23994 q^2 r^2 s^4 t u v^2 w^3-1719 q^2 r^2
   s^4 t v^4 w^2-4695 q^2 r^2 s^4 u^3 v w^3-3942 q^2 r^2 s^4 u^2 v^3 w^2+567 q^2 r^2
   s^4 u v^5 w+162 q^2 r^2 s^4 v^7-6960 q^2 r^2 s^3 t^3 u w^4-9984 q^2 r^2 s^3 t^3
   v^2 w^3-30960 q^2 r^2 s^3 t^2 u^2 v w^3+18432 q^2 r^2 s^3 t^2 u v^3 w^2-13536 q^2
   r^2 s^3 t^2 v^5 w+41160 q^2 r^2 s^3 t u^4 w^3-32400 q^2 r^2 s^3 t u^2 v^4 w+23328
   q^2 r^2 s^3 t u v^6-22869 q^2 r^2 s^3 u^5 v w^2+28350 q^2 r^2 s^3 u^4 v^3 w-12960
   q^2 r^2 s^3 u^3 v^5+3456 q^2 r^2 s^2 t^5 w^4+28800 q^2 r^2 s^2 t^4 u v w^3-192 q^2
   r^2 s^2 t^4 v^3 w^2+9120 q^2 r^2 s^2 t^3 u^3 w^3-63360 q^2 r^2 s^2 t^3 u v^4
   w+52224 q^2 r^2 s^2 t^3 v^6-110160 q^2 r^2 s^2 t^2 u^4 v w^2+200160 q^2 r^2 s^2 t^2
   u^3 v^3 w-123120 q^2 r^2 s^2 t^2 u^2 v^5+5418 q^2 r^2 s^2 t u^6 w^2-20736 q^2 r^2
   s^2 t u^5 v^2 w+38070 q^2 r^2 s^2 t u^4 v^4+5103 q^2 r^2 s^2 u^7 v w-10368 q^2 r^2
   s^2 u^6 v^3-4608 q^2 r^2 s t^6 v w^3-33984 q^2 r^2 s t^5 u^2 w^3-3072 q^2 r^2 s
   t^5 v^4 w+83040 q^2 r^2 s t^4 u^3 v w^2+5760 q^2 r^2 s t^4 u^2 v^3 w-23040 q^2 r^2
   s t^4 u v^5+22536 q^2 r^2 s t^3 u^5 w^2-92880 q^2 r^2 s t^3 u^4 v^2 w+53760 q^2 r^2
   s t^3 u^3 v^4-45504 q^2 r^2 s t^2 u^6 v w+10152 q^2 r^2 s t^2 u^5 v^3+26568 q^2 r^2
   s t u^8 w+1134 q^2 r^2 s t u^7 v^2-5508 q^2 r^2 s u^9 v+9216 q^2 r^2 t^7 u
   w^3-16128 q^2 r^2 t^6 u^2 v w^2-6144 q^2 r^2 t^6 u v^3 w+49152 q^2 r^2 t^6
   v^5-9936 q^2 r^2 t^5 u^4 w^2+15360 q^2 r^2 t^5 u^3 v^2 w-170496 q^2 r^2 t^5 u^2
   v^4+31248 q^2 r^2 t^4 u^5 v w+245280 q^2 r^2 t^4 u^4 v^3-14136 q^2 r^2 t^3 u^7
   w-203592 q^2 r^2 t^3 u^6 v^2+84024 q^2 r^2 t^2 u^8 v-12960 q^2 r^2 t u^{10}-72
   q^2 r s^7 w^5-1416 q^2 r s^6 t v w^4+2082 q^2 r s^6 u^2 w^4+363 q^2 r s^6 u v^2
   w^3+4725 q^2 r s^6 v^4 w^2+3972 q^2 r s^5 t^2 u w^4+504 q^2 r s^5 t^2 v^2 w^3+13032
   q^2 r s^5 t u^2 v w^3-76428 q^2 r s^5 t u v^3 w^2+14040 q^2 r s^5 t v^5 w-8610 q^2
   r s^5 u^4 w^3+22869 q^2 r s^5 u^3 v^2 w^2+11988 q^2 r s^5 u^2 v^4 w-7614 q^2 r s^5
   u v^6-1920 q^2 r s^4 t^4 w^4-8400 q^2 r s^4 t^3 u v w^3+20400 q^2 r s^4 t^3 v^3
   w^2-39900 q^2 r s^4 t^2 u^3 w^3+110160 q^2 r s^4 t^2 u^2 v^2 w^2+83340 q^2 r s^4
   t^2 u v^4 w-55080 q^2 r s^4 t^2 v^6-125820 q^2 r s^4 t u^3 v^3 w+65610 q^2 r s^4 t
   u^2 v^5+5508 q^2 r s^4 u^6 w^2+3807 q^2 r s^4 u^5 v^2 w-6885 q^2 r s^4 u^4 v^4-384
   q^2 r s^3 t^5 v w^3+45600 q^2 r s^3 t^4 u^2 w^3-83040 q^2 r s^3 t^4 u v^2 w^2+17280
   q^2 r s^3 t^4 v^4 w-200640 q^2 r s^3 t^3 u^2 v^3 w+127680 q^2 r s^3 t^3 u v^5+18756
   q^2 r s^3 t^2 u^5 w^2+97920 q^2 r s^3 t^2 u^4 v^2 w-141480 q^2 r s^3 t^2 u^3
   v^4+36504 q^2 r s^3 t u^6 v w+60264 q^2 r s^3 t u^5 v^3-24300 q^2 r s^3 u^8 w-6723
   q^2 r s^3 u^7 v^2-11520 q^2 r s^2 t^6 u w^3+16128 q^2 r s^2 t^6 v^2 w^2+31488 q^2 r
   s^2 t^5 u v^3 w-119808 q^2 r s^2 t^5 v^5-54480 q^2 r s^2 t^4 u^4 w^2+213120 q^2 r
   s^2 t^4 u^3 v^2 w+123840 q^2 r s^2 t^4 u^2 v^4-137088 q^2 r s^2 t^3 u^5 v w+10560
   q^2 r s^2 t^3 u^4 v^3+31500 q^2 r s^2 t^2 u^7 w-150984 q^2 r s^2 t^2 u^6 v^2+93960
   q^2 r s^2 t u^8 v-16200 q^2 r s^2 u^{10}+12288 q^2 r s t^7 v^3 w+33792 q^2 r s t^6
   u^3 w^2-124416 q^2 r s t^6 u^2 v^2 w+233472 q^2 r s t^6 u v^4+41856 q^2 r s t^5 u^4
   v w-604416 q^2 r s t^5 u^3 v^3-2400 q^2 r s t^4 u^6 w+661536 q^2 r s t^4 u^5
   v^2-315120 q^2 r s t^3 u^7 v+54000 q^2 r s t^2 u^9-6144 q^2 r t^8 u^2 w^2+12288
   q^2 r t^8 u v^2 w-73728 q^2 r t^8 v^4+3072 q^2 r t^7 u^3 v w+208896 q^2 r t^7 u^2
   v^3-2880 q^2 r t^6 u^5 w-234624 q^2 r t^6 u^4 v^2+114720 q^2 r t^5 u^6 v-20400
   q^2 r t^4 u^8+360 q^2 s^8 v w^4-1128 q^2 s^7 t u w^4-1092 q^2 s^7 t v^2 w^3-7026
   q^2 s^7 u^2 v w^3+9855 q^2 s^7 u v^3 w^2-8667 q^2 s^7 v^5 w+464 q^2 s^6 t^3
   w^4+7548 q^2 s^6 t^2 u v w^3+4320 q^2 s^6 t^2 v^3 w^2+16186 q^2 s^6 t u^3 w^3-5418
   q^2 s^6 t u^2 v^2 w^2+25488 q^2 s^6 t u v^4 w+20250 q^2 s^6 t v^6-5508 q^2 s^6 u^4
   v w^2-18009 q^2 s^6 u^3 v^3 w-5670 q^2 s^6 u^2 v^5-1536 q^2 s^5 t^4 v w^3-18912
   q^2 s^5 t^3 u^2 w^3-22536 q^2 s^5 t^3 u v^2 w^2-40152 q^2 s^5 t^3 v^4 w-18756 q^2
   s^5 t^2 u^3 v w^2-35136 q^2 s^5 t^2 u^2 v^3 w-105300 q^2 s^5 t^2 u v^5+122364 q^2
   s^5 t u^4 v^2 w+73305 q^2 s^5 t u^3 v^4-27540 q^2 s^5 u^6 v w-18549 q^2 s^5 u^5
   v^3+6384 q^2 s^4 t^5 u w^3+9936 q^2 s^4 t^5 v^2 w^2+54480 q^2 s^4 t^4 u^2 v
   w^2+166320 q^2 s^4 t^4 u v^3 w+76800 q^2 s^4 t^4 v^5-139320 q^2 s^4 t^3 u^3 v^2
   w+136500 q^2 s^4 t^3 u^2 v^4-93780 q^2 s^4 t^2 u^5 v w-248400 q^2 s^4 t^2 u^4
   v^3+47925 q^2 s^4 t u^7 w+130005 q^2 s^4 t u^6 v^2-22275 q^2 s^4 u^8 v-512 q^2 s^3
   t^7 w^3-33792 q^2 s^3 t^6 u v w^2-57344 q^2 s^3 t^6 v^3 w-131712 q^2 s^3 t^5 u^2
   v^2 w-314880 q^2 s^3 t^5 u v^4+272400 q^2 s^3 t^4 u^4 v w+320000 q^2 s^3 t^4 u^3
   v^3-91600 q^2 s^3 t^3 u^6 w+8160 q^2 s^3 t^3 u^5 v^2-99900 q^2 s^3 t^2 u^7 v+27000
   q^2 s^3 t u^9+6144 q^2 s^2 t^8 v w^2+129024 q^2 s^2 t^7 u v^2 w+92160 q^2 s^2 t^7
   v^4-168960 q^2 s^2 t^6 u^3 v w+168960 q^2 s^2 t^6 u^2 v^3+52800 q^2 s^2 t^5 u^5
   w-482400 q^2 s^2 t^5 u^4 v^2+319200 q^2 s^2 t^4 u^6 v-66000 q^2 s^2 t^3 u^8-24576
   q^2 s t^9 v^2 w+30720 q^2 s t^8 u^2 v w-184320 q^2 s t^8 u v^3-9600 q^2 s t^7 u^4
   w+337920 q^2 s t^7 u^3 v^2-206400 q^2 s t^6 u^5 v+42000 q^2 s t^5 u^7+32768 q^2
   t^{10} v^3-61440 q^2 t^9 u^2 v^2+38400 q^2 t^8 u^4 v-8000 q^2 t^7 u^6-9 q r^7 s
   w^6+24 q r^7 t v w^5-3 q r^7 u^2 w^5-12 q r^7 u v^2 w^4+135 q r^6 s^2 v w^5-48 q
   r^6 s t u w^5-336 q r^6 s t v^2 w^4+33 q r^6 s u^2 v w^4+144 q r^6 s u v^3 w^3-128
   q r^6 t^3 w^5+48 q r^6 t^2 u v w^4+512 q r^6 t^2 v^3 w^3+534 q r^6 t u^3 w^4-1056 q
   r^6 t u^2 v^2 w^3-54 q r^6 u^4 v w^3+216 q r^6 u^3 v^3 w^2-141 q r^5 s^3 u
   w^5-822 q r^5 s^3 v^2 w^4+336 q r^5 s^2 t^2 w^5+1188 q r^5 s^2 t u v w^4+1104 q r^5
   s^2 t v^3 w^3-363 q r^5 s^2 u^3 w^4+432 q r^5 s^2 u^2 v^2 w^3-648 q r^5 s^2 u v^4
   w^2-864 q r^5 s t^3 v w^4+36 q r^5 s t^2 u^2 w^4+5184 q r^5 s t^2 u v^2 w^3-4608 q
   r^5 s t^2 v^4 w^2-10272 q r^5 s t u^3 v w^3+8208 q r^5 s t u^2 v^3 w^2+2781 q r^5 s
   u^5 w^3-567 q r^5 s u^4 v^2 w^2-1296 q r^5 s u^3 v^4 w+3456 q r^5 t^4 u w^4-4608
   q r^5 t^4 v^2 w^3-6336 q r^5 t^3 u^2 v w^3+9216 q r^5 t^3 u v^3 w^2-1020 q r^5 t^2
   u^4 w^3+13536 q r^5 t^2 u^3 v^2 w^2-13824 q r^5 t^2 u^2 v^4 w-14040 q r^5 t u^5 v
   w^2+13608 q r^5 t u^4 v^3 w+8667 q r^5 u^7 w^2-11259 q r^5 u^6 v^2 w+2916 q r^5 u^5
   v^4-159 q r^4 s^4 t w^5+1590 q r^4 s^4 u v w^4+2862 q r^4 s^4 v^3 w^3+480 q r^4 s^3
   t^2 v w^4-1965 q r^4 s^3 t u^2 w^4-16440 q r^4 s^3 t u v^2 w^3+1728 q r^4 s^3 t
   v^4 w^2+7110 q r^4 s^3 u^3 v w^3-4212 q r^4 s^3 u^2 v^3 w^2+1296 q r^4 s^3 u v^5
   w-12240 q r^4 s^2 t^3 u w^4+23040 q r^4 s^2 t^3 v^2 w^3+27720 q r^4 s^2 t^2 u^2 v
   w^3-58176 q r^4 s^2 t^2 u v^3 w^2+13824 q r^4 s^2 t^2 v^5 w-615 q r^4 s^2 t u^4
   w^3+32400 q r^4 s^2 t u^3 v^2 w^2-11988 q r^4 s^2 u^5 v w^2-3483 q r^4 s^2 u^4 v^3
   w+1944 q r^4 s^2 u^3 v^5+1536 q r^4 s t^5 w^4-34560 q r^4 s t^4 u v w^3-1536 q r^4
   s t^4 v^3 w^2+41280 q r^4 s t^3 u^3 w^3+63360 q r^4 s t^3 u^2 v^2 w^2-83340 q r^4 s
   t^2 u^4 v w^2-93312 q r^4 s t^2 u^3 v^3 w+41472 q r^4 s t^2 u^2 v^5-25488 q r^4 s
   t u^6 w^2+167832 q r^4 s t u^5 v^2 w-69984 q r^4 s t u^4 v^4-21951 q r^4 s u^7 v
   w+8991 q r^4 s u^6 v^3+12288 q r^4 t^6 v w^3-19968 q r^4 t^5 u^2 w^3+3072 q r^4 t^5
   u v^2 w^2-17280 q r^4 t^4 u^3 v w^2-66048 q r^4 t^4 u^2 v^3 w+49152 q r^4 t^4 u
   v^5+40152 q r^4 t^3 u^5 w^2+192000 q r^4 t^3 u^4 v^2 w-147456 q r^4 t^3 u^3
   v^4-207576 q r^4 t^2 u^6 v w+152928 q r^4 t^2 u^5 v^3+54999 q r^4 t u^8 w-47466 q
   r^4 t u^7 v^2+2187 q r^4 u^9 v+36 q r^3 s^6 w^5-1284 q r^3 s^5 t v w^4-93 q r^3
   s^5 u^2 w^4-4554 q r^3 s^5 u v^2 w^3-6885 q r^3 s^5 v^4 w^2+17760 q r^3 s^4 t^2 u
   w^4-27720 q r^3 s^4 t^2 v^2 w^3+2640 q r^3 s^4 t u^2 v w^3+95904 q r^3 s^4 t u v^3
   w^2-13608 q r^3 s^4 t v^5 w-4635 q r^3 s^4 u^4 w^3-28350 q r^3 s^4 u^3 v^2
   w^2+3483 q r^3 s^4 u^2 v^4 w-972 q r^3 s^4 u v^6-2880 q r^3 s^3 t^4 w^4+78720 q
   r^3 s^3 t^3 u v w^3-18304 q r^3 s^3 t^3 v^3 w^2-88880 q r^3 s^3 t^2 u^3
   w^3-200160 q r^3 s^3 t^2 u^2 v^2 w^2+93312 q r^3 s^3 t^2 u v^4 w-13824 q r^3 s^3
   t^2 v^6+125820 q r^3 s^3 t u^4 v w^2-45360 q r^3 s^3 t u^2 v^5+18009 q r^3 s^3 u^6
   w^2-39690 q r^3 s^3 u^5 v^2 w+31590 q r^3 s^3 u^4 v^4-37632 q r^3 s^2 t^5 v
   w^3-5760 q r^3 s^2 t^4 u^2 w^3-5760 q r^3 s^2 t^4 u v^2 w^2+66048 q r^3 s^2 t^4
   v^4 w+200640 q r^3 s^2 t^3 u^3 v w^2-230400 q r^3 s^2 t^3 u v^5+35136 q r^3 s^2 t^2
   u^5 w^2-339120 q r^3 s^2 t^2 u^4 v^2 w+544320 q r^3 s^2 t^2 u^3 v^4+17820 q r^3 s^2
   t u^6 v w-270216 q r^3 s^2 t u^5 v^3+13527 q r^3 s^2 u^8 w+53946 q r^3 s^2 u^7
   v^2+49152 q r^3 s t^6 u w^3+6144 q r^3 s t^6 v^2 w^2-31488 q r^3 s t^5 u^2 v
   w^2-98304 q r^3 s t^5 v^5-166320 q r^3 s t^4 u^4 w^2-226560 q r^3 s t^4 u^3 v^2
   w+652800 q r^3 s t^4 u^2 v^4+641280 q r^3 s t^3 u^5 v w-1235520 q r^3 s t^3 u^4
   v^3-207144 q r^3 s t^2 u^7 w+841752 q r^3 s t^2 u^6 v^2-284148 q r^3 s t u^8
   v+43740 q r^3 s u^{10}-12288 q r^3 t^8 w^3-12288 q r^3 t^7 u v w^2+57344 q r^3 t^6
   u^3 w^2+116736 q r^3 t^6 u^2 v^2 w-98304 q r^3 t^6 u v^4-247680 q r^3 t^5 u^4 v
   w+200704 q r^3 t^5 u^3 v^3+81456 q r^3 t^4 u^6 w-101184 q r^3 t^4 u^5 v^2+20736 q
   r^3 t^3 u^7 v-3240 q r^3 t^2 u^9+324 q r^2 s^7 v w^4-8808 q r^2 s^6 t u w^4+22398
   q r^2 s^6 t v^2 w^3+2835 q r^2 s^6 u^2 v w^3-4860 q r^2 s^6 u v^3 w^2+11259 q r^2
   s^6 v^5 w+792 q r^2 s^5 t^3 w^4-109800 q r^2 s^5 t^2 u v w^3+30600 q r^2 s^5 t^2
   v^3 w^2+58884 q r^2 s^5 t u^3 w^3+20736 q r^2 s^5 t u^2 v^2 w^2-167832 q r^2 s^5 t
   u v^4 w+16848 q r^2 s^5 t v^6-3807 q r^2 s^5 u^4 v w^2+39690 q r^2 s^5 u^3 v^3
   w+13851 q r^2 s^5 u^2 v^5+50400 q r^2 s^4 t^4 v w^3+81360 q r^2 s^4 t^3 u^2
   w^3+92880 q r^2 s^4 t^3 u v^2 w^2-192000 q r^2 s^4 t^3 v^4 w-97920 q r^2 s^4 t^2
   u^3 v w^2+339120 q r^2 s^4 t^2 u^2 v^3 w+265680 q r^2 s^4 t^2 u v^5-122364 q r^2
   s^4 t u^5 w^2-329670 q r^2 s^4 t u^3 v^4+30861 q r^2 s^4 u^6 v w+59778 q r^2 s^4
   u^5 v^3-103296 q r^2 s^3 t^5 u w^3-15360 q r^2 s^3 t^5 v^2 w^2-213120 q r^2 s^3
   t^4 u^2 v w^2+226560 q r^2 s^3 t^4 u v^3 w+284160 q r^2 s^3 t^4 v^5+139320 q r^2 s^3
   t^3 u^4 w^2-1434240 q r^2 s^3 t^3 u^2 v^4-48168 q r^2 s^3 t^2 u^5 v w+1464480 q
   r^2 s^3 t^2 u^4 v^3+11340 q r^2 s^3 t u^7 w-534438 q r^2 s^3 t u^6 v^2+54675 q r^2
   s^3 u^8 v+24576 q r^2 s^2 t^7 w^3+124416 q r^2 s^2 t^6 u v w^2-116736 q r^2 s^2 t^6
   v^3 w+131712 q r^2 s^2 t^5 u^3 w^2-248832 q r^2 s^2 t^5 u v^4-498960 q r^2 s^2 t^4
   u^4 v w+1610880 q r^2 s^2 t^4 u^3 v^3+190920 q r^2 s^2 t^3 u^6 w-1849824 q r^2 s^2
   t^3 u^5 v^2+874800 q r^2 s^2 t^2 u^7 v-153900 q r^2 s^2 t u^9-12288 q r^2 s t^8 v
   w^2-129024 q r^2 s t^7 u^2 w^2+196608 q r^2 s t^7 v^4+390144 q r^2 s t^6 u^3 v
   w-866304 q r^2 s t^6 u^2 v^3-155520 q r^2 s t^5 u^5 w+868608 q r^2 s t^5 u^4
   v^2-387600 q r^2 s t^4 u^6 v+70200 q r^2 s t^3 u^8+24576 q r^2 t^9 u w^2-73728 q
   r^2 t^8 u^2 v w+49152 q r^2 t^8 u v^3+30720 q r^2 t^7 u^4 w-36864 q r^2 t^7 u^3
   v^2+11520 q r^2 t^6 u^5 v-2400 q r^2 t^5 u^7+1584 q r s^8 u w^4-5364 q r s^8 v^2
   w^3+336 q r s^7 t^2 w^4+48912 q r s^7 t u v w^3-45684 q r s^7 t v^3 w^2-13428 q r
   s^7 u^3 w^3-5103 q r s^7 u^2 v^2 w^2+21951 q r s^7 u v^4 w-8748 q r s^7 v^6-22896
   q r s^6 t^3 v w^3-61764 q r s^6 t^2 u^2 w^3+45504 q r s^6 t^2 u v^2 w^2+207576 q r
   s^6 t^2 v^4 w-36504 q r s^6 t u^3 v w^2-17820 q r s^6 t u^2 v^3 w-83268 q r s^6 t
   u v^5+27540 q r s^6 u^5 w^2-30861 q r s^6 u^4 v^2 w+29889 q r s^6 u^3 v^4+63216 q r
   s^5 t^4 u w^3-31248 q r s^5 t^4 v^2 w^2+137088 q r s^5 t^3 u^2 v w^2-641280 q r
   s^5 t^3 u v^3 w-259488 q r s^5 t^3 v^5+93780 q r s^5 t^2 u^4 w^2+48168 q r s^5 t^2
   u^3 v^2 w+735156 q r s^5 t^2 u^2 v^4-338904 q r s^5 t u^4 v^3-6075 q r s^5 u^7
   w+53460 q r s^5 u^6 v^2-14784 q r s^4 t^6 w^3-41856 q r s^4 t^5 u v w^2+247680 q r
   s^4 t^5 v^3 w-272400 q r s^4 t^4 u^3 w^2+498960 q r s^4 t^4 u^2 v^2 w+713280 q r
   s^4 t^4 u v^4-1759680 q r s^4 t^3 u^3 v^3-48600 q r s^4 t^2 u^6 w+1095120 q r s^4
   t^2 u^5 v^2-275400 q r s^4 t u^7 v+30375 q r s^4 u^9-3072 q r s^3 t^7 v w^2+168960
   q r s^3 t^6 u^2 w^2-390144 q r s^3 t^6 u v^2 w-331776 q r s^3 t^6 v^4+206592 q r
   s^3 t^5 u^2 v^3+44400 q r s^3 t^4 u^5 w+638400 q r s^3 t^4 u^4 v^2-565200 q r s^3
   t^3 u^6 v+121500 q r s^3 t^2 u^8-30720 q r s^2 t^8 u w^2+73728 q r s^2 t^8 v^2
   w+454656 q r s^2 t^7 u v^3-9600 q r s^2 t^6 u^4 w-898560 q r s^2 t^6 u^3
   v^2+561600 q r s^2 t^5 u^5 v-114000 q r s^2 t^4 u^7-98304 q r s t^9 v^3+184320 q r
   s t^8 u^2 v^2-115200 q r s t^7 u^4 v+24000 q r s t^6 u^6-144 q s^9 t w^4-7488 q
   s^9 u v w^3+13932 q s^9 v^3 w^2+3504 q s^8 t^2 v w^3+14280 q s^8 t u^2 w^3-26568 q
   s^8 t u v^2 w^2-54999 q s^8 t v^4 w+24300 q s^8 u^3 v w^2-13527 q s^8 u^2 v^3
   w+21870 q s^8 u v^5-12960 q s^7 t^3 u w^3+14136 q s^7 t^3 v^2 w^2-31500 q s^7 t^2
   u^2 v w^2+207144 q s^7 t^2 u v^3 w+61560 q s^7 t^2 v^5-47925 q s^7 t u^4 w^2-11340
   q s^7 t u^3 v^2 w-163215 q s^7 t u^2 v^4+6075 q s^7 u^5 v w+42525 q s^7 u^4
   v^3+2960 q s^6 t^5 w^3+2400 q s^6 t^4 u v w^2-81456 q s^6 t^4 v^3 w+91600 q s^6 t^3
   u^3 w^2-190920 q s^6 t^3 u^2 v^2 w-213120 q s^6 t^3 u v^4+48600 q s^6 t^2 u^4 v
   w+468720 q s^6 t^2 u^3 v^3-230850 q s^6 t u^5 v^2+30375 q s^6 u^7 v+2880 q s^5 t^6
   v w^2-52800 q s^5 t^5 u^2 w^2+155520 q s^5 t^5 u v^2 w+101376 q s^5 t^5 v^4-44400
   q s^5 t^4 u^3 v w+6960 q s^5 t^4 u^2 v^3-316800 q s^5 t^3 u^4 v^2+243000 q s^5 t^2
   u^6 v-50625 q s^5 t u^8+9600 q s^4 t^7 u w^2-30720 q s^4 t^7 v^2 w+9600 q s^4 t^6
   u^2 v w-184320 q s^4 t^6 u v^3+357600 q s^4 t^5 u^3 v^2-222000 q s^4 t^4 u^5
   v+45000 q s^4 t^3 u^7+40960 q s^3 t^8 v^3-76800 q s^3 t^7 u^2 v^2+48000 q s^3 t^6
   u^4 v-10000 q s^3 t^5 u^6+2 r^9 w^6-36 r^8 s v w^5+18 r^8 u^2 v w^4+45 r^7 s^2 u
   w^5+270 r^7 s^2 v^2 w^4-72 r^7 s t u v w^4-45 r^7 s u^3 w^4-216 r^7 s u^2 v^2
   w^3+768 r^7 t^3 v w^4-960 r^7 t^2 u^2 w^4-1536 r^7 t^2 u v^2 w^3+2736 r^7 t u^3 v
   w^3-756 r^7 u^5 w^3-162 r^7 u^4 v^2 w^2-30 r^6 s^3 t w^5-639 r^6 s^3 u v
   w^4-1080 r^6 s^3 v^3 w^3-1656 r^6 s^2 t^2 v w^4+1665 r^6 s^2 t u^2 w^4+3168 r^6
   s^2 t u v^2 w^3-1242 r^6 s^2 u^3 v w^3+972 r^6 s^2 u^2 v^3 w^2+1920 r^6 s t^3 u
   w^4-6144 r^6 s t^3 v^2 w^3+2784 r^6 s t^2 u^2 v w^3+13824 r^6 s t^2 u v^3 w^2+720
   r^6 s t u^4 w^3-23328 r^6 s t u^3 v^2 w^2+7614 r^6 s u^5 v w^2+972 r^6 s u^4 v^3
   w-2816 r^6 t^5 w^4+13824 r^6 t^4 u v w^3+8192 r^6 t^4 v^3 w^2-5760 r^6 t^3 u^3
   w^3-52224 r^6 t^3 u^2 v^2 w^2+55080 r^6 t^2 u^4 v w^2+13824 r^6 t^2 u^3 v^3
   w-20250 r^6 t u^6 w^2-16848 r^6 t u^5 v^2 w+8748 r^6 u^7 v w-1458 r^6 u^6 v^3+9
   r^5 s^5 w^5+1674 r^5 s^4 t v w^4-225 r^5 s^4 u^2 w^4+2754 r^5 s^4 u v^2 w^3+2430
   r^5 s^4 v^4 w^2-3720 r^5 s^3 t^2 u w^4+13728 r^5 s^3 t^2 v^2 w^3-12240 r^5 s^3 t
   u^2 v w^3-24624 r^5 s^3 t u v^3 w^2+675 r^5 s^3 u^4 w^3+12960 r^5 s^3 u^3 v^2
   w^2-1944 r^5 s^3 u^2 v^4 w+8640 r^5 s^2 t^4 w^4-43200 r^5 s^2 t^3 u v w^3-10752
   r^5 s^2 t^3 v^3 w^2+6600 r^5 s^2 t^2 u^3 w^3+123120 r^5 s^2 t^2 u^2 v^2 w^2-41472
   r^5 s^2 t^2 u v^4 w-65610 r^5 s^2 t u^4 v w^2+45360 r^5 s^2 t u^3 v^3 w+5670 r^5
   s^2 u^6 w^2-13851 r^5 s^2 u^5 v^2 w-1458 r^5 s^2 u^4 v^4+6144 r^5 s t^5 v
   w^3+23040 r^5 s t^4 u v^2 w^2-49152 r^5 s t^4 v^4 w-127680 r^5 s t^3 u^3 v
   w^2+230400 r^5 s t^3 u^2 v^3 w+105300 r^5 s t^2 u^5 w^2-265680 r^5 s t^2 u^4 v^2
   w-41472 r^5 s t^2 u^3 v^4+83268 r^5 s t u^6 v w+68040 r^5 s t u^5 v^3-21870 r^5 s
   u^8 w-15309 r^5 s u^7 v^2-49152 r^5 t^6 v^2 w^2+119808 r^5 t^5 u^2 v w^2+98304 r^5
   t^5 u v^3 w-76800 r^5 t^4 u^4 w^2-284160 r^5 t^4 u^3 v^2 w-73728 r^5 t^4 u^2
   v^4+259488 r^5 t^3 u^5 v w+241920 r^5 t^3 u^4 v^3-61560 r^5 t^2 u^7 w-280584 r^5
   t^2 u^6 v^2+126846 r^5 t u^8 v-19683 r^5 u^{10}-441 r^4 s^6 v w^4+1575 r^4 s^5 t u
   w^4-13356 r^4 s^5 t v^2 w^3+1458 r^4 s^5 u^2 v w^3-2430 r^4 s^5 u v^3 w^2-2916
   r^4 s^5 v^5 w-11100 r^4 s^4 t^3 w^4+54840 r^4 s^4 t^2 u v w^3-6480 r^4 s^4 t^2 v^3
   w^2+1350 r^4 s^4 t u^3 w^3-38070 r^4 s^4 t u^2 v^2 w^2+69984 r^4 s^4 t u v^4 w+6885
   r^4 s^4 u^4 v w^2-31590 r^4 s^4 u^3 v^3 w+1458 r^4 s^4 u^2 v^5-21120 r^4 s^3 t^4 v
   w^3+22800 r^4 s^3 t^3 u^2 w^3-53760 r^4 s^3 t^3 u v^2 w^2+147456 r^4 s^3 t^3 v^4
   w+141480 r^4 s^3 t^2 u^3 v w^2-544320 r^4 s^3 t^2 u^2 v^3 w+41472 r^4 s^3 t^2 u
   v^5-73305 r^4 s^3 t u^5 w^2+329670 r^4 s^3 t u^4 v^2 w-29889 r^4 s^3 u^6 v
   w-15309 r^4 s^3 u^5 v^3-7680 r^4 s^2 t^5 u w^3+170496 r^4 s^2 t^5 v^2 w^2-123840
   r^4 s^2 t^4 u^2 v w^2-652800 r^4 s^2 t^4 u v^3 w+73728 r^4 s^2 t^4 v^5-136500 r^4
   s^2 t^3 u^4 w^2+1434240 r^4 s^2 t^3 u^3 v^2 w-735156 r^4 s^2 t^2 u^5 v w-278640
   r^4 s^2 t^2 u^4 v^3+163215 r^4 s^2 t u^7 w+147987 r^4 s^2 t u^6 v^2-24057 r^4 s^2
   u^8 v-233472 r^4 s t^6 u v w^2+98304 r^4 s t^6 v^3 w+314880 r^4 s t^5 u^3
   w^2+248832 r^4 s t^5 u^2 v^2 w-713280 r^4 s t^4 u^4 v w-345600 r^4 s t^4 u^3
   v^3+213120 r^4 s t^3 u^6 w+774144 r^4 s t^3 u^5 v^2-441936 r^4 s t^2 u^7 v+76545
   r^4 s t u^9+73728 r^4 t^8 v w^2-92160 r^4 t^7 u^2 w^2-196608 r^4 t^7 u v^2
   w+331776 r^4 t^6 u^3 v w+147456 r^4 t^6 u^2 v^3-101376 r^4 t^5 u^5 w-309504 r^4
   t^5 u^4 v^2+179424 r^4 t^4 u^6 v-31860 r^4 t^3 u^8-180 r^3 s^7 u w^4+3618 r^3 s^7
   v^2 w^3+7800 r^3 s^6 t^2 w^4-20412 r^3 s^6 t u v w^3+24084 r^3 s^6 t v^3 w^2-945
   r^3 s^6 u^3 w^3+10368 r^3 s^6 u^2 v^2 w^2-8991 r^3 s^6 u v^4 w+1458 r^3 s^6
   v^6+30048 r^3 s^5 t^3 v w^3-49560 r^3 s^5 t^2 u^2 w^3-10152 r^3 s^5 t^2 u v^2
   w^2-152928 r^3 s^5 t^2 v^4 w-60264 r^3 s^5 t u^3 v w^2+270216 r^3 s^5 t u^2 v^3
   w-68040 r^3 s^5 t u v^5+18549 r^3 s^5 u^5 w^2-59778 r^3 s^5 u^4 v^2 w+15309 r^3
   s^5 u^3 v^4+4800 r^3 s^4 t^4 u w^3-245280 r^3 s^4 t^4 v^2 w^2-10560 r^3 s^4 t^3
   u^2 v w^2+1235520 r^3 s^4 t^3 u v^3 w-241920 r^3 s^4 t^3 v^5+248400 r^3 s^4 t^2 u^4
   w^2-1464480 r^3 s^4 t^2 u^3 v^2 w+278640 r^3 s^4 t^2 u^2 v^4+338904 r^3 s^4 t u^5 v
   w-42525 r^3 s^4 u^7 w-4374 r^3 s^4 u^6 v^2+5120 r^3 s^3 t^6 w^3+604416 r^3 s^3 t^5
   u v w^2-200704 r^3 s^3 t^5 v^3 w-320000 r^3 s^3 t^4 u^3 w^2-1610880 r^3 s^3 t^4
   u^2 v^2 w+345600 r^3 s^3 t^4 u v^4+1759680 r^3 s^3 t^3 u^4 v w-468720 r^3 s^3 t^2
   u^6 w-321408 r^3 s^3 t^2 u^5 v^2+160380 r^3 s^3 t u^7 v-18225 r^3 s^3 u^9-208896
   r^3 s^2 t^7 v w^2-168960 r^3 s^2 t^6 u^2 w^2+866304 r^3 s^2 t^6 u v^2 w-147456 r^3
   s^2 t^6 v^4-206592 r^3 s^2 t^5 u^3 v w-6960 r^3 s^2 t^4 u^5 w-344160 r^3 s^2 t^4
   u^4 v^2+300240 r^3 s^2 t^3 u^6 v-64800 r^3 s^2 t^2 u^8+184320 r^3 s t^8 u
   w^2-49152 r^3 s t^8 v^2 w-454656 r^3 s t^7 u^2 v w+184320 r^3 s t^6 u^4 w+374784
   r^3 s t^6 u^3 v^2-301440 r^3 s t^5 u^5 v+61200 r^3 s t^4 u^7-32768 r^3 t^{10}
   w^2+98304 r^3 t^9 u v w-40960 r^3 t^8 u^3 w-73728 r^3 t^8 u^2 v^2+61440 r^3 t^7
   u^4 v-12800 r^3 t^6 u^6-2880 r^2 s^8 t w^4+1836 r^2 s^8 u v w^3-8586 r^2 s^8 v^3
   w^2-26856 r^2 s^7 t^2 v w^3+29340 r^2 s^7 t u^2 w^3-1134 r^2 s^7 t u v^2 w^2+47466
   r^2 s^7 t v^4 w+6723 r^2 s^7 u^3 v w^2-53946 r^2 s^7 u^2 v^3 w+15309 r^2 s^7 u
   v^5+15720 r^2 s^6 t^3 u w^3+203592 r^2 s^6 t^3 v^2 w^2+150984 r^2 s^6 t^2 u^2 v
   w^2-841752 r^2 s^6 t^2 u v^3 w+280584 r^2 s^6 t^2 v^5-130005 r^2 s^6 t u^4
   w^2+534438 r^2 s^6 t u^3 v^2 w-147987 r^2 s^6 t u^2 v^4-53460 r^2 s^6 u^5 v w+4374
   r^2 s^6 u^4 v^3-11136 r^2 s^5 t^5 w^3-661536 r^2 s^5 t^4 u v w^2+101184 r^2 s^5
   t^4 v^3 w-8160 r^2 s^5 t^3 u^3 w^2+1849824 r^2 s^5 t^3 u^2 v^2 w-774144 r^2 s^5
   t^3 u v^4-1095120 r^2 s^5 t^2 u^4 v w+321408 r^2 s^5 t^2 u^3 v^3+230850 r^2 s^5 t
   u^6 w-18225 r^2 s^5 u^7 v+234624 r^2 s^4 t^6 v w^2+482400 r^2 s^4 t^5 u^2
   w^2-868608 r^2 s^4 t^5 u v^2 w+309504 r^2 s^4 t^5 v^4-638400 r^2 s^4 t^4 u^3 v
   w+344160 r^2 s^4 t^4 u^2 v^3+316800 r^2 s^4 t^3 u^5 w-145800 r^2 s^4 t^2 u^6
   v+30375 r^2 s^4 t u^8-337920 r^2 s^3 t^7 u w^2+36864 r^2 s^3 t^7 v^2 w+898560 r^2
   s^3 t^6 u^2 v w-374784 r^2 s^3 t^6 u v^3-357600 r^2 s^3 t^5 u^4 w+133200 r^2 s^3
   t^4 u^5 v-27000 r^2 s^3 t^3 u^7+61440 r^2 s^2 t^9 w^2-184320 r^2 s^2 t^8 u v
   w+73728 r^2 s^2 t^8 v^3+76800 r^2 s^2 t^7 u^3 w-28800 r^2 s^2 t^6 u^4 v+6000 r^2
   s^2 t^5 u^6+432 r s^{10} w^4+12384 r s^9 t v w^3-5400 r s^9 u^2 w^3+5508 r s^9 u
   v^2 w^2-2187 r s^9 v^4 w-15120 r s^8 t^2 u w^3-84024 r s^8 t^2 v^2 w^2-93960 r
   s^8 t u^2 v w^2+284148 r s^8 t u v^3 w-126846 r s^8 t v^5+22275 r s^8 u^4
   w^2-54675 r s^8 u^3 v^2 w+24057 r s^8 u^2 v^4+7440 r s^7 t^4 w^3+315120 r s^7 t^3 u
   v w^2-20736 r s^7 t^3 v^3 w+99900 r s^7 t^2 u^3 w^2-874800 r s^7 t^2 u^2 v^2
   w+441936 r s^7 t^2 u v^4+275400 r s^7 t u^4 v w-160380 r s^7 t u^3 v^3-30375 r s^7
   u^6 w+18225 r s^7 u^5 v^2-114720 r s^6 t^5 v w^2-319200 r s^6 t^4 u^2 w^2+387600 r
   s^6 t^4 u v^2 w-179424 r s^6 t^4 v^4+565200 r s^6 t^3 u^3 v w-300240 r s^6 t^3 u^2
   v^3-243000 r s^6 t^2 u^5 w+145800 r s^6 t^2 u^4 v^2+206400 r s^5 t^6 u w^2-11520 r
   s^5 t^6 v^2 w-561600 r s^5 t^5 u^2 v w+301440 r s^5 t^5 u v^3+222000 r s^5 t^4 u^4
   w-133200 r s^5 t^4 u^3 v^2-38400 r s^4 t^8 w^2+115200 r s^4 t^7 u v w-61440 r s^4
   t^7 v^3-48000 r s^4 t^6 u^3 w+28800 r s^4 t^6 u^2 v^2-2160 s^{11} v w^3+3600
   s^{10} t u w^3+12960 s^{10} t v^2 w^2+16200 s^{10} u^2 v w^2-43740 s^{10} u v^3
   w+19683 s^{10} v^5-1600 s^9 t^3 w^3-54000 s^9 t^2 u v w^2+3240 s^9 t^2 v^3
   w-27000 s^9 t u^3 w^2+153900 s^9 t u^2 v^2 w-76545 s^9 t u v^4-30375 s^9 u^4 v
   w+18225 s^9 u^3 v^3+20400 s^8 t^4 v w^2+66000 s^8 t^3 u^2 w^2-70200 s^8 t^3 u v^2
   w+31860 s^8 t^3 v^4-121500 s^8 t^2 u^3 v w+64800 s^8 t^2 u^2 v^3+50625 s^8 t u^5
   w-30375 s^8 t u^4 v^2-42000 s^7 t^5 u w^2+2400 s^7 t^5 v^2 w+114000 s^7 t^4 u^2 v
   w-61200 s^7 t^4 u v^3-45000 s^7 t^3 u^4 w+27000 s^7 t^3 u^3 v^2+8000 s^6 t^7
   w^2-24000 s^6 t^6 u v w+12800 s^6 t^6 v^3+10000 s^6 t^5 u^3 w-6000 s^6 t^5 u^2
   v^2$.
\fi
\noindent
The representatives have been chosen so that $I_4$ has vanishing coefficient for $q^2w^2$, $I_6$ for
$q^3w^3$ and $q^2suw^2$ and $I_{10}$ for $q^5w^5$, $q^4suw^4$, $q^4t^2w^4$, $q^3s^2u^2w^3$ and $q^3st^2uw^3$.

These invariants fulfill the following polynomial relation at order 30:
\begin{align}\label{15'2}
0&=I_{15}^2+4I_{10}^3 +396I_{10}^2I_6I_4-I_{10}^2I_6I_2^2 +120I_{10}^2I_4^2I_2 -18I_{10}I_6^3I_2 +12528I_{10}I_6^2I_4^2 \nn\\
&-120I_{10}I_6^2I_4I_2^2 +6768I_{10}I_6I_4^3I_2 -84I_{10}I_6I_4^2I_2^3 +6912I_{10}I_4^5 +48I_{10}I_4^4I_2^2 \nn\\
&-16I_{10}I_4^3I_2^4 +27I_6^5 -432I_6^4I_4I_2 +4I_6^4I_2^3 +124848I_6^3I_4^3 -2376I_6^3I_4^2I_2^2\\
&+12I_6^3I_4I_2^4 +88992I_6^2I_4^4I_2 -2096I_6^2I_4^3I_2^3  +12I_6^2I_4^2I_2^5 +297216I_6I_4^6 \nn\\
&-4752I_6I_4^5I_2^2 -372I_6I_4^4I_2^4 +4I_6I_4^3I_2^6 +179712I_4^7I_2 -7520I_4^6I_2^3 +96I_4^5I_2^5\;. \nn
\end{align}
{From this relation it follows that the invariant $I_{15}$ is functionally dependent.}

\section{List of useful identities involving $10D$ self-dual 5-forms}\label{id5f}
All the identities are obtained by replacing (anti)self-dual 5-forms  with their Hodge duals and using the expression of the product of the two Levi-Civita symbols as the generalized Kronecker delta
$$
\frac 1{m!\,p!}\varepsilon_{\kappa_1\ldots\kappa_m\mu_1\ldots \mu_{p}}\varepsilon^{\kappa_1\ldots\kappa_m\nu_1\ldots \nu_{p}}=-\delta^{[\nu_1}_{\mu_1}\ldots \delta^{\nu_p]}_{\mu_p} \equiv -\delta_{\mu_1 \dots \mu_p}^{[\nu_1 \dots \nu_p]}\,.
$$
For two different self-dual $F^1_{\mu_1\ldots \mu_5}$ and $F^2_{\mu_1\ldots \mu_5}$ we thus have
\bea\label{id4final*}
&F^{1}_{\mu_1\mu_2\ldots\mu_5}F^{2\,\nu_1\ldots\nu_5}
= F^2_{\mu_1\mu_2\ldots\mu_5}F^{1\,\nu_1\ldots\nu_5}-25\delta^{[\nu_5}_{[\mu_5}\,F^2_{\mu_1\mu_2\mu_3\mu_4]\lambda}F^{1\,\nu_1\nu_2\nu_3\nu_4]\lambda} &\\
&
+ 100\,\delta^{[\nu_5}_{[\mu_5}\delta^{\nu_1}_{\mu_1}F^2_{\mu_2\mu_3\mu_4]\lambda_1\lambda_2}F^{1\,\nu_2\nu_3\nu_4]\lambda_1\lambda_2} &\nonumber\\
& - 150\,\delta^{[\nu_5}_{[\mu_5}\delta^{\nu_1}_{\mu_1}\delta^{\nu_2}_{\mu_2}F^2_{\mu_3\mu_4]\lambda_1\lambda_2\lambda_3}F^{1\,\nu_3\nu_4]\lambda_1\lambda_2\lambda_3}+50\,\delta^{[\nu_5}_{[\mu_5}\delta^{\nu_1}_{\mu_1}\delta^{\nu_2}_{\mu_2}\delta^{\nu_3}_{\mu_3}F^2_{\mu_4]\lambda_1\lambda_2\lambda_3\lambda_4}F^{1\, \nu_4]\lambda_1\lambda_2\lambda_3\lambda_4}&\,,\nonumber
 \eea
\bea\label{id3cfinal*}
F^1_{\mu_1\mu_2\mu_3\mu_4\lambda}F^{2\,\nu_1\nu_2\nu_3\nu_4\lambda}=-\,F^2_{\mu_1\mu_2\mu_3\mu_4\lambda}F^{1\,\nu_1\nu_2\nu_3\nu_4\lambda}
+ 8\,\delta^{[\nu_1}_{[\mu_1}F^2_{\mu_2\mu_3\mu_4]\lambda_1\lambda_2}F^{1\,\nu_2\nu_3\nu_4]\lambda_1\lambda_2}\nonumber\\
-12\,\delta^{[\nu_1}_{[\mu_1}\delta^{\nu_2}_{\mu_2}F^2_{\mu_3\mu_4]\lambda_1\lambda_2\lambda_3}F^{1\,\nu_3\nu_4]\lambda_1\lambda_2\lambda_3}+4\,\delta^{[\nu_1}_{[\mu_1}\delta^{\nu_2}_{\mu_2}\delta^{\nu_3}_{\mu_3}F^2_{\mu_4]\lambda_1\lambda_2\lambda_3\lambda_4}F^{1\,\nu_4]\lambda_1\lambda_2\lambda_3\lambda_4}\,.
\eea
\bea\label{id2c**}
F^1_{\mu_1\mu_2\mu_3\lambda_1\lambda_2}F^{2\nu_1\nu_2\nu_3\lambda_1 \lambda_2}
=F^2_{\mu_1\mu_2\mu_3\lambda_1\lambda_2}F^{1\,\nu_1\nu_2\nu_3\lambda_1 \lambda_2}- 3\,\delta^{[\nu_1}_{[\mu_1}F^2_{\mu_2\mu_3]\kappa_1\kappa_2\kappa_3}\,F^{1\nu_2\nu_3]\kappa_1\kappa_2\kappa_3}\nonumber\\
+\frac 32\,\delta^{[\nu_1}_{[\mu_1}\delta^{\nu_2}_{\mu_2}F^2_{\mu_3]\kappa_1\kappa_2\kappa_3\kappa_4}\,F^{1\nu_3]\kappa_1\kappa_2\kappa_3\kappa_4}\,.
\eea
\bea\label{id3*}
F^1_{\mu_1\mu_2\lambda_1\lambda_2\lambda_3}F^{2\nu_1\nu_2\lambda_1 \lambda_2\lambda_3}=
-F^2_{\mu_1\mu_2\lambda_1\lambda_2\lambda_3}F^{1\,\nu_1\nu_2\lambda_1 \lambda_2\lambda_3}+\delta_{[\mu_1}^{[\nu_1}F^{1\,\nu_2]\lambda_1 \lambda_2\lambda_3\lambda_4}F^2_{\mu_2]\lambda_1\lambda_2\lambda_3\lambda_4}
\eea
From the above identities  for a single $F_5$ we have
\bea\label{id2c*}
N_{\mu_1\mu_2\mu_3}{}^{\nu_1\nu_2\nu_3}
=N_{\mu_1\mu_2\mu_3}{}^{\nu_1\nu_2\nu_3}- 3\,\delta^{[\nu_1}_{[\mu_1}F_{\mu_2\mu_3]\kappa_1\kappa_2\kappa_3}\,F^{\nu_2\nu_3]\kappa_1\kappa_2\kappa_3}+\frac 32\,\delta^{[\nu_1}_{[\mu_1}\delta^{\nu_2}_{\mu_2}M_{\mu_3]}{}^{\nu_3]},
\eea
and hence
\be\label{id1*}
N_{\mu\rho\lambda}{}^{\nu\kappa\lambda}=\frac 12M_{[\mu}{}^{[\nu}\delta^{\kappa]}_{\rho]},
\ee
where
\be\label{M,N}
M_{\mu}{}^\nu=F_{\mu\lambda_1\ldots\lambda_4} F^{\nu\lambda_1\ldots\lambda_4}\,, \qquad
N_{\mu_1\mu_2\mu_3}{}^{\nu_1\nu_2\nu_3}=F_{\mu_1\mu_2\mu_3\lambda_1\lambda_2}F^{\nu_1\nu_2\nu_3\lambda_1\lambda_2}\,.
\ee
Then
\be\label{id3ccf*}
F_{\mu_1\mu_2\mu_3\mu_4\lambda}F^{\nu_1\nu_2\nu_3\nu_4\lambda}=4\,\delta^{[\nu_1}_{[\mu_1}N_{\mu_2\mu_3\mu_4]}{}^{\nu_2\nu_3\nu_4]}-\delta^{[\nu_1}_{[\mu_1}\delta^{\nu_2}_{\mu_2}\delta^{\nu_3}_{\mu_3}M_{\mu_4]}{}^{\nu_4]}\,.
\ee
Other identities are:
\be\label{trx*}
\delta_{\nu_{p+1}}^{\mu_{p+1}}\delta^{\nu_1\ldots\nu_{p}\nu_{p+1}}_{[\mu_1\ldots\mu_{p}\mu_{p+1}]}=\frac{10-p}{p+1}\delta^{\nu_1\ldots\nu_{p}}_{[\mu_1\ldots\mu_{p}]}\,,\qquad \delta_{\nu_3}^{\mu_3}\,\left(\delta^{[\nu_1}_{[\mu_1}\delta^{\nu_2}_{\mu_2}M_{\mu_3]}{}^{\nu_3]}\right)=\frac{14}{9} \delta^{[\nu_1}_{[\mu_1} M_{\mu_2]}{}^{\nu_2]}\,,
\ee
\be\label{trx4*}
\delta^{\mu_4}_{\nu_4}\delta^{[\nu_1}_{[\mu_1}\delta^{\nu_2}_{\mu_2}\delta^{\nu_3}_{\mu_3}M_{\mu_4]}{}^{\nu_4]}=\frac 98\delta^{[\nu_1}_{[\mu_1}\delta^{\nu_2}_{\mu_2}M_{\mu_3]}{}^{\nu_3]}\,.
\ee
\be\label{asdMF*}
M_{[\mu_1}{}^{\lambda}F_{\mu_2\mu_3\mu_4\mu_5]\lambda}=-\frac1{5!}\varepsilon_{\mu_1\ldots\mu_5}{}^{\nu_1\ldots \nu_5}M_{\nu_1}{}^{\lambda}F_{\nu_2\nu_3\nu_4\nu_5\lambda},
\ee
\be\label{MFF*}
M_{[\mu_1}{}^{\lambda}F_{\mu_2\mu_3\mu_4\mu_5]\lambda}F^{\mu_2\mu_3\mu_4\mu_5\kappa}=\frac 1{10}\delta_{\mu_1}^\kappa \tr M^2\,.
\ee
\bea\label{1050+*}
\frac 1{5!}\varepsilon_{\mu_1 \mu_2 \mu_3 \nu_1 \nu_2}{}^{\rho_1\ldots\rho_5}N_{[\rho_1 \rho_2 \rho_3,\rho_4 \rho_5]\nu_3} = N_{[\mu_1 \mu_2 \mu_3, \nu_1 \nu_2] \nu_3}~.
\eea
\be\label{[6]}
N_{[\mu_1 \mu_2 \mu_3, \nu_1 \nu_2 \nu_3]}=0\,.
\ee
From the latter identity it follows that
\be\label{ID6*}
N^{[\alpha_1 \alpha_2\alpha_3, {\color{red}[}\mu_1\mu_2]\mu_3{\color{red}]}}=N^{[\mu_1\mu_2\mu_3,{\color{red}[}\alpha_1\alpha_2]\alpha_3{\color{red}]}}\,,
\ee
(where the indices within the red brackets are antisymmetrized after the antisymmetrization of the indices within the black brackets)
\bea\label{4+2}
N_{\mu_1 [\mu_2 \nu_1, \nu_2 \nu_3\nu_4]}-N_{\mu_2 [\mu_1 \nu_1, \nu_2 \nu_3\nu_4]}
&=&-N_{\nu_1 [\mu_1\mu_2, \nu_2 \nu_3\nu_4]}+N_{\nu_2 [\mu_1 \mu_2,\nu_1 \nu_3\nu_4]}\nonumber\\
&&-N_{\nu_3 [\mu_1\mu_2, \nu_1 \nu_2\nu_4]}+N_{\nu_4 [\mu_1\mu_2 ,\nu_1 \nu_2 \nu_3]}\,.
\eea
or
\bea\label{4+2*}
N_{{\color{red}[}\mu_1 [\mu_2{\color{red}]} \nu_1, \nu_2 \nu_3\nu_4]}
&=&-2N_{{\color{red}[}\nu_1 [\nu_2 \nu_3,\nu_4{\color{red}]}\mu_1\mu_2]}.
\eea
or
\bea\label{4+2**}
N_{[\nu_1\nu_2 \nu_3\nu_4,{\color{red}[}\mu_1]\mu_2{\color{red}]}}
&=&-2N_{[\mu_1\mu_2,{\color{red}[}\nu_1\nu_2 \nu_3]\nu_4{\color{red}]}}.
\eea
\bea\label{N4=N5*}
& N_{[\mu_1 \mu_2 \mu_3,\nu_1 ]\nu_2\nu_3}= \frac{1}{48} \varepsilon_{\mu_1 \mu_2 \mu_3\nu_1[\nu_2}{}^{\lambda_1\ldots\lambda_5}N_{\nu_3]\lambda_1\lambda_2,\lambda_3\lambda_4\lambda_5} &\nonumber\\
%&= \frac{5!}{96} (N_{[\mu_1 \mu_2 \mu_3,\nu_1\nu_2]\nu_3}-N_{[\mu_1 \mu_2 \mu_3,\nu_1\nu_3]\nu_2})\,.&\nonumber\\
&=  \frac{5}{4} (N_{[\mu_1 \mu_2 \mu_3,\nu_1\nu_2]\nu_3}-N_{[\mu_1 \mu_2 \mu_3,\nu_1\nu_3]\nu_2})\,.&
\eea
\bea\label{N4=N51*}
& N_{\mu_1 [\mu_2 \mu_3,\nu_1 \nu_2]\nu_3}= \frac{1}{72}\varepsilon_{\mu_2 \mu_3 \nu_1\nu_2(\nu_3}{}^{\lambda_1\ldots\lambda_5}N_{\mu_1)\lambda_1\lambda_2,\lambda_3\lambda_4\lambda_5}, &
\nonumber\\
%&=\frac{5!}{144}(N_{\mu_1[\mu_2 \mu_3,\nu_1 \nu_2\nu_3]}+N_{\nu_3[\mu_2 \mu_3,\nu_1 \nu_2\mu_1]})&\nonumber\\
&=\frac{5}6(N_{\mu_1[\mu_2 \mu_3,\nu_1 \nu_2\nu_3]}+N_{\nu_3[\mu_2 \mu_3,\nu_1 \nu_2\mu_1]})\,.&
\eea
\be\label{3antisym*}
N_{\mu_1 [\mu_2 \mu_3,\nu_1] \nu_2 \nu_3} =\frac 43 N_{[\mu_1 \mu_2 \mu_3,\nu_1]\nu_2 \nu_3} + \frac 13N_{\nu_1 \mu_2 \mu_3,\mu_1 \nu_2 \nu_3}
\ee
We also have
\bea\label{2antisym*}
N_{\mu_1 \mu_2 [\mu_3,\nu_1] \nu_2 \nu_3} &=&2N_{[\mu_1 \mu_2 \mu_3,\nu_1]\nu_2 \nu_3} -N_{\mu_3\nu_1 [\mu_1 ,\mu_2] \nu_2 \nu_3}\nonumber\\
&=& 3 N_{\mu_1 [ \mu_2 \mu_3,\nu_1\nu_2] \nu_3}-\frac 12 N_{\mu_1\mu_3\nu_1,\mu_2\nu_2\nu_3} + \frac 32 N_{\mu_1 \nu_2 [\mu_2, \mu_3 \nu_1] \nu_3}\,.
\eea

From \eqref{3antisym*} and \eqref{N4=N5*} it follows that
\bea \label{1050+**}
N^{[\mu_1\mu_2\mu_3,{\color{red}[}\alpha_1\alpha_2]\alpha_3{\color{red}]}}=\frac 1{10} (N^{\mu_1\mu_2\mu_3,\alpha_1\alpha_2\alpha_3}-3N^{{\color{blue}[}\mu_1\mu_2{\color{red}[}\alpha_3,\alpha_1\alpha_2{\color{red}]}\mu_3{\color{blue}]}}),
\eea
where on the l.h.s the antisymmetrization of the indices $\alpha$ is taken after the antisymmetrization of the five indices, while in the second term on the r.h.s the two sets of three indices are antisymmetrized separately.

The above identities tell us that the tensor $N_{\mu_1\mu_2\mu_3,\nu_1\nu_2\nu_3}$ takes values in the reducible representation of $SO(1,9)$ which is decomposable into three irreducible representations as given in \eqref{4125irrep}.

Identities involving products of two $N$-tensors (as a consequence of their composite nature in terms of $FF$, eq. \eqref{M,N}):
\bea
\label{NN3*}
N_{\mu_1 \mu_2 \mu_3}{}^{\nu_1 \nu_2 \nu_3} N_{\nu_1 \nu_2 \nu_3}{}^{\rho_1 \rho_2 \rho_3}&=&
-\frac{1}{32} \delta_{[\mu_1}{}^{[\rho_1} \delta_{\mu_2}{}^{\rho_2} \delta_{\mu_3]}{}^{\rho_3]} \tr M^2
+ \frac{3}{16}\delta_{[\mu_1}{}^{[\rho_1} M_{\mu_2}{}^{\rho_2} M_{\mu_3]}{}^{\rho_3]} \nonumber\\
&&+ \frac{3}{16}\delta_{[\mu_1}{}^{[\rho_1} \delta_{\mu_2}{}^{\rho_2} M_{\mu_3]}{}^{|\lambda|} M_{\lambda}{}^{\rho_3]}
+ \frac{9}{8} N_{\lambda [\mu_1 \mu_2}{}^{\nu[\rho_1 \rho_2} \delta_{\mu_3]}{}^{\rho_3]}M_{\nu}{}^{\lambda}
\nonumber\\
&&-\frac{3}{8}M_{\nu}{}^{[\rho_1}N_{\mu_1 \mu_2 \mu_3}{}^{\rho_2 \rho_3]\nu}
-\frac{3}{8}M_{[\mu_1}{}^{\lambda}N_{\mu_2 \mu_3] \lambda}{}^{\rho_1 \rho_2 \rho_3}~,
\eea
\be\label{NN4*1}
N_{\mu_1 \mu_2 \gamma}{}^{\delta_2 \delta_3 \delta_4} N_{\delta_2 \delta_3 \delta_4}{}^{\nu_1 \nu_2 \gamma}=\frac 12 M^{\alpha}{}_{\beta}N_{\alpha \mu_1 \mu_2}{}^{\nu_1 \nu_2 \beta} \,,
\ee
\be\label{NN=I-MM*}
N_{\nu_1 \nu_2 \sigma}{}^{\mu_1 \mu_2 \mu_3} N_{\mu_1 \mu_2 \mu_3}{}^{\nu_1 \nu_2 \lambda}  = \frac{1}{16} \delta_{\sigma}{}^{\lambda} \tr M^2 - \frac{1}{8} (MM)_{\sigma}{}^{\lambda}~,
\ee
\bea\label{hat M=MM}
N^{^{(1050)}}_{[\mu\rho_1\rho_2\rho_3,\rho_4]\rho_5}N_{_{(1050)}}^{[\nu\rho_1\rho_2\rho_3,\rho_4]\rho_5} = -\frac{1}{60} \Big[ (MM)_{\mu}{}^{\nu} -\frac{1}{10} \delta_{\mu}{}^{\nu} \tr M^2\Big]~.
\eea
\be\label{NNscalar}
N_{\nu_1 \nu_2 \nu_3}{}^{\mu_1 \mu_2 \mu_3} N_{\mu_1 \mu_2 \mu_3}{}^{\nu_1 \nu_2 \nu_3}  =\frac 12 \tr M^2\,.
\ee
\be\label{NNscalar-4125}
N^{^{(4125)}}_{\nu_1 \nu_2 \nu_3}{}^{\mu_1 \mu_2 \mu_3} N^{^{(4125)}}_{\mu_1 \mu_2 \mu_3}{}^{\nu_1 \nu_2 \nu_3}  =\frac{5}{28} \tr M^2\,.
\ee
\bea\label{NN4*}
N_{\delta\mu_3 \mu_4 }{}^{\rho \nu_3 \nu_4} N_{\kappa\nu_3 \nu_4}{}^{\lambda\mu_3 \mu_4}
 &=&  F_{\delta \mu_3 \mu_4 \alpha_1 \alpha_2} N_{\kappa \beta_1 \beta_2}{}^{\rho \alpha_1 \alpha_2} F^{\lambda\mu_3 \mu_4  \beta_1 \beta_2}\nonumber\\
 &=& N_{\delta \mu_3 \mu_4}{}^{\lambda \nu_3 \nu_4}\,N_{\kappa \nu_3 \nu_4}{}^{\rho\mu_3 \mu_4}\,.
\eea
 \bea\label{NN3ar*}
2N_{\mu_1 \mu_2 \rho}{}^{\rho_1 \rho_2 \lambda} N_{\rho_1 \rho_2 \kappa}{}^{\nu_1 \nu_2 \rho} &=&
\frac 32
N_{\rho[\mu_1 \mu_2 }{}^{\rho_1 \rho_2 \lambda} N_{\kappa]\rho_1 \rho_2 }{}^{\nu_1 \nu_2 \rho}
+\frac 32 N_{\mu_1 \mu_2 \rho}{}^{\rho_1 \rho_2 [\lambda} N_{\rho_1 \rho_2 \kappa}{}^{\nu_1 \nu_2] \rho}\nonumber\\
&+&\delta_{[\mu_1}^{[\nu_1}N_{\mu_2]\rho_1\rho_2}{}^{\nu_2]\nu_3\nu_4}N_{\kappa \nu_3 \nu_4}{}^{\l \rho_1 \rho_2}+
{ \frac 18} \delta_\kappa^\lambda\,N_{\mu_1\mu_2\rho}{}^{\nu_1\nu_2\nu}M_{\nu}{}^\rho \nonumber\\
&-& { \frac 18} N_{\mu_1\mu_2\rho}{}^{\nu_1\nu_2\lambda}M_\kappa{}^{\rho}
-{ \frac 18} N_{\mu_1\mu_2\kappa}{}^{\nu_1\nu_2\rho}M_\rho{}^{\lambda}\nonumber\\
&+& { \frac{1}{16}} M_{\kappa}{}^\lambda\,\delta_{[\mu_1}^{[\nu_1} M_{\mu_2]}{}^{\nu_2]}-\delta^{[\nu_1}_{[\mu_1}\delta^{\nu_2}_{\mu_2}\delta^{\nu_3}_{\mu_3}M_{\mu_4]}{}^{\nu_4]}\,N_{\kappa \nu_3 \nu_4}{}^{\l \mu_3 \mu_4}~,~~~~~~~~~
\eea
\be
N_{[\rho_2\rho_3\rho_4,\rho_1][\mu_1}{}^{[\nu_1}N_{\mu_2]}{}^{\nu_2][\rho_1,\rho_2\rho_3\rho_4]}=  N_{[\rho_2\rho_3\rho_4,\rho_1]\mu_1\mu_2}N^{[\rho_2\rho_3\rho_4,\rho_1]\nu_1\nu_2}\,,
\ee
\be\label{NNsim}
N^{^{(1050)}}_{[\rho_1\rho_2\rho_3,\rho_4,\mu_1]}{}^{\nu_1}N^{[\rho_1\rho_2\rho_3,\rho_4\nu_2]}{}_{\mu_2}=N^{[\rho_1\rho_2\rho_3,\rho_4,\nu_2]}{}^{\nu_1}N_{[\rho_1\rho_2\rho_3,\rho_4\mu_1]}{}_{\mu_2}\,.
\ee

\section{Spinor formalism in ten dimensions}
\label{spin10d-app}

In this appendix we give a summary  of the ten-dimensional spinor formalism available in the literature (see e.g.
Refs. \cite{Gliozzi:1976qd, Scherk:1978fh, Bergshoeff:1981um, Bergshoeff:1982az, Howe:1983sra, Gates:1986is, Park:2022pjv}) and derive identities used for the construction of independent invariants of $F_5$ in Subsection \ref{spinorform} .

Let $\g_\m  $ be the gamma matrices in ten dimensions,
\begin{subequations} \label{gamma}
\bea
\{ \g_\m , \g_\n \} = -2 \eta_{\m\n} {\mathbbm 1}_{32}~,
\label{gamma.a}
\eea
with $\eta_{\m\n}$ the mostly plus Minkowski metric. We assume $\g_\m$ to obey the standard Hermiticity condition
\bea
(\g^\m)^\dagger = \g^0 \g^\m \g^0 = -\eta_{\m\n} \g^\n = - \g_\m~, \qquad \g^\m = (\g^0, \g^i)~.
\label{gamma.b}
\eea
\end{subequations}
We introduce  matrices $B$ and $C$ as solutions of the equations
\bea
(\g_\m)^* &=& - B \g_\m B^{-1}~, \label{B-matrix}\\
(\g_\m)^{\rm T} &=& - C\g_\m C^{-1} ~,
\label{C-matrix}
\eea
where $(\g_\m)^* $ denotes the complex conjugate of $\g_\m$.
The matrices $B$ and $C$ can always be chosen to be unitary, and prove to be  symmetric and antisymmetric, respectively.
In summary, their properties are:
\bea
B^\dagger B &=& {\mathbbm 1}_{32} ~, \qquad B^{\rm T} = B~,
\label{B-matrix-symmetry} \\
C^\dagger C &=& {\mathbbm 1}_{32} ~, \qquad C^{\rm T} = -C~,
\label{C-matrix-symmetry}
\eea
see \cite{Gliozzi:1976qd,Scherk:1978fh} for more details.
One can choose $C= B \g^0$.

Introducing the matrices
\bea
\g^{\m(k)} = \g^{\m_1 \dots \m_{k}}= \g^{[\m_1} \g^{\m_2}\dots \g^{\m_k]} ~, \qquad 1 \leq k \leq 10
\eea
one observes that the matrices $\g_{\m(k)} C^{-1}$ are (anti)symmetric,
%$\g_\m C^{-1}$ and $\g_{\m (5)} C^{-1}$ are symmetric, while  $\g_{\m(3)}C^{-1}$ is antisymmetric.
\bea
\big(\g_{\m(k)} C^{-1}\big)^{\rm T} = - (-1)^{\hf k(k+1)} \g_{\m(k)} C^{-1}~.
\label{anti-symmetry}
\eea
When working with Weyl spinors, it suffices to deal with $\g_{\m(k)}$ with $k\leq 5$,
since
\bea
\g_{\m_1\dots \m_k} \propto \g^*_{\m_1 \dots \m_k} \g_{11}~,
 \qquad
 \g^*_{\m_1 \dots \m_k} :=
\frac{1}{(10-k)!} \ve_{\m_1 \dots \m_k \n_1 \dots \n_{10-k} }
\g^{\n_1 \dots \n_{10-k}}
%\g_*
~,
\eea
where $\g_{11}$ denotes
the ten-dimensional counterpart of the four-dimensional matrix $\g_5$,
\bea
%\g_* \equiv
\g_{11} = \g^0 \g^1 \dots \g^9  = (\g_{11})^\dagger~, \qquad (\g_{11})^2 ={\mathbbm 1}_{32} ~,
\qquad \{ \g_{11} , \g_\m\} =0~.
\eea
In particular, it holds that
\bea
\g_{\m_1 \dots \m_5} = - \frac{1}{5!} \ve_{\m_1 \dots \m_5 \n_1 \dots \n_5}
\g^{\n_1 \dots \n_5 } \g_{11}~,
\eea
where we have assumed the following normalisation of the Levi-Civita tensor
$\ve_{01\dots 9} =1$.

In what follows, we work in the Weyl representation for $\g_\m$  in which
\bea
\g_{11} = \left(\begin{array}{cc}{\mathbbm 1}_{16} ~&0 \\0~& -{\mathbbm 1}_{16}\end{array}\right)~,
\eea
and the matrices $\g_\m$ are block off-diagonal,
\bea
\g_\m = \left(\begin{array}{cc}0~& (\s_\m)_{a \dot b} \\
(\tilde{\s}_\m)^{\dot a b} ~&0\end{array}\right)~.
\label{Weyl-gamma}
\eea
Here the $\s$-matrices obey the anti-commutation relations
\bea
 \s_\m \tilde{\s}_\n + \s_\n \tilde{\s}_\m = - 2 \eta_{\m\n} {\mathbbm 1}_{16} ~,\qquad
\tilde{\s}_\m {\s}_\n + \tilde{\s}_\n {\s}_\m = - 2 \eta_{\m\n} {\mathbbm 1}_{16} ~ .
 \eea
 In the Weyl representation the Lorentz generators are block diagonal
 \bea
 M_{\m \n} = - \frac 14 [\g_\m , \g_\n] = - \hf \g_{\m\n}
 = -\hf \left(\begin{array}{cc}
 (\s_{\m\n})_a{}^b  ~& 0 \\
0 ~ &(\tilde{\s}_{\m \n})^{\dot a }{}_{\dot b} \end{array}\right)
=\big( (M_{\m\n})_A{}^B\big)~.
 \eea

 A Dirac spinor $\J$ has the form
 \bea
 \J = \left(\begin{array}{c}\vf_{a } \\
\bar \c^{\dot a}
\end{array}\right) \equiv \big( \J_A \big)~,
\eea

In any representation for the $\gamma$-matrices, eq. \eqref{gamma},
it holds that  $(\g_{11})^{\rm T} = - C\g_{11} C^{-1}$. This relation turns into  $\{\g_{11} , C\} =0$ in the Weyl representation since $\g_{11}$ is symmetric, and thus the charge conjugation matrix $C$ is block off-diagonal,
\bea
C = \left(\begin{array}{cc}0~& \mathfrak{c}^a{}_ {\dot b} \\
-\mathfrak{c}_{\dot a}{}^b ~&0\end{array}\right) = (C^{AB})~, \qquad
\mathfrak{c}_{\dot a}{}^b = \mathfrak{c}^b{}_{\dot a}~.
\eea
Due to the identity
\bea
M_{\m\n} C^{-1} + C^{-1} M^{\rm T}_{\m\n} =0~,
\eea
the matrix $C=( C^{AB})$ and its inverse $C^{-1}$ are invariant tensors of $\sSpin (9,1)$. Therefore,
we can use the components of $C$ and $C^{-1}$ to convert all dotted indices into undotted ones, following the definitions:
\begin{subequations}\label{dotted-to-undotted}
\bea
\bar \c^{\dot a} &\to& \bar \c^{a} := \mathfrak{c}^a{}_ {\dot b}  \bar \c^{\dot b} ~, \\
\qquad (\s_\m)_{a \dot b} &\to& (\s_\m)_{a  b} := (\s_\m)_{a \dot c} (\mathfrak{c}^{-1})^{\dot c}{}_b~, \\
(\tilde{\s}_\m)^{\dot a b} &\to & (\tilde{\s}_\m)^{a b}
:= \mathfrak{c}^a{}_ {\dot c} (\tilde{\s}_\m)^{\dot c b}~.
\eea
 \end{subequations}
 The matrices
 \bea
\underline{\g}{}_\m = \left(\begin{array}{cc}0~& (\s_\m)_{a  b} \\
(\tilde{\s}_\m)^{ a b} ~&0\end{array}\right)
\eea
are related to the $\g$-matrices \eqref{Weyl-gamma} by the rule
\bea
\underline{\g}{}_\m = M \g_\m M^{-1} ~, \qquad
M =  \left(\begin{array}{cc}{\mathbbm 1}_{16} ~& 0 \\
0~ & \mathfrak{c}
\end{array}\right)~, \qquad M^\dagger M = {\mathbbm 1}_{32}~,
\eea
hence the matrices $\underline{\g}{}_\m $ obey the same algebra and Hermiticity condition as the matrices $\g_\m$, eq. \eqref{gamma}.
Relation \eqref{C-matrix} turns into
\bea
(\underline{\g}{}_\m)^{\rm T} = - \underline{C} \underline{\g}{}_\m \underline{C}^{-1}~,
\qquad
\underline{C} =  \left(
\begin{array}{cc}0 ~& {\mathbbm 1}_{16} \\
- {\mathbbm 1}_{16}~ & 0
\end{array}\right)~.
\eea
In what follows, we will work with $\underline{\g}{}_\m $ and their descendants
$\underline{\g}{}_{\m (k) }$ and refer to them simply
as ${\g}{}_\m $ and ${\g}{}_{\m (k)} $.

It follows from \eqref{anti-symmetry} that the matrices $\s_\m$ and
$\s_{\m(5)} := \s_{[\m_1} \tilde{\s}_{\m_2} \s_{\m_3} \tilde{\s}_{\m_4} \s_{\m_5]} $
are symmetric, while $\s_{\m(3)} := \s_{[\m_1} \tilde{\s}_{\m_2} \s_{\m_3] } $ is antisymmetric
\begin{subequations}
\bea
(\s_\m)_{a b} &=& (\s_\m)_{ba}~, \qquad (\s_{\m(5)} )_{a b} = (\s_{\m (5)})_{ba}~, \\
%\qquad
(\s_{\m(3)} )_{a b} &=&- (\s_{\m (3)})_{ba}~.
\eea
Similar properties hold for the matrices $\tilde{\s}_\m$,
$\tilde{\s}_{\m(3)} := \tilde{\s}_{[\m_1} {\s}_{\m_2} \tilde{\s}_{\m_3] } $ and
$\tilde{\s}_{\m(5)} := \tilde{\s}_{[\m_1} {\s}_{\m_2}  \tilde{\s}_{\m_3} {\s}_{\m_4} \tilde{\s}_{\m_5]} $,
specifically:
\bea
(\tilde{\s}_\m)^{a b} &=& (\tilde{\s}_\m)^{ba}~, \qquad
(\tilde{\s}_{\m(5)} )^{a b} = (\tilde{\s}^{\m (5)})^{ba}~, \\
%\qquad
(\tilde{\s}_{\m(3)} )^{a b} &=&- ( \tilde{\s}_{\m (3)})^{ba}~.
\eea
\end{subequations}
The matrices ${\s}_{\m (5)}$ and $\tilde{\s}_{\m (5)} $ are (anti) self-dual,
\bea
{\s}^*_{\m (5)} = {\s}_{\m (5)}~, \qquad \tilde{\s}^*_{\m (5)} = - \tilde{\s}_{\m (5)}~.
\eea

Matrices $\g_{\m(k)} $ are characterised by the properties
\bea
{\rm tr}\, \big( \g_{\m(k)} \g_{\n(l)} \big)= 0~, \qquad k \neq l~.
\eea
In particular,
\bea
{\rm tr}\, \big( \g_{\m} \g_{\n(5)} \big)= 0~, \qquad {\rm tr}\, \big( \g_{\m} \g_{\n(5)} \g_{11} \big)= 0~.
\eea
In the Weyl representation, these identities are equivalent to
\bea
(\tilde{\s}_\m)^{ab} (\s_{\n(5)})_{ab} =0~, \qquad ({\s}_\m)_{ab} (\tilde{\s}_{\n(5)})^{ab} =0~.
\eea

Matrices $\g_{\m(k)} $ are traceless,
\bea
{\rm tr}\, \big( \g_{\m(k)} \big)= 0~, \qquad {\rm tr}\, \big( \g_{\m(k)} \g_{11} \big)= 0~, \qquad 0< k <10~.
\eea
In the Weyl representation, these identities lead to
\bea
{\rm tr}\, \big( \s_{\m(2)} \big)= {\rm tr}\, \big( \tilde{\s}_{\m(2)} \big)=0~, \qquad {\rm tr}\, \big( \s_{\m(4)} \big)=
{\rm tr}\, \big( \tilde{\s}_{\m(4)} \big)=0~.
\eea
The following completeness relations hold
\begin{subequations}  \label{CompletenessRel}
\bea
\d_{a_1}{}^{[b_1} \d_{a_2}{}^{b_2]} &=& - \frac{1}{3! \cdot 16 }
(\tilde{\s}^{\m(3)})^{b_1 b_2} (\s_{\m(3)})_{a_1 a_2} ~, \label{CompletenessRel-a}\\
\d_{a_1}{}^{(b_1} \d_{a_2}{}^{b_2)} &=& - \frac{1}{32} \Big\{ 2
(\tilde{\s}^\m)^{b_1 b_2} (\s_\m)_{a_1 a_2}
+ \frac{1}{5! } (\tilde{\s}^{\m(5)} )^{b_1 b_2} (\s_{\m(5)})_{a_1 a_2} \Big\}~.
 \label{CompletenessRel-b}
\eea
\end{subequations}

In $d$ spacetime dimensions it holds that
\bea
\g^\n \g_{\m (k)} \g_\n = - (-1)^k (d-2k) \g_{\m(k)}~.
\label{ddimensions}
\eea
In the $d=10$ case, this implies the identities
\begin{subequations}
\bea
\g^\n \g_{\m} \g_\n &=& 8 \g_\m~, \label{gammam1}\\
\g^\n \g_{\m(3)} \g_\n &=& 4 \g_{\m (3)}~, \label{gammam3}\\\
\g^\n \g_{\m(5)} \g_\n &=&0~, \label{gammam5}
\eea
\end{subequations}
which are equivalent, in the Weyl basis, to
\begin{subequations} \label{3sigmas}
\bea
(\s^\n)_{ac} (\tilde{\s}_\m)^{cd} (\s_\n)_{db} = 8 (\s_\m)_{ab} ~,
%\label{lowering}
& \qquad &
(\tilde{\s}^\n)^{ac} ({\s}_\m)_{cd} (\tilde{\s}_\n)^{db} = 8 (\tilde{\s}_\m)^{ab} ~, \label{raising} \\
(\s^\n)_{ac} (\tilde{\s}_{\m(3)})^{cd} (\s_\n)_{db} = 4 (\s_{\m (3)})_{ab} ~,
%\label{lowering}
& \qquad &
(\tilde{\s}^\n)^{ac} ({\s}_{\m (3)})_{cd} (\tilde{\s}_\n)^{db} = 4 (\tilde{\s}_{\m (3)})^{ab} \,, \\
\s^\m \tilde{\s}_{\n(5)} \s_\m =0~,  & \qquad &
\tilde{\s}^\m \s_{\n(5)} \tilde{\s}_\m =0~.
\eea
\end{subequations}
Using the definition of the invariant tensors introduced  in \eqref{I-property1} and \eqref{I-property1-c},
the relations \eqref{3sigmas} can be rewritten as
\begin{subequations}
\bea
 (\tilde{\s}_\m)^{cd} I_{ac, db} = 8 (\s_\m)_{ab} ~,
 & \qquad &
({\s}_\m)_{cd} \tilde{I}^{ac,db}  = 8 (\tilde{\s}_\m)^{ab} ~,  \\
 (\tilde{\s}_{\m(3)})^{cd} I_{ac, db} = 4 (\s_{\m(3)} )_{ab} ~,
 & \qquad &
({\s}_{\m (3)})_{cd} \tilde{I}^{ac,db}  = 4 (\tilde{\s}_{\m (3)})^{ab} ~,  \\
 (\tilde{\s}_{\m(5)})^{cd} I_{ac, db} = 0 ~,
 & \qquad &
({\s}_{\m (5)})_{cd} \tilde{I}^{ac,db}  = 0 ~.
\eea
\end{subequations}
Contracting the completeness relations \eqref{CompletenessRel} with $I_{c_1 b_1 , b_2, c_2}$ leads to
\begin{subequations}
\bea
- 4 I_{c_1 [ a_1, a_2] c_2} &=& \frac{1}{3! } (\s^{ \m (3) } )_{a_1 a_2} (\s_{\m(3)})_{c_1 c_2} ~, \\
3 I_{c_1(a_1, a_2c_2)} &\equiv& I_{c_1 a_1, a_2 c_2} + I_{c_1 a_2, c_2 a_1} + I_{c_1 c_2, a_1 a_2}=0~.
\label{I-property2}
\eea
\end{subequations}
Another identity used in Subsection \ref{spinorform} is
 \bea \label{IIJJ}
&&- \frac 78  \tilde{I}^{f a, b_1 b_2} I_{f c, d_1 d_2}  =
\d^a_c J^{b_1b_2}_{d_1d_2} - \hf \big( \d^{b_1}_c J^{ab_2}_{d_1d_2}
+ \d^{b_2}_c J^{ab_1}_{d_1d_2} \big) \nonumber \\
&&\quad- \hf \Big[
\d^a_{d_1} J^{b_1b_2}_{cd_2} - \hf \big( \d^{b_1}_{d_1} J^{ab_2}_{cd_2}
+ \d^{b_2}_{d_1} J^{ab_1}_{cd_2} \big) \Big]
- \hf \Big[
\d^a_{d_2} J^{b_1b_2}_{cd_1} - \hf \big( \d^{b_1}_{d_2} J^{ab_2}_{cd_1}
+ \d^{b_2}_{d_2} J^{ab_1}_{cd_1} \big) \Big] \nonumber \\
&&\quad +\frac 29 \bigg\{
\d^a_c \big( \d^{b_1}_{d_1} \d^{b_2}_{d_2} +\d^{b_2}_{d_1}\d^{b_1}_{d_2} \big)
- \hf \d^{b_1}_c \big( \d^a_{d_1} \d^{b_2}_{d_2} +\d^{b_2}_{d_1} \d^a_{d_2} \big)
- \hf \d^{b_2}_c \big( \d^{b_1}_{d_1} \d^{a}_{d_2} +\d^{a}_{d_1} \d^{b_1}_{d_2} \big)
\nonumber \\
&&\quad - \hf \Big[
\d^a_{d_1} \big( \d^{b_1}_{d_2} \d^{b_2}_{c} +\d^{b_2}_{d_2}\d^{b_1}_{c} \big)
- \hf \d^{b_1}_{d_1} \big( \d^{a}_{d_2} \d^{b_2}_{c} +\d^{b_2}_{d_2}\d^{a}_{d_2} \big)
- \hf \d^{b_2}_{d_1} \big( \d^{b_1}_{d_2} \d^{a}_{c} +\d^{a}_{d_2}\d^{b_1}_{c} \big)
\Big]
\nonumber \\
&&\quad - \hf \Big[
\d^a_{d_2} \big( \d^{b_1}_{c} \d^{b_2}_{d_1} +\d^{b_2}_{c}\d^{b_1}_{d_1} \big)
- \hf \d^{b_1}_{d_2} \big( \d^{a}_{c} \d^{b_2}_{d_1} +\d^{b_2}_{c}\d^{a}_{d_1} \big)
- \hf \d^{b_2}_{d_2} \big( \d^{b_1}_{c} \d^{a}_{d_1} +\d^{a}_{c}\d^{b_1}_{d_1} \big)
\Big] \bigg\}~.~~~~~~~~
\eea

Now, let $X_{ab}= X_{ba}$ be a symmetric rank-two spinor. Applying the completeness relation
\eqref{CompletenessRel-b} allows us to represent
\bea
X_{ab} = - \frac{1}{16} {\rm tr} \big(\tilde{\s}^\m X\big) (\s_\m)_{a b}
- \frac{1}{5! \cdot 32} {\rm tr} \big(\tilde{\s}^{\m(5)} X\big) (\s_{\m(5)})_{a b} ~.
\eea
It follows that the space of symmetric rank-two spinors $X_{ab}$ is the direct sum of two Lorentz-invariant subspaces
\bea
T_{ab} = V_{ab} + G_{ab}, \qquad (\tilde{\s}_{\m(5)})^{ab} V_{ab} =0~, \qquad
(\tilde{\s}_{\m})^{ab} G_{ab} =0~.
\eea
A similar decomposition exists for symmetric rank-two spinors $Y^{ab}$.

Given a ten-vector $V^\m$, it can equivalently be described by a symmetric rank-two spinor $V_{ab}$ defined by
\bea
V_{ab}=V^\m (\s_\m)_{ab}  \quad \implies \quad V_{ab} (\tilde{\s}_{\m(5)})^{ab} =0~.
\eea
The indices of $V_{ab}$ may be raised by applying \eqref{raising}
\bea
(\tilde{\s}^\n)^{ac} V_{cd} (\tilde{\s}_\n)^{db} = 8 V^{ab} ~, \qquad V^{ab} =V^\m (\tilde{\s}_\m)^{ab} ~.
\eea
Given a self-dual five-form $F_{\m(5)}$, it is equivalently described by a symmetric rank-two spinor $F^{ab}$ defined by
\bea
F^{ab} = \frac{1}{5!} F_{\m(5)} (\tilde{\s}^{\m(5)} )^{ab} \quad \implies \quad F^{ab} (\s_\m)_{ab} =0~.
\label{Fab}
\eea
The spin-tensor $F^{ab}$ has vanishing contractions with the tensor \eqref{I-property1},
\bea
I_{ab, cd} F^{cd} =0 ~, \qquad I_{a c, d b} F^{cd} =0~.
\label{D41}
\eea

%%%%%%%%%%%%%%%%%%%%%%%%%%%%%%%%%%%%%%
%%%%%%%%%%%%%%%%%%%%%%%%%%%%%%%%%%%%%%

\section{Spinor formalism in six dimensions}
\label{AppendixE}

 In this Appendix we briefly describe the spinor formalism in six dimensions\footnote{More detailed description of this formalism can be found, e.g., in \cite{LT-M12, BKNT}.}
 and concentrate on emphasising the differences from the ten-dimensional case. The algebra of the $d=6$ gamma matrices is
 \bea
\{ \g_\m , \g_\n \} = -2 \eta_{\m\n} {\mathbbm 1}_{8}~, \qquad \m,\n= 0,1, \dots, 5~,
%\label{gamma.a}
\eea
and their Hermiticity properties are given by \eqref{gamma.b}. The $B$ and $C$ matrices are universally defined by the  relations \eqref{B-matrix} and \eqref{C-matrix}, respectively. Unlike the $d=10$ case, however,
the algebraic properties of the Hermitian matrices $B$ and $C$ differ from those in \eqref{B-matrix-symmetry}
and \eqref{C-matrix-symmetry}, specifically
\bea
B^{\rm T} = - B~, \qquad C^{\rm T} = C~.
\eea
The symmetry of $C$ implies that the $d=10$ relation \eqref{anti-symmetry} is replaced with
\bea
\big(\g_{\m(k)} C^{-1}\big)^{\rm T} =  (-1)^{\hf k(k+1)} \g_{\m(k)} C^{-1}~.
\label{6Dsymmetry}
\eea
This relation tells us that the matrices $\g_\m C^{-1}$ , $\g_{\m\n} C^{-1}$ , $\g_\m \g_7 C^{-1}$ and $\g_{7} C^{-1}$ are anti-symmetric, while the matrices $C^{-1}$, $\g_{\m\n \l }C^{-1}$ and $\g_{\m\n} \g_{7} C^{-1}$ are symmetric, where $\g_7$ is the $d=6$ analogue of the matrix $\g_5$ in four dimensions,
\bea
\g_7 := \g^0 \g^1 \dots \g^5~, \qquad (\g_7)^2 = {\mathbbm 1}_{8}~, \qquad (\g_7)^\dagger = \g_7~,
\qquad \big\{ \g_7, \g_\m\big\} =0~.
\eea

In the  Weyl representation for $\g_\m$ defined by
\bea
\g_{7} = \left(\begin{array}{cc}{\mathbbm 1}_{4} ~&0 \\0~& -{\mathbbm 1}_{4}\end{array}\right)~,
\eea
the matrices $\g_\m$ become block off-diagonal and have the same form as in eq. \eqref{Weyl-gamma}. Since $\big\{ \g_7 , C \big\} =0$ in this representation, the charge conjugation matrix also becomes block off-diagonal,
\bea
C = \left(\begin{array}{cc}0~& \mathfrak{c}^a{}_ {\dot b} \\
\mathfrak{c}_{\dot a}{}^b ~&0\end{array}\right)
%= (C^{AB})
~, \qquad
\mathfrak{c}_{\dot a}{}^b = \mathfrak{c}^b{}_{\dot a}~.
\eea
The Lorentz-invariant tensor  $ \mathfrak{c}^a{}_ {\dot b} $ and its inverse can be used to convert all dotted indices into undotted ones following the rules \eqref{dotted-to-undotted}.
As a result, one ends up with the gamma matrices
 \bea
{\g}{}_\m = \left(\begin{array}{cc}0~& (\s_\m)_{a  b} \\
(\tilde{\s}_\m)^{ a b} ~&0\end{array}\right)~,
\eea
and  for their off-diagonal blocks eq. \eqref{6Dsymmetry} implies
\bea
(\s_\m)_{a  b} = - (\s_\m)_{b a} ~, \qquad (\tilde{\s}_\m)^{ a b} = - (\tilde{\s}_\m)^{b a} ~.
\eea
It follows that the off-diagonal blocks of the matrices
 \bea
{\g}{}_{\m(3)} = \left(\begin{array}{cc}0~& (\s_{\m(3)})_{a  b} \\
(\tilde{\s}_{\m(3)})^{ a b} ~&0\end{array}\right)
\eea
are symmetric,
\bea
(\s_{\m(3)})_{a  b} =  (\s_{\m(3)})_{b a} ~, \qquad (\tilde{\s}_{\m(3)})^{ a b}
=  (\tilde{\s}_{\m(3)})^{b a} ~.
\eea
Due to the identity
\bea
\g_{\m(3)} = \frac{1}{3!} \ve_{\m(3) \n(3)} \g^{\n(3)} \g_7~,
\eea
the matrices  $ {\s}_{\m(3)}$ and $\tilde{\s}_{\m(3)}$ are (anti) self-dual,
\bea
\frac{1}{3!} \ve_{\m(3) \n(3)} \s^{\n(3)} = - \s_{\m(3)} ~, \qquad
\frac{1}{3!} \ve_{\m(3) \n(3)} \tilde{\s}^{\n(3)} =  \tilde{\s}_{\m(3)}~.
\eea

In the $d=6$ case, the relation \eqref{ddimensions} has the following important implications:
\begin{subequations}
\bea
\s^\n \tilde{\s}_{\m(3)} \s_\n =0~,
& \qquad & \tilde{\s}^\n \s_{\m(3)} \tilde{\s}_\n =0 ~, \label{A.13a} \\
\s^\n \tilde{\s}_{\m} \s_\n =4\s_\m~,  &\qquad &
 \tilde{\s}^\n \s_{\m} \tilde{\s}_\n = 4 \tilde{\s}_\m~. \label{A.13b}
\eea
\end{subequations}
Introducing Lorentz-invariant tensors
\begin{subequations}
\bea
I_{ab, cd} &:=& (\s^\n)_{ab} (\s_\n)_{cd} = I_{[ab], [cd]} = I_{cd, ab}  ~, \\
\tilde{I}^{ab, cd} &:=& (\tilde{\s}^\n)^{ab} (\tilde{\s}_\n)^{cd} = \tilde{I}^{[ab], [cd]} = \tilde{I}^{cd, ab}  ~,
\eea
\end{subequations}
they allow us to raise and lower the spinor indices of $\s_\m$ and $\tilde{\s}_\m$,
\bea
\frac14 \tilde{I}^{a c, d b} (\s_\m)_{cd} = \tilde{\s}^{ab}~, \qquad
\frac 14 {I}_{a c, d b} (\tilde{\s}_\m)^{cd} = {\s}^{ab}~,
\label{sigma-tilde-sigma}
\eea
in accordance with \eqref{A.13b}. On the other hand, eq. \eqref{A.13a} tells us that the invariant tensors
$I_{ab, cd} $ and $\tilde{I}^{ab, cd} $ are completely antisymmetric,
\bea
I_{ab, cd} = I_{[ab, cd]}~, \qquad \tilde{I}^{ab, cd} = \tilde{I}^{[ab, cd]}~.
\eea

Let $\ve^{a_1 a_2 a_3 a_4} $ and $\ve_{b_1 b_2 b_3 b_4}$ be the spinor Levi-Civita tensor and its inverse, respectively,
 \bea
 \ve^{a_1 a_2 a_3 a_4}  \ve_{b_1 b_2 b_3 b_4 }
 = 4! \d^{a_1}{}_{ [b_1} \d^{a_2}{}_{b_2} \d^{a_3}{}\d_{ b_3} \d^{a_4}{}_{b_4]} ~\implies ~
 \ve^{a_1 a_2 ac_1 c_2}  \ve_{b_1 b_2 c_1 c_2 }
 = 4 \d^{a_1}{}_{ [b_1} \d^{a_2}{}_{b_2]}~.
 \eea
These tensors are used to raise and lower the antisymmetric rank-2 spinors associated with a six-vector $V^\m$
\begin{subequations}
\bea
V^\m ~\to ~ \tilde{V}{}^{a_1 a_2} := V^\m (\tilde{\s}_\m)^{a_1 a_2} ~, & \qquad &
V^\m ~\to ~ {V}_{a_1 a_2} := V^\m ({\s}_\m)_{a_1 a_2} ~, \\
\tilde{V}{}^{a_1 a_2} = \hf \ve^{a_1a_2 c_1 c_2} V_{c_1 c_2} ~, &  \qquad &
 {V}{}_{a_1 a_2} = \hf \ve_{a_1a_2 c_1 c_2} \tilde{V}^{c_1 c_2} ~. \label{A18.b}
\eea
\end{subequations}
Comparing \eqref{A18.b} with \eqref{sigma-tilde-sigma} gives
\bea
\hf I_{a_1a_2, a_3 a_4} = \ve_{a(4)}~, \qquad
\hf \tilde{I}^{a_1a_2, a_3 a_4} = \ve^{a(4)}
~.
\eea
These relations imply that every invariant tensor with upper spinor indices may be expressed as a product of several $\tilde{I}$s.

The following completeness relations hold
\begin{subequations}
\bea
\d_a{}^{[c}\, \d_b{}^{d]} &=& \frac 14 (\tilde{\s}^\m)^{cd} (\s_\m)_{ab}~,  \\
\d_a{}^{(c}\, \d_b{}^{d)} &=& \frac{1}{3! \cdot 8} (\tilde{\s}^{\m (3)})^{cd} (\s_{\m(3)})_{ab}~.
\eea
\end{subequations}
Given a self-dual three-form $F_{\m(3)}$,  such that $\star F_{\m(3)} = F_{\m(3)}$,  it can equivalently be
described by a symmetric rank-2 spinor
\bea
F_{ab} := \frac{1}{3!} F_{\m(3)} ({\s}^{\m (3)})_{ab} ~.
\eea
The three-form is reconstructed from $F_{ab}$ by making use of the identity
\bea
\frac{1}{3! \cdot 4} {\rm tr} \big( \tilde{\s}^{\m(3)} \s_{\n(3)} \big)
= \d^{\m_1}{}_{[\n_1} \d^{\m_2}{}_{\n_2} \d^{\m_3}{}_{\n_3]} +\frac{1}{3!} \ve^{\m(3)}{}_{\n(3)}~.
\eea
The unique independent invariant in the $D=6$ case is
$$
I_4=\varepsilon^{a_1a_2a_3a_4}\varepsilon^{b_1b_2b_3b_4}F_{a_1b_1}F_{a_2b_2}F_{a_3b_3}F_{a_4b_4}\,.
$$

As a simple application of the above formalism, we construct independent invariants of a generic three-form $F_{\m(3)}$. Let $F^{(+)}$ and $F^{(-)}$ be its self-dual and anti-self-dual parts, $F= F^{(+)} +F^{(-)}$, and let $F^{(+)}_{ab}$ and $F^{(-)\,ab}$ be their spinor counterparts. There is only one invariant constructed solely from $F^{(+)}_{ab}$:
\bea\label{I4+}
I^{(+)}_4 = {\rm det} \,\big(F^{(+)}_{ab}\big)~.
\eea
And there is only one invariant constructed solely from $F^{(-) \,ab}$:
\bea
I^{(-)}_4 = {\rm det} \,\big(F^{(-)\,ab}\big)~.
\eea
Let us introduce the following $4\times 4$ matrix
\bea
G= (G_a{}^b), \qquad G_a{}^b := F^{(+)}_{ac} F^{(-) \, cb}~.
\eea
Three independent invariants can be constructed from $G$:
\bea
I_2 = {\rm tr} \, G~, \qquad
I^{(0)}_4 = {\rm tr} \,\big( G^2\big) ~, \qquad I_6 = {\rm tr} \,\big( G^3\big) ~.
\eea
The Cayley-Hamilton theorem tells us that
\bea\label{CHt}
G^4 +c_3 G^3 +c_2 G^2 +c_1 G +{\rm det}\, G \,{\mathbbm 1}
=0~,
\eea
where ${\rm det} \,\big( G-\l {\mathbbm 1} \big)
= \l^4 +c_3 \l^3 +c_2\l^2 +c_1 \l +{\rm det} \, G  $ is the characteristic polynomial. Since ${\rm det} \, G=I_4^{(+)}I_4^{(-)}$, we conclude that any invariant of the generic three-form $F_{\m(3)}$ can be expressed in terms of the invariants $I_2$, $I^{(0)}_4$, $I^{(\pm)}_4$ and $I_6$.
%%%%%%%%%%%%%%%%%%%%%%%%%%%%%%%%%%%%%

\end{appendix}

%\bibliographystyle{utphysmod2}

%\bibliography{biblio}

\providecommand{\href}[2]{#2}\begingroup\raggedright\endgroup

\end{document}